\RequirePackage{fix-cm}
\documentclass[natbib,smallextended]{svjour3}       
\smartqed  
\usepackage{graphicx}
 \journalname{SSR}
\newcommand{\aap}{{Astron. Astrophys.}}
\newcommand{\apj}{{Astrophys. J.}}
\newcommand{\apjl}{{Astrophys. J. Lett.}}
\newcommand{\apjs}{{Astrophys. J. Supp.}}

\newcommand{\aj}{Astron. J.} 
\newcommand{\mnras}{MNRAS} 
\newcommand{\nat}{Nature} 
\newcommand{\araa}{ARA\&A}

\newcommand{\pasj}{PASJ} 
 
\newcommand{\physrep}{Physics Reports}

\newcommand{\Zsun}{Z_\odot}

\def\lesssim{\mathrel{\hbox{\rlap{\hbox{\lower4pt\hbox{$\sim$}}}\hbox{$<$}}}} 
\def\gtrsim{\mathrel{\hbox{\rlap{\hbox{\lower4pt\hbox{$\sim$}}}\hbox{$>$}}}}

\begin{document}

\title{Long-Duration Gamma-Ray Burst Host Galaxies in Emission and Absorption}



\author{Daniel A. Perley    \and
        Yuu Niino           \and
        Nial R. Tanvir      \and
        Susanna D. Vergani  \and
        Johan P. U. Fynbo}


\institute{D. A. Perley \at
              Dark Cosmology Centre \\
              Niels Bohr Institute, University of Copenhagen\\
              Juliane Maries Vej 30, 2100 Copenhagen \O, Denmark\\
              \email{dperley@dark-cosmology.dk}           
           \and
           Y. Niino \at
              Division of Optical \& Infrared Astronomy  \\
              National Astronomical Observatory of Japan \\
              2-21-1 Osawa, Mitaka, Tokyo, 181-8588, Japan\\
              \email{yuu.niino@nao.ac.jp}           
           \and
           N. R. Tanvir \at
              Department of Physics and Astronomy  \\ 
              University of Leicester \\ 
              University Road, Leicester, LE1 7RH, UK \\
              \email{nrt3@leicester.ac.uk}           
           \and
           S. Vergani \at
             GEPI, Observatoire de Paris \\
             PSL Research University, CNRS, Univ. Paris Diderot \\
             Sorbonne Paris Cit\'e, Place Jules Janssen, 92195, Meudon, France \\
              \email{susanna.vergani@obspm.fr}           
           \and
           J. P. U. Fynbo \at
              Dark Cosmology Centre \\
              Niels Bohr Institute, University of Copenhagen\\
              Juliane Maries Vej 30, 2100 Copenhagen \O, Denmark\\
              \email{jfynbo@dark-cosmology.dk}
}

\date{Accepted to SSR 2015-01-20}

\maketitle

\begin{abstract}
The galaxy population hosting long-duration GRBs provides a means to constrain the progenitor and an opportunity to use these violent explosions to characterize the nature of the high-redshift universe.   Studies of GRB host galaxies in emission reveal a population of star-forming galaxies with great diversity, spanning a wide range of masses, metallicities, and redshifts.  However, as a population GRB hosts are significantly less massive and poorer in metals than the hosts of other core-collapse transients, suggesting that GRB production is only efficient at metallicities significantly below Solar.  GRBs may also prefer compact galaxies, and dense and/or central regions of galaxies, more than other types of core-collapse explosion.   Meanwhile, studies of hosts in absorption against the luminous GRB optical afterglow provide a unique means of unveiling properties of the ISM in even the faintest and most distant galaxies; these observations are helping to constrain the chemical evolution of galaxies and the properties of interstellar dust out to very high redshifts.  New ground- and space-based instrumentation, and the accumulation of larger and more carefully-selected samples, are continually enhancing our view of the GRB host population.
\end{abstract}

\section{Introduction}
\label{sec:introduction}

The study of long-duration gamma-ray burst (GRB) host galaxies is, in essence, an attempt to use each of these two different astronomical objects (GRBs and galaxies) to understand the other.    By studying the population of galaxies that produces GRBs and the locations of the GRBs inside their hosts, we hope to identify the GRB progenitor and how it is formed.  The ubiquitously star-forming nature of the GRB host population is one of several lines of evidence linking GRBs to massive stars---but how does the progenitor star evolve to its final state and why does it explode?  

The field is equally driven by interest in reversing this line of investigation to use GRBs to understand distant galaxies.  Spectroscopy of a GRB optical afterglow provides rich detail on the properties of the absorbing system in a way that is not possible with other observational methods.  In addition, since GRBs can be detected from galaxies of arbitrarily faint luminosities (even if their hosts cannot be detected in emission), they may provide a way to quantify and characterize the contributions to star formation of galaxy populations too faint, distant, or dusty to be easily studied by more traditional means.

In the following chapter we will discuss a number of ways in which GRB hosts are used to bridge the gap in understanding between GRBs and high-$z$ galaxies on both sides.  We will first present what is known about the GRB host population in emission via studies of their host galaxy population after the burst had faded, and compare the population of known GRB hosts to other distant galaxy populations.  Afterwards, we will outline the properties of GRB hosts in absorption, illustrating some of the ways in which spectroscopy of GRB afterglows can uniquely reveal the properties of distant galaxies.

\section{Long GRB Hosts in Emission}
\label{sec:grbhostemission}


\subsection{Background and Theoretical Predictions}
\label{sec:background}

Since nearly any observational technique or wavelength range that can be used to study other populations of distant galaxies can also be applied to GRB hosts, and because known GRBs span a vast redshift range ($0.008 < z < 9.4$; \citealt{Galama+1998,Cucchiara+2011}), the observational study of GRB hosts involves many different kinds of data and similar challenges as high-$z$ galaxy astronomy generally.  For most of its history these investigations have been limited to small samples of objects and subject to uncertain selection effects, but in the past five years the scene has changed dramatically as the large sample of positions provided by the Swift satellite (nearly 1000 GRBs localized within 2$^{\prime\prime}$ [\citealt{Gehrels+2004,Butler2007,Evans+2009}], and several hundred within 0.5$^{\prime\prime}$ and with known redshift [\citealt{Perley+2008grbox}]) is systematically exploited with a range of modern observatories, including powerful new instruments such as WFC3 on the Hubble Space Telescope and X-shooter at the Very Large Telescope.

The link between GRBs and the explosions of massive stars is now well-established \citep{Hjorth+2006}, so we expect GRBs to be found only in galaxies that are undergoing active star-formation (with the exception of short GRBs, which are not addressed in this review chapter).  Of course, the diversity of star-forming galaxies is vast, and any attempt to establish the expectations for \emph{which} star-forming galaxies GRBs should be found in at what frequency requires establishing assumptions about how the environment influences their rate.

It is helpful to start by imagining the simplest possible case: the assumption that the GRB rate is independent of all factors other than the overall rate of star-formation itself, such that a fixed fraction of all newly-formed stars explode as GRBs without regard to any of the chemical, physical, or other properties of the galaxy in which those stars formed.  Observationally, this implies that GRBs should stochastically sample the locations of cosmic star-formation throughout the volume of the Universe in which they can be observed.  The probability that any given galaxy will host a GRB during some period of time is proportional to its SFR.

Theoretically, there are various reasons to expect that reality may be more complicated than this.  The most frequently-invoked possibility is a dependence of the GRB rate on metallicity.  A fundamental challenge in producing a GRB is the need to eliminate the prognitor star's hydrogen envelope (which would smother the jet, and produce emission lines in the associated SN that are not observed) without spinning down the angular momentum that the central engine---a fast-rotating black hole or neutron star---needs to launch the jet in the first place.  The (initial) metallicity of a star can affect this process in several ways: a higher metal abundance produces stronger stellar winds, greater mass loss, and less interior mixing \citep{Crowther+2002,Heger+2003,Vink+2005,Hirschi+2008}.  Low metal abundance will both discourage mass (angular momentum) loss and encourage mixing of the hydrogen layer into the core, and single-star models \citep{MacFadyen+1999,Hirschi+2005,Yoon+2005,Yoon+2006,Langer+2006,Woosley+2006} robustly predict that GRBs should occur exclusively or preferentially in very metal-poor ($Z < 0.2$--$0.3\Zsun$) environments.  Binary channels offer alternative means of exchanging mass and/or angular momentum \citep{Izzard+2004,Fryer+2005,Podsiadlowski+2010}, although these may also be metal-sensitive to a lesser extent.

Other environmental factors aside from metallicity may also be expected to impact the likelihood of producing a gamma-ray burst.  Variations in the stellar IMF or close-binary fraction could lead to more massive stars and/or more close binaries, and a higher rate of GRBs, in some environments \citep{Dave2008,Wang+2011}.   Some GRB progenitor models appeal to dynamical effects between multiple stars \citep{vandenHeuvel+2013}; these would only be important in very dense environments and would cause GRBs to prefer the most intensely star-forming galaxies with many dense young star clusters.

In any case, from the observer's standpoint the general methodology to distinguish the possibilities above is straightforward: catalog the known population of GRB hosts as thoroughly as possible and measure its parameter distribution across as many metrics as possible---and then compare these to the distribution expected under the environment-independent null hypothesis (and to the predictions of specific theories, if possible) to identify what factors do and do not matter for GRB production.

Calculating this expected distribution requires detailed knowledge of galaxy evolution and cosmic star-formation at a variety of redshifts, since a detailed mapping of the star-formation history as a function of a wide variety of observables is required.  This knowledge can come directly from observations, or from theory.   Many studies simply use the observed distribution of total star-formation rate as a function of galaxy parameters in known surveys and apply empirical laws relating galaxy properties to each other (e.g. the mass-metallicity relation) to calculate the cosmic GRB rate as a function of various observable parameters \citep{Wolf+2007,Kocevski+2009,Trenti+2015} and determine the best-fit rate-dependence model.  Theoretical techniques have also been employed to avoid the dependence on observations from field surveys, which can be incomplete (or lacking entirely) at high redshifts and low galaxy luminosities.  In particular, semi-analytic models of galaxy formation \citep{Lapi+2008a,Campisi+2009a,Chisari+2010a}, and cosmological hydrodynamic simulations \citep{Nuza+2007a,Niino+2011a,Salvaterra+2013a,Elliott+2015a} have been employed. 
 However, these techniques are still limited by the uncertain physics surrounding star formation and feedback, as well as the difficulty in achieving sufficient resolution to compute internal structure within galaxies and sufficient volume to reproduce the cosmological population of galaxies at the same time. 


During this process, great care must be taken to be sure that the observed GRB host sample being analyzed is chosen without selection bias.  There are at least two obvious potential sources of bias when constructing samples.  The first is dust extinction: significant star-formation occurs in heavily dust-obscured regions \citep{Casey+2014}; if these stars produce GRBs, then their optical afterglows will be obscured and the afterglow position will need to be measured to $\sim$arcsecond precision at another wavelength (radio or X-rays) to localize the host; these data are not always available.  Furthermore, luminous GRB hosts are easier to detect and more likely to be reported (and scrutinized in the literature) than low-luminosity hosts, especially at high redshift.  For these reasons, the means by which a sample was constructed should always be kept in mind when interpreting observations.

\begin{figure*}
  \includegraphics[width=0.98\textwidth]{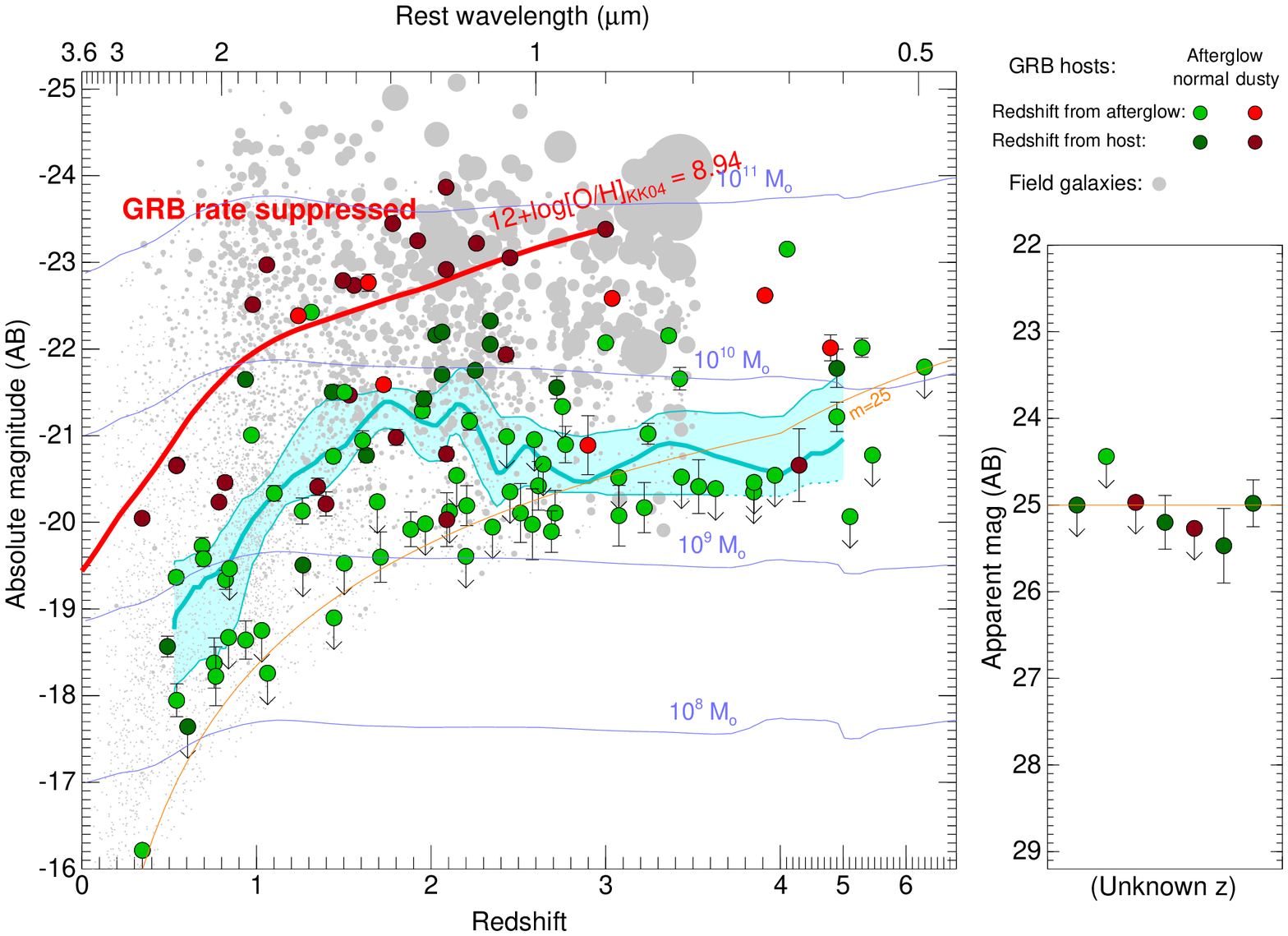}
\caption{Near-infrared luminosites of GRB hosts as a function of redshift for a large and unbiased sample of 119 GRB hosts from the SHOALS sample \citep{Perley+2015b,Perley+2015c}, compared to star-forming galaxies (gray, from \citealt{Kajisawa+2011}).  The NIR luminosity can be used as a stellar mass proxy: the horizontal blue curves indicate equivalent stellar masses.  GRBs sample galaxies of all masses and redshifts, but rarely occur in the most luminous galaxies, especially at low redshift ($z<1.5$).  This is probably because GRBs are strongly suppressed in metal-rich galaxies, leading to a soft ``upper limit'' on the host stellar mass that increases with $z$ due to the evolving mass-metallicity relation: the red curve shows the luminosity of a galaxy at the metallicity threshold of 12+log[O/H]=8.94 (the value inferred by \citealt{Perley+2015c}) as calculated using the mass-metallicity relation of \cite{Zahid+2014}.  The strong correlation between host luminosity and the degree of attenuation of the afterglow can also be seen: nearly all GRBs with dusty or ``dark'' afterglows are hosted within galaxies at the upper end of the mass distribution.}
\label{fig:spitzermag}
\end{figure*}

\subsection{Photometric Properties}

It was noticed early on that the GRB host population is relatively faint in comparison to other star-forming galaxies found at the same redshift \citep{LeFloch+2003} with a curious lack of massive, luminous galaxies with large populations of older stars among the pre-Swift (1997--2004) host population.  Of course, the number of small galaxies in the Universe is much larger than the number of massive galaxies, and the manner in which GRBs select galaxies means that their hosts are frequently expected to be faint compared to galaxies found by flux-limited surveys, even if GRB formation is environmentally-independent.   A more quantitative treament was adopted by \cite{Wolf+2007}, who used archival data from a sample of hosts at $0.2<z<1.0$ to calculate the rest-frame optical luminosity distribution for GRB hosts and compare it to a model of the star-formation weighted galaxy luminosity distibution.  Consistent with earlier reports, they found a deficiency of GRBs in the brightest galaxies: although the difference is not enormous, and if interpreted as a metallicity ``cap'' on GRB production they estimated that the limit would have to be approximately Solar (12+log[O/H]$_{\rm KK04}$ $<$ 8.7 $\pm$ 0.30).  While these studies primarily examined nearby GRBs, optical-afterglow-selected high-redshift samples from Swift \citep{Laskar+2011} also showed very few luminous hosts.

The colors and SEDs of the population have also attracted interest; the same pre-Swift samples used in the above studies appear to exhibit relatively blue colors, due to low extinctions and star-formation rates that are quite high for galaxies of their mass \citep{Chary+2002,CastroCeron+2006,CastroCeron+2010,Savaglio+2009}.  While not quantified in detail, the apparent absence (or at least rarity) of more mature or dusty systems again suggested a trend towards a galaxy population that is metal-poor.

The pre-Swift GRB sample was localized primarily via optical afterglow, and the results above were subject to the significant caveat that they omitted dust-obscured ``dark'' GRBs (see \S \ref{sec:extinction}).  Such events were already known to exist in significant numbers in the Swift sample \citep{Cenko+2009,Perley+2010}, and a handful of individual cases suggested their hosts may indeed be quite different from host galaxies selected optically (e.g., \citealt{Levan+2006,Berger+2007,CastroTirado+2007,Rol+2007,Hashimoto+2010,Hunt+2011,Svensson+2012}).  Systematic studies of large numbers of optically-reddened or optically-undetected bursts confirmed this, establishing that most (albeit not all) dust-obscured GRBs are hosted within luminous, massive, reddened systems of exactly the type that GRBs were previously found to be deficient in \citep{Kruehler+2011,Rossi+2012,Perley+2013a}.
Even so, heavily dust-obscured GRBs are not large contributors to the GRB population as a whole: events with $A_V > 1$ mag represent perhaps 20\% of the population in total \citep{Greiner+2011}.  \cite{Perley+2013a} therefore argued that they were not enough to eliminate the apparent deficiency of GRBs in luminous, red hosts.  

Given the clear impact of the selection method on the nature of the host galaxy population that is probed, any proper quantitative assessment of the host population requires selecting a single sample (of both ordinary and obscured bursts) in a uniform manner, free of afterglow-related selection biases.  These and similar considerations motivated the construction of several large uniform host-galaxy samples known by various acronyms: TOUGH (\citealt{Hjorth+2012}, containing 69 hosts), BAT6 (\citealt{Salvaterra+2012}; 58 hosts), and SHOALS (\citealt{Perley+2015b}; 119 hosts).  In each case, a subset of GRBs from the full Swift catalog is targeted for extensive host follow-up independent of afterglow properties (using the Swift XRT to localize the host if ground-based observations were not acquired) to provide as complete and unbiased a sample as possible.  The selection criteria employed in constructing the three samples are similar, usually combining a Sun-distance constraint, a foreground-extinction limit, and a requirement that Swift slewed to the position within a short timeframe in order to acquire XRT observations.  BAT6 and SHOALS also require a minimum peak flux and fluence (respectively) to exclude faint bursts although TOUGH does not, while TOUGH and BAT6 (but not SHOALS) exclude bursts near the celestial poles.  By eliminating bursts with poor-observability in this way and pursuing spectroscopic redshift measurements for any unknown-redshift GRBs which pass the cuts (which are typically dark GRBs), these techniques have been able to raise the redshift completeness from an initial value of $\sim$30\% (for the unrestricted Swift sample) to 90\% or better.

\cite{Vergani+2015} analyzed the $K$-band luminosity distribution of low-$z$ GRBs from BAT6, and found that even in this unbiased sample a strong deficiency of massive galaxies remained.  Since the NIR luminosity is a tracer of a galaxy's stellar mass, which in turn is strongly correlated with metallicity, this can be interpreted as evidence in favor of a metallicity bias, although does not necessarily rule out other models.  The role of metallicity specifically can be tested by extending this analysis to higher redshifts: thanks to the evolving mass-metallicity relation \citep{Tremonti+2004,Erb+2006}, higher-$z$ galaxies are more metal-poor at a given mass than lower-$z$ galaxies, and the mass- and luminosity trends should therefore evolve with redshift in a consistent way \citep{Fynbo+2006a,Kocevski+2009}.  

This behavior has recently been confirmed by \cite{Perley+2015c}, who measured the luminosity/mass ``ceiling'' for efficient GRB production within a galaxy and found a significant increase in this value with redshift, to a degree consistent with a simple model in which the GRB rate is uniform with respect to metallicity below a critical value of log[O/H]=8.94 (under the \citealt{Kobulnicky+2004} system) but drops by about an order of magnitude in galaxies more metal rich than this (Figure \ref{fig:spitzermag}).  This critical value is somewhat higher than, but consistent with, the estimate by \cite{Wolf+2007} obtained many years previously.  A redshift-dependent luminosity cutoff has also been observed using rest-frame UV luminosities, which should also correlate with stellar mass and metallicity (albeit with significantly more scatter).  \cite{Schulze+2015} analyzed the distribution of UV luminosities for galaxies within TOUGH and found a strong preference towards faint galaxies at low redshift but not at high redshifts.  Consistently, \cite{Greiner+2015} found good agreement between the GRB host and LBG luminosity functions at $z\sim3$.  Curiously, however, at \emph{very} high redshifts of $z>5$, \cite{Schulze+2015} again found a trend towards fainter galaxies: this is difficult to interpret and will need to be confirmed with larger samples.

\subsection{Long-Wavelength Observations}

Galaxies emit radiation beyond the UV/optical/near-infrared range as well: in particular, many of the most luminous galaxies contain copious amounts of dust that reprocesses nearly all of the UV starlight into FIR/submillimeter radiation (we will loosely refer to these types of galaxies as DSFGs, for dusty star-forming galaxies; \citealt{Casey+2014}).  The star-forming conditions in these systems are among the most extreme in the Universe, and the GRB fraction originating from these galaxies represents another way to investigate the impact of local conditions on the GRB efficiency.

The literature on this topic can be confusing:  the importance of selection bias, the limited sensitivities of observations at these wavelengths and small samples, and the controversies over the exact contribution of luminous and dusty galaxies to cosmic star-formation in the first place often lead to contradictory conclusions \citep{Berger+2003,Tanvir+2004,LeFloch+2006,Michalowski+2012}.  Studies using large samples (including dark bursts) and sensitive instruments \citep{Perley+2013a,Perley+2015a,Hunt+2014,Schady+2014,Kohn+2015} indicate that DSFG hosts are not uncommon, but still constitute a small minority ($\sim$10--30\% ) of the entire GRB host population.  This is similar to the fraction of cosmic star-formation found in these systems, so GRBs neither strongly prefer nor avoid DSFGs overall.  On the other hand, GRB-selected DSFGs do not look like field-selected DSFGs: they are much lower in mass, but younger, higher in sSFR and in dust temperature \citep{Michalowski+2008,Perley+2015a}.  This could indicate that the apparent agreement in overall rate is in fact a coincidence: GRBs avoid metal-rich, dusty DSFGs but are strongly preferred in intensely star-forming environments within the less-massive and younger subset of the population.  

More detailed investigation will be needed to confirm this trend.  The Atacama Large Millimeter Array (ALMA) is expected to play major role in characterizing obscured star formation within GRB hosts in the future, since it is able to detect submillimeter emission from even ``normal'' ($\sim L_*$) galaxies out to very high redshifts.  Indeed, ALMA observations have already detected submillimeter emission from a few GRB hosts, even out to $z\sim3$ \citep{Wang+2012,Hatsukade+2014}.  Observations of a larger sample will permit the full demographics of the population at these wavelengths to be studied in the near future.

Recently, several observers have also explored the possibility of studying GRB host galaxies in long-wavelength line emission, in particular in HI \citep{Michalowski+2015} and CO \citep{Hatsukade+2014}.  Samples of objects studied via these methods remain very small, but appear to be consistent with a general picture in which GRB hosts, and GRB sites within their hosts, are typically gas-rich but poor in metals and deficient in molecular gas.

\subsection{Spectroscopic Properties}
\label{sec:spectroscpic}

Spectroscopic measurements are more time-intensive than photometric ones and until recently have been limited to only the lowest-redshift GRBs, so the available catalog of spectroscopic host measurements is significantly smaller than what is known from photometric studies.  Nevertheless, given the theoretical arguments presented earlier, the observed metallicity of the GRB host (which is estimated via emission-line ratios measured from optical spectra) is a key parameter of interest from the point of view of distinguishing GRB models, and so spectroscopy is critical to establish a firm understanding of GRB formation.

Some GRB hosts have been known to be very metal poor almost since the earliest studies.  \cite{Stanek+2006} analyzed a small sample and proposed a metallicity-dependent limit on the GRB luminosity (i.e., luminous GRBs can be produced only by very metal-poor galaxies, although less luminous GRBs can form at higher metallicity).  This specific model has not been borne out by subsequent studies \citep{Levesque+2010eiso}: nevertheless, the more general notion that the GRB rate is lower in metal-rich galaxies does appear to be supported.  In particular, \cite{Modjaz+2008} compared the metallicities of the hosts of GRBs to those of SNe Ic-BL without GRBs (GRBs are always or almost always associated with SN Ic-BL, but SN Ic-BL are observed without associated GRBs, even off-axis ones: \citealt{Soderberg+2006}).  They found that GRBs were hosted in significantly metal-poorer systems than the SNe Ic-BL: in fact, the two populations seemed to be divided at critical value (of approximately log[O/H]=8.5 on the \citealt{Kewley+2002} scale) above which SNe Ic-BL were never associated with GRBs and below which they were always associated with GRBs.   Both the GRB and SN samples were small, however, and follow-up of larger samples of GRBs including dark bursts has demonstrated that it is not quite this simple.  In particular, several dark GRB hosts, as well as some more recent ordinary GRB hosts, have been found to be metal-rich \citep{Levesque+2010zrich,Elliott+2013} and SN Ib/c without GRBs have been found in very metal-poor environments as well \citep{Modjaz+2011}, so the most up-to-date GRB vs. star-formation and GRB vs. SN metallicity comparisons \citep{Graham+2013} do show significant overlap.  Nevertheless, a strong difference in metallicity between GRB hosts and all other classes of transient host remains---suggesting that metallicity plays a key role in encouraging the successful launch of the GRB jet.

Emission-line studies are currently undergoing a renaissance thanks largely to the prolific capabilities of a single new instrument: X-shooter at the VLT \citep{Vernet+2011}.  X-shooter has observed over 100 GRB hosts, and for the majority of these targets, multiple line detections permit chemical abundance analysis.  The existing X-shooter sample is not unbiased, but it includes both obscured/dark and unobscured GRBs, and the derived properties present a picture that is quite similar to what is inferred from the photometric properties of GRB hosts \citep{Kruehler+2015,Piranomonte+2015}.  At every redshift, GRBs in very metal-rich (super-Solar) galaxies are uncommon but present, representing about 10\% of the sample.   The median GRB host is moderately enriched (log[O/H] = 8.5), and very metal-poor hosts are again uncommon.
These properties further support the conclusion that the metallicity sensitivity of GRBs exhibits a sharp transition at around Solar metallicity that suppresses (but does not necessarily prevent) their formation in metal-rich (super-Solar) galaxies, but does not exert much impact on the rate in low versus moderate metallicity galaxies.  It does not necessarily rule out the influence of other parameters on the GRB rate---however, the small number of luminous GRB hosts with measured metallicities tend to be metal-poorer than field-selected star-forming galaxies of the same mass \citep{Levesque+2010mz,Graham+2013,Hashimoto+2015}, providing some evidence that metallicity is a dominant factor.


\begin{figure}
\includegraphics{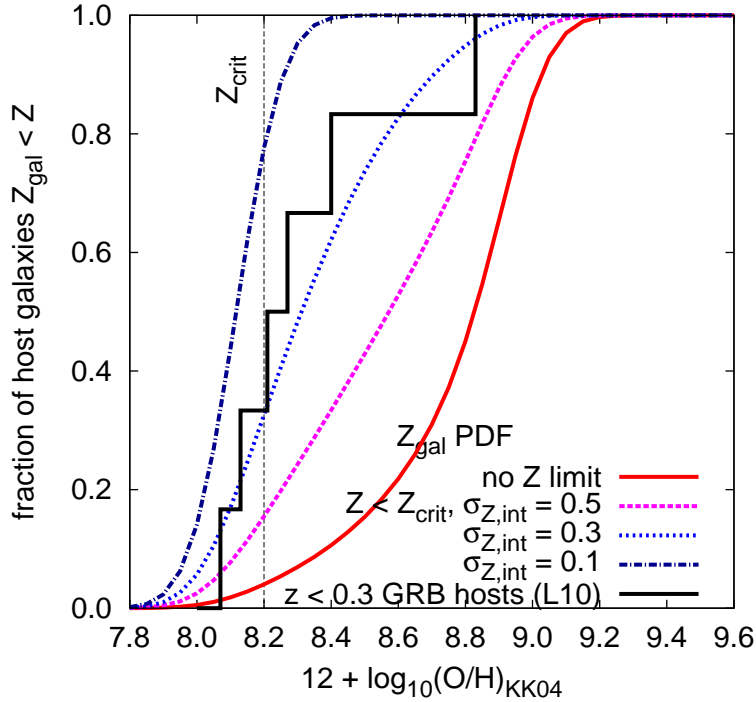}
\caption{
The metallicity (expressed as the oxygen abundance, 12+log(O/H)) distribution of GRB host galaxies predicted in \citet{Niino+2011b}.  The red solid curve represents the case  in which long GRBs occur at the same rate regardless of metallicity.  Dashed, dotted, and dot-dashed curves (magenta, blue, and dark-blue, respectively) represent cases in which only low metallicity stars  below the assumed metallicity threshold of 12+log(O/H) = 8.2 (shown with vertical dashed line) produce GRBs when they collapse, 
and stars formed in each galaxy have variation of metallicity with $\sigma = $ 0.5, 0.3, and 0.1 dex around the mean value of the galaxy, respectively.  Due to the internal metallicity variation within each galaxy,  GRB host galaxies frequently have higher metallicity than the assumed threshold.   Metallicities of observed GRB host galaxies at $z < 0.3$ (from \citealt{Levesque+2010mz}, using the calibration of \citet[][KK04]{Kobulnicky+2004}) are also shown.}
\label{fig:metaldist} 
\end{figure}

When interpreting these results, it should be kept in mind that individual galaxies also have significant internal chemical heterogeneity \citep[e.g.,][]{Afflerbach+1997a,Smartt+2001a,Cioni+2009a,Sanders+2012c}, and the measured metallicity of a GRB host galaxy may therefore be different from the actual metallicity of the GRB progenitor.  \citet{Niino+2011b} predicted the expected metallicity distribution of long GRB host galaxies under the assumption in which long GRBs trace \emph{only} low-metallicity stars using a model in which the internal variation of metallicity within galaxies is taken into account, assuming that a similar internal metallicity variation as seen in local galaxies such as the Milky Way also exists within GRB host galaxies.  They found that both the existence of a few high metallicity hosts as well as the systematically low metallicities of more typical GRB hosts can be explained at the same time, even if the GRB progenitor is exclusively a moderately metal-poor ($< 0.3 Z_\odot$) star (Figure~\ref{fig:metaldist}).  While these results have not yet been updated to incorporate data from the large, unbiased samples mentioned earlier, they clearly suggest that the ``true'' metallicity threshold for GRB production is likely to be somewhat lower than the $\sim$1 $Z_\odot$ value preferred if all galaxies are assumed to be chemically homogeneous.  Determining this value will require a careful accounting of the metallicity distribution within and between galaxies at a variety of masses and redshifts in the future.  Spatially-resolved spectroscopy (which we will discuss in the next section) will also be informative.

Spectroscopic properties other than strong-line metallicity diagnostics have also been used to investigate the GRB progenitor and rate dependence, although so far the results of these analyses have been relatively inconclusive.  For example, early samples suggested that strong Lyman-$\alpha$ emission might be much more common in GRB hosts than in star-forming galaxies generally \citep{Fynbo+2003,Jakobsson+2005}, although this was not confirmed by larger and more uniform samples \citep{MilvangJensen+2012}.  Other authors have searched for Wolf-Rayet and other very short-lived starburst features in nearby hosts in an attempt to constrain the population age \citep{Han+2010} and found some indication that the progenitor is likely to be very young, but the small and possibly biased nature of the samples suitable for this analysis makes it difficult to make firm conclusions about the entire GRB population.  Finally, detailed analysis of the rich data set provided by the large X-shooter host spectroscopic sample---which also provided information on specific star-formation rates, ionization states, and velocities--is likely to be a promising avenue for host investigations in the future (see also \citealt{Kruehler+2015}), although comparison of this type of spectroscopic data against field galaxy samples is not straightforward, since the samples of both GRB hosts and high-redshift field galaxies chosen to be targeted for deep spectroscopic follow-up are subject to complex biases.

\begin{figure*}
  \includegraphics[width=0.98\textwidth]{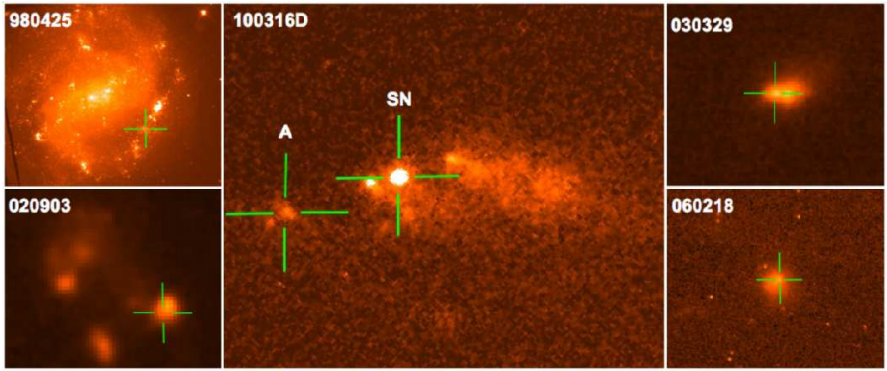}
\caption{Host galaxies of five of the nearest GRB host galaxies (from \citealt{Starling+2011}).  Most GRB hosts at $z\sim0$ are irregular, compact, star-forming galaxies---quite unlike the host population of core-collapse supernovae, which is dominated by more massive spiral galaxies \citep{Fruchter+2006}.}
\label{fig:nearbyhosts}
\end{figure*}

\subsection{Spatially Resolved Analysis}

Most GRBs are sufficiently distant that their host galaxies appear only as marginally-extended point sources in typical ground-based seeing conditions and it is difficult to extract much information about their internal structure without spaceborne observations.  However, a small number of GRB hosts are close enough that some spatially-resolved analysis has been possible \citep{Christensen+2008a,Thoene+2008,Levesque+2010zrich,Levesque+2011}; Figure \ref{fig:nearbyhosts} shows five of the nearest and best-resolved examples.  Observations of these systems have given some support to the conclusions of \cite{Niino+2011b}, showing that in at least some instances the metallicity at the GRB site may be (slightly) lower than the metallicity of the host overall.   However, even in these cases, the spatial resolution of these observations is limited to kpc scales.  Significant metallicity variation within a galaxy is observed even on scales of $< 1$ kpc in the local universe \citep[33\% of neighboring pairs of HII regions with distance $< 0.5$ kpc in M31 have different oxygen abundance by more than 0.3 dex, ][]{Sanders+2012c}.  If such small scale variation of metallicity exists in GRB host galaxies, the metallicities observed on a few kpc scale may still be different from that of the direct environment (and the progenitor) of the long GRBs \citep{Niino+2015}, a possibility that should be incorporated in future efforts.

For the vast majority of GRB hosts, only space-based observations (specifically, observations from the Hubble Space Telescope) provide a hope of spatially resolving the galaxy.  While it is not yet practical to perform spatially resolved spectroscopy of GRB hosts from space, HST observations have nevertheless been instrumental in constraining the identity of the GRB progenitor in other ways---helping to constrain its age, and providing some evidence for influences beyond metallicity on the GRB rate.


\cite{Fruchter+2006} and \cite{Wainwright+2007} analyzed the morphologies of GRB hosts within a moderately large sample of pre-Swift observations with HST, and noted that GRB hosts are smaller than other galaxies and have a higher abundance of irregular and merging systems compared to the population expected for a ``uniform'' tracer.  This behavior is expected for a metallicity-biased tracer, although morphologies are difficult to quantify, making it difficult to directly distinguish between progenitor models using this observation.  More recently, \cite{Kelly+2014} carried out a similar but more quantitative analysis, measuring the characteristic sizes (half-light radii) of galaxies in the GRB host sample in comparison to a variety of massive-stellar explosions (SN Ib/c and SN II).  Consistent with earlier work, they found that the GRB hosts are significantly smaller.  In particular, they found that they are smaller even at the same stellar mass.  As the size of a galaxy at a fixed stellar mass does not correlate with metallicity significantly, this seems to provide strong evidence that the GRB progenitor must prefer dense stellar environments \emph{in addition to} any preference towards low metallicity.

The location of a GRB within a resolved map of its host galaxy can also be used to constrain the progenitor, by studying whether or not it exhibits a preference or aversion for different types of environment on subgalactic scales.  Such an analysis was first carried out by \cite{Bloom+2002} using the nucleus-site offset distribution for a sample of 20 GRBs; they concluded that GRBs trace the galaxy light.  Analysis based on the offset distribution alone is not sensitive to azimuthal structure of the source, however, and these data were subsequently re-analyzed by \cite{Fruchter+2006} (and, later, by \citealt{Svensson+2010} and \citealt{Blanchard+2015} with expanded samples), who compared the UV surface brightness at the GRB host position compared to that of the galaxy as a whole, pixel-by-pixel.  They found a significant preference for brighter regions (pixels) of the galaxy.  Importantly, the same trend is not seen among ordinary core-collapse supernovae (but \emph{is} seen in Type Ib/c supernovae; \citealt{Kelly+2008}).

This result has been interpreted as suggestive of a very short progenitor lifetime (and, possibly, a large initial mass): on subgalactic scales, the sites of the most recent star-formation within a galaxy are expected to be brighter than locations for which a slightly longer time has passed since their formation episode, since the most massive stars have not yet exploded.  If the GRB progenitor exploded faster than the ``typical'' star responsible for a galaxy's UV luminosity---as would be expected if the progenitor star was particularly massive---a concentration towards the brightest (youngest) concentrations of young stars is expected.  In this model, the observational constraints suggest an initial mass in the range of $>20-45 M_\odot$ \citep{Larsson+2007,Raskin+2008}.

However, it is also possible that is the actual density and not age-brightness effects that are responsible: regions of particularly intense star-formation may be more likely to produce very massive stars or close massive binaries, or interactions in dense stellar nurseries could also play a fundamental role.   Indeed, the compact sizes of nearby GRB hosts noted by \cite{Kelly+2014} is not well-explained by an age effect, but this result and the concentration result are naturally explained if GRBs preferentially occur in regions of vigorous star-formation.



\subsection{GRB hosts at high redshift}
\label{sec:highzhosts}

Although studies of GRB hosts in emission are typically oriented towards using the galaxy properties to better understand the progenitor (e.g., measuring its metallicity tolerance), at the highest redshifts the direction of understanding reverses: instead, it is hoped that by observing the properties of the GRB host distribution we can better understand the properties of star-forming galaxies at that time.  Indeed, a majority of the star-formation in the universe at $z\gtrsim6$ is thought to occur in galaxies too faint to detect even in deep HST observations.  However, GRBs from arbitrarily faint and distant galaxies are readily detectable, providing us a way to confirm and constrain these predictions independently.  Of course, the GRB rate at high-$z$ will also be affected to some degree by the progenitor's dependence on metallicity.  But as high-$z$ galaxies are thought to be safely well below the 0.3--1.0 $Z_\odot$ threshold discussed above, there is reason to think that the impact of this limitation will be rather small in practice.


Until recently, searches for GRB hosts in emission had failed to find any above $z\sim5$ (e.g., \citealt{Stanway+2011,Berger+2014}).
This is consistent with the bulk of star formation occurring in very faint galaxies in this era \citep{Tanvir+2012}, indicative of a steep faint end slope of the galaxy luminosity function, as has also been found in successively deeper {\em HST} deep field campaigns (e.g., \citealt{Bouwens+2015}).  Since a steep luminosity function would allow a global star formation rate factors of several
above that seen in galaxies detected to {\em Hubble} Ultra-Deep Field depths at $z\sim8$, this reduces the tension between the far-ultraviolet radiation density required to maintain a reionized intergalactic medium, and the production of radiation from massive stars.  It should be noted that addressing this problem via GRB hosts has the important advantage that no particular form of the galaxy luminosity function is required, since it ultimately depends only on the fraction of hosts found above the detection threshold (e.g., \citealt{Trenti+2012}).

\begin{figure}
\includegraphics[width=6cm]{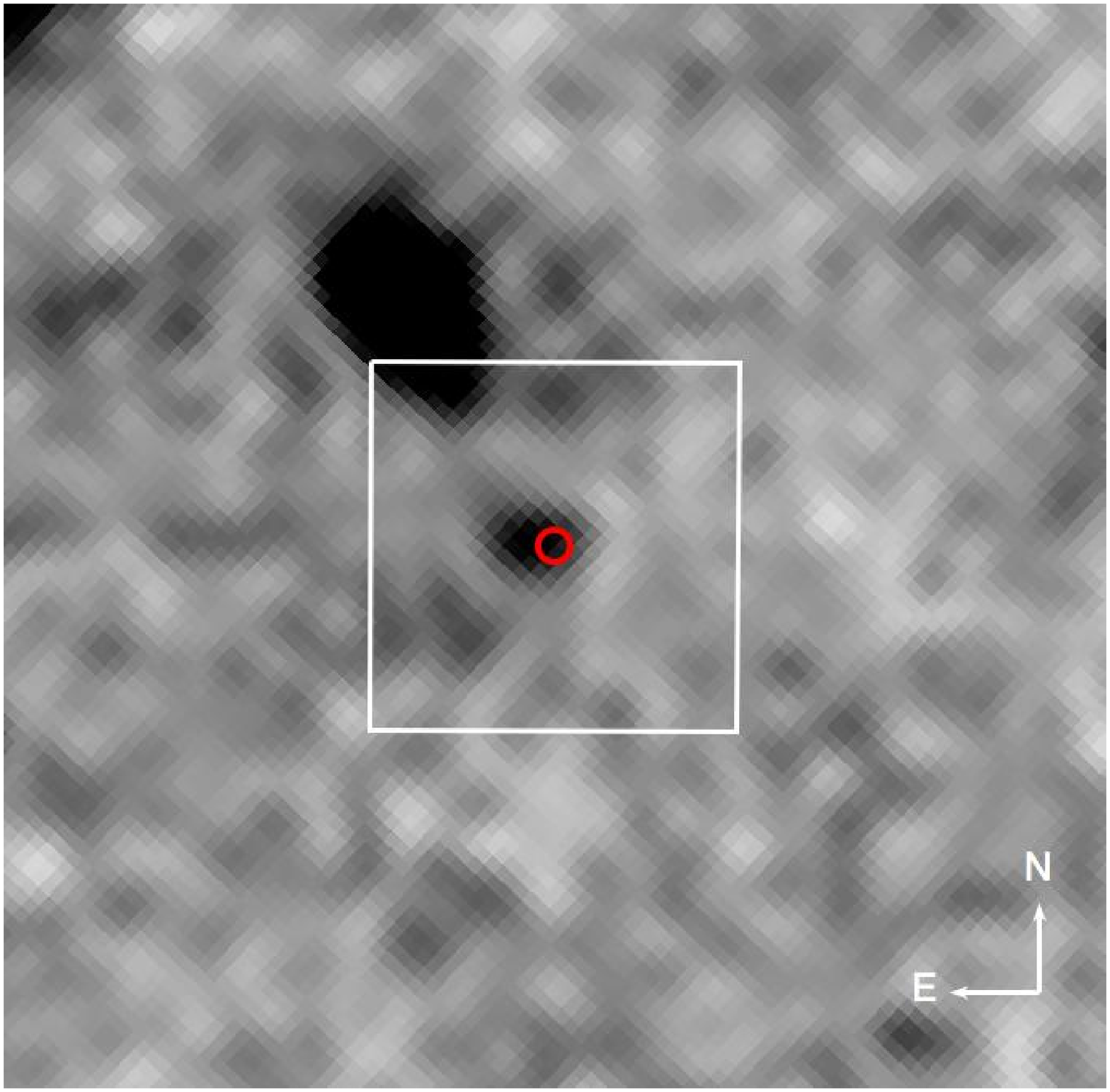}
\includegraphics[width=6cm]{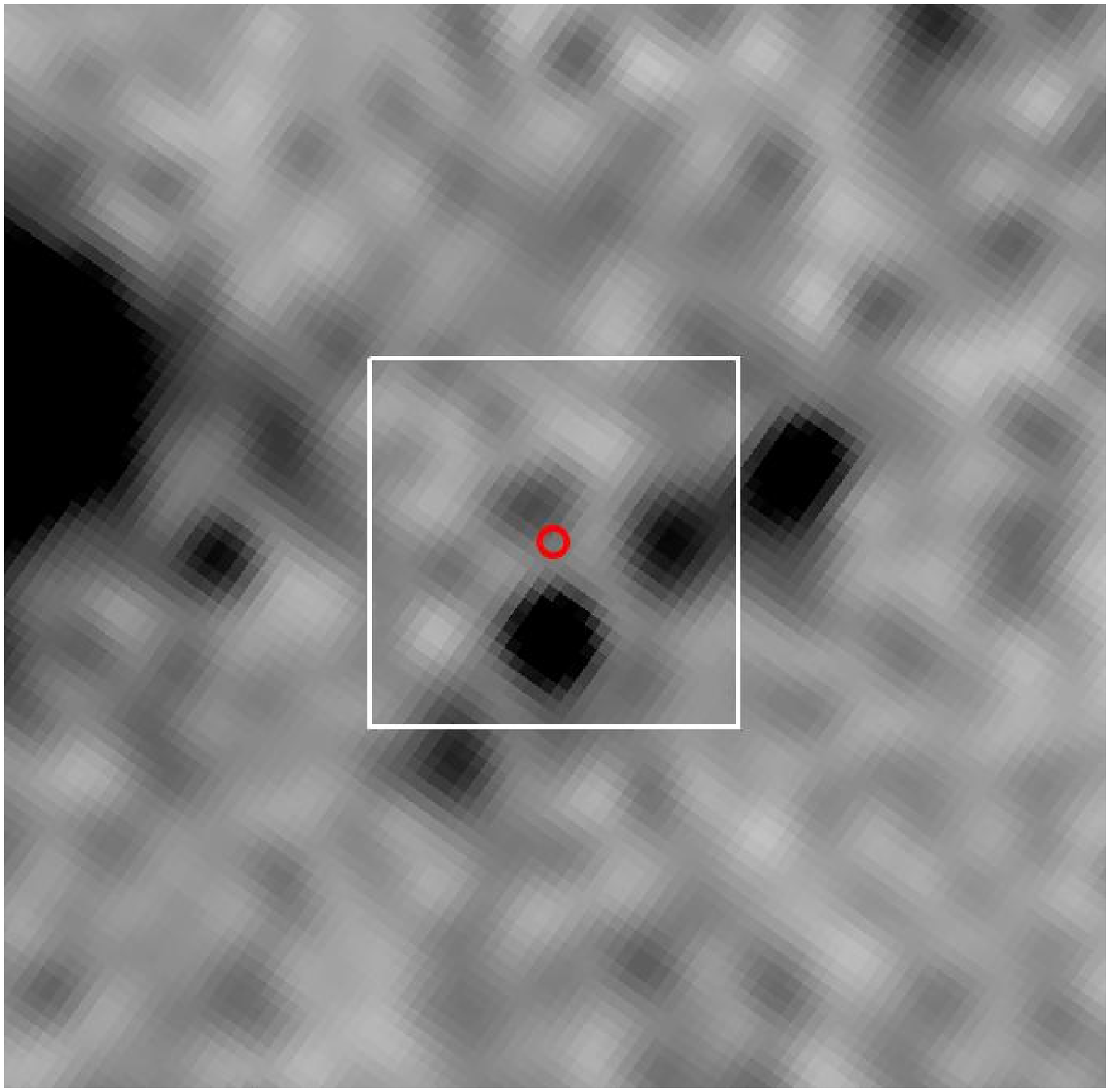}
\caption{Deep {\em HST} F140W filter images (lightly smoothed) of the locations of GRBs 050904 (left) and
140515A (right).  Both bursts had redshifts of $z=6.3$ and
in each case the host is detected at AB magnitude $\approx28$, underlying the GRB positions (indicated by red circles).
These are the first GRB host galaxies found in emission at $z>5$. The white boxes are 2 arcsec on a side.  Modified from \citealt{McGuire+2015}.}
\label{fig:highzhost}       
\end{figure}

Recently, new, deeper {\em HST} observations (program GO13831, PI: Tanvir) have for the first time 
identified two high-redshift GRB hosts, specifically of GRBs 050904 and 140515A, at $z\approx6.3$ (Fig.~\ref{fig:highzhost}; \citealt{McGuire+2015}).   The luminosity of these hosts suggests they are consistent with being faint examples of the Lyman-break galaxies found in other deep {\em HST} surveys, roughly 20\% and 10\% of $L_*$ at that redshift respectively (taking the luminosity function of \citealt{Bouwens+2015}).  In each case the afterglow spectroscopy had shown that the metallicities are  $<10$\% of Solar (\citealt{Thoene+2013,Hartoog+2015,Chornock+2013}), confirming that galaxies at this epoch are well below the threshold on GRB production inferred by lower-redshift studies.

\section{Long GRB Host Galaxies in Absorption}
\label{sec:grbhostabsorption}

Thanks to their exceptional luminosities, GRB afterglows can also be used as probes of their host galaxies in \emph{absorption}.
Gas and dust in front of the GRB selectively absorb certain wavelengths of light, imprinting their signatures onto the optical and X-ray spectra of the afterglow where they can be revealed by spectroscopic observations.  Other matter along the line of sight is also unveiled, including the inter-galactic medium (IGM) and any foreground galaxies along the sightline.

\begin{figure}
\includegraphics[width=0.7\linewidth]{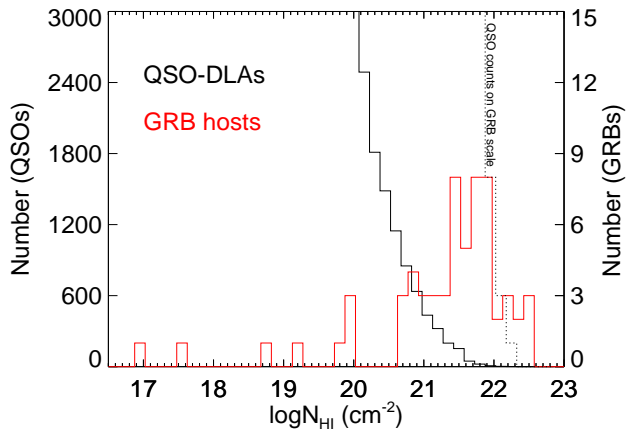}
\caption{HI column density measurements for more than 12000 QSO DLAs
and for 52 GRB absorbers. The QSO histogram is plotted with a full-drawn line relative to the left-hand
ordinate axis and with a dotted line relative to the right-hand ordinate axis. As seen, typical GRBs probe
column densities that are extremely rare or absent among QSO-DLAs.}
\label{fig:nhifig} 
\end{figure}

While this type of study can also be performed with other extragalactic continuum sources (in particular, QSOs), there are many factors that make GRBs particularly powerful absorption probes of galaxy evolution and the high redshift universe, even into the reionization epoch.  First of all, GRBs and their afterglows can be extremely luminous (much moreso than QSOs), making them accessible to current-day optical spectrographs even out to very high redshifts \citep{Lamb+2000,Bloom+2009}.  Because GRBs are associated with the death of massive stars, they are indeed expected to exist at extremely high redshifts of $10 < z < 20$ (whereas QSOs, which require time for supermassive black holes to assemble, may not), so they uniquely probe chemical enrichment far into the epoch of reionization.  But even at lower redshifts, GRBs have advantages.  Their association with massive stars also means they are located within the disks of typical star-forming galaxies and provide a more effective means to select high-density gas columns (Figure \ref{fig:nhifig}; \citealt{Pontzen+2010}).  Because the afterglow continuum is intrinsically expected to be a simple power-law, it can be used to constrain broad absorption features (such as DLA wings or, especially, extinction profiles) more reliably than QSOs, which have complex intrinsic spectra.   Finally, because the afterglow emission fades away over time, the same GRB host galaxy characterized in absorption can also be studied in emission using the techniques outlined in the previous section.  In particular, GRBs offer a unique opportunity to systematically investigate both the neutral gas using absorption lines in the the afterglow spectra \emph{and} the ionized gas using the emission lines in the host galaxy spectra.  


The transient nature of GRBs is, on the other hand, also the main disadvantage of the use of these sources as cosmological probes.  The optical/near-infrared afterglow observations necessary to provide a position of sufficient accuracy ($<0.5''$) to point an optical spectrograph, as well as the spectroscopic observations themselves, must take place as early as possible after the GRB prompt emission detection.   And unlike with QSOs, it is not possible to carry out observations over long integration times (more than a few hours) in order to increase the signal-to-noise (S/N) for faint objects.  To date, less than 30\% of GRBs have a measured redshift, and less than $\sim$10\% have a spectrum with a high S/N sufficient for detailed analysis of faint lines and uncommon chemical species.


To achieve the scientific goals mentioned above, medium/high resolution afterglow spectra with good S/N are necessary.  Observations starting as soon after the GRB trigger as possible---when the afterglow is brightest---provide the best signal and are richer in information than observations taken later, so fast response to new GRBs is essential.  For these reasons, target-of-opportunity afterglow follow-up programs exist at a variety of telescopes around the world, including Keck, Gemini, GTC, Subaru, and Magellan.   However, it is the ESO Very Large Telescope (VLT) that has been the driving force in GRB afterglow spectroscopy for much of the past decade, for a variety of reasons.  First, year-round target-of-opportunity observations (including dedicated rapid response-mode observations executed by the Observatory staff) are available.   Even more importantly, X-shooter has been just as revolutionary in the study of GRB afterglows in absorption as it has been in studies of GRB hosts in emission: its wide wavelength coverage and ideal combination of resolution and sensitivity have permitted detailed studies of the environments of even faint and high-redshift afterglows.  

\subsection{Damped Lyman Alpha Systems}
The most prominent absorption line seen in high-redshift GRB spectra is, unsurprisingly, the Lyman-$\alpha$ line of neutral hydrogen (HI) at 1217\AA, which is accessible to ground-based spectrographs at redshifts higher than approximately $z\gtrsim1.9$.  The strength of this line causes it to inevitably saturate, but the absorption column can be derived by fitting the wings of the profile using the same techniques derived for QSOs, producing a direct and accurate measurement of the neutral hydrogen column density $N_{HI}$.  In addition to this Lyman-alpha absorption, the afterglow spectra also typically reveal many metal lines at the same redshift, present in both their low and high ionization state. It is also occasionally possible to detect the absorption line from molecules such as H$_2$, as well as fine structure lines associated with gas excited by the GRB afterglow itself.  The combination of a neutral gas column and measurements of numerous metal lines can be used for a variety of purposes, described in the following sections.


\subsection{Metallicity}
\begin{figure}
\includegraphics[width=0.8\linewidth]{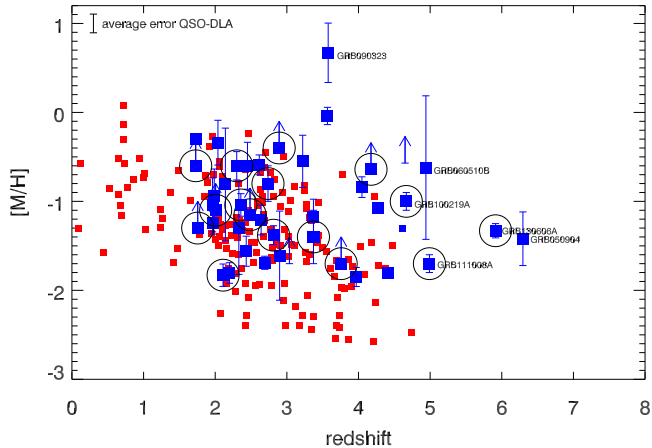}
\caption{Absorption-based metallicity measurements as a function of redshift for QSO intervening-system DLAs (red symbols) and GRB host DLAs (blue symbols).  The two populations evolve differently with redshift, probably because they probe different locations within the galaxy:  with GRB DLAs the metallicity of the inner regions of star-forming galaxies are probed, while QSO DLAs typically probe more outlying regions of foreground galaxies.  Adapted from \cite{Thoene+2013} and \cite{Sparre+2014}. }
\label{fig:dlagrbz} 
\end{figure}

The metallicity of the intervening gas towards a GRB can be directly measured using the ratio of the strengths of metal lines and HI in absorption,  even in galaxies up to the highest redshift.  For example, X-shooter observations have succeeded in measuring the metallicity of GRB\,130606A at $z=5.9$ (\citealt{Hartoog+2015}; see also \citealt{Chornock+2013} and \citealt{Totani+2014a}). By comparing systems at high redshift to others at more moderate redshifts, GRB DLAs allow us to probe the metallicity evolution of the ISM of star-forming galaxies.  The data collected to date show that GRBs explode generally in sub-solar metallicity environments---but measured metallicities are not extremely low, even at high redshift, demonstrating that galaxies were already metal enriched (see Fig.\,\ref{fig:dlagrbz}).

\subsection{Dust Depletion}
\label{sec:dustdepletion}
Refractory elements, such as Fe, Ni, and Cr, can be heavily depleted into dust grains (e.g. \citealt{Savage1996}), and thus can be missing from the gas-phase abundances. To estimate the level of depletion in the ISM (and therefore the presence of dust), the relative abundance of heavily depleted species into dust grains, as Fe, towards those undergoing little depletion, as Zn, is used. Then, using different methods (e.g.\,\citealt{Vladilo2006,Bohlin1978,Prochaska2002,De-Cia2013}), it is possible to estimate the iron dust-phase column density, the dust-to-gas ratio, and the flux attenuation.  The depletion patterns can be compared to the Galactic ones to estimate the origin of the gas and of dust, and/or to look for evolution of the dust-to-metal ratio (e.g.\,\citealt{De-Cia2013,Zafar2013}).

GRB afterglows can also be used to study the shape of the dust extinction curve, as will be discussed in Section \ref{sec:extinction}

\subsection{Molecules}
\label{sec:molecules}
\begin{figure}
\includegraphics[width=1.0\linewidth]{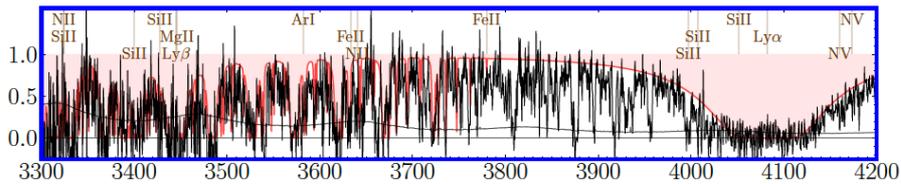}
\caption{X-shooter spectrum of GRB\,120815A between 3300 and 4200\AA\, illustrating the presence of H$_2$ absorption. The solid red line denotes the synthetic H$_2$ model (including HI Lyman-alpha and Lyman-beta absorptions). Also marked are prominent metal absorption lines previously detected in GRB-DLAs and foreground absorbers.  From \cite{Kruhler2013}; reproduced with permission from Astronomy \& Astrophysics.}
\label{fig:h2}  
\end{figure}

As GRB DLAs should probe the gas associated with star-forming regions, a direct detection of absorption lines due to molecules is expected in their afterglow spectra.  The detection of absorption lines from rotationally and vibrationally-excited H$_2$ transitions, however, has proven to be difficult:  these H$_2$ transitions are typically blended with the Lyman-alpha forest, and disentangling the often-faint H$_2$ features from the forest is much easier in spectra with high spectral resolution and S/N. This favours bright afterglows.  Physically, however, the H$_2$ column density is correlated with metallicity and dust depletion, and a large H$_2$ column is more likely for sightlines that are significantly obscured (and therefore much fainter in the rest-frame UV).  As a result, afterglows that are both bright and heavily H$_2$-absorbed are uncommon.
The combination of resolution and sensitivity available with the X-shooter spectrograph should make the detections of molecules more frequent than in the past.   Indeed, three detections of H$_2$ and one of CH$^+$ in X-shooter afterglow spectra have been published to date (Fig.\,\ref{fig:h2}; \citealt{Kruhler2013,DElia+2014,Friis+2015,Fynbo+2014}), compared to only one robust detection of molecules in GRB DLAs before the advent of X-shooter \citep{Prochaska+2009}.  

In general, the fraction of H$_2$ found along GRB sightlines seems to be quite low compared to the HI content. The comparison of the content of molecules and neutral hydrogen may be useful to better understand the processes triggering star-formation, especially if extended to HI and CO detections (or limits) obtained at millimeter and radio wavelengths with ALMA and ATCA (e.g.\,\citealt{Hatsukade+2014,Michalowski+2015}).

\subsection{Distance of the gas from the GRB}
\label{sec:gasdistance}
The GRB afterglow radiation is intense enough to have an important impact on its environment at the time of the explosion. In particular, the UV radiation ionizes the neutral gas (e.g. \citealt{Perna+2002}) and destroys molecules and dust grains up to tens of parsecs away \citep{Waxman2000,Draine2002}.  The metastable states of existing species (O I, Si II, Fe II) are populated by UV pumping followed by radiative cascade \citep{Prochaska2006,Vreeswijk2008}. 

From the presence or absence of neutral, low and high ionization metal line absorptions, and thanks to the different fine-structure and metastable line transitions (due to UV pumping followed by radiative cascade) it is possible to establish the distances of the gas clouds absorbing the afterglow light. Indeed, because the GRB afterglow fades rapidly, recombination prevails and the populations of the metastable levels change, making these absorptions vary and disappear. The detection of these time-dependent processes, with timescales ranging from seconds to days in the observer frame, can lead to interesting information on the burst itself and on the ISM of the host, in particular on the distance of the absorbing clouds.

\begin{figure}
\includegraphics[width=0.7\linewidth]{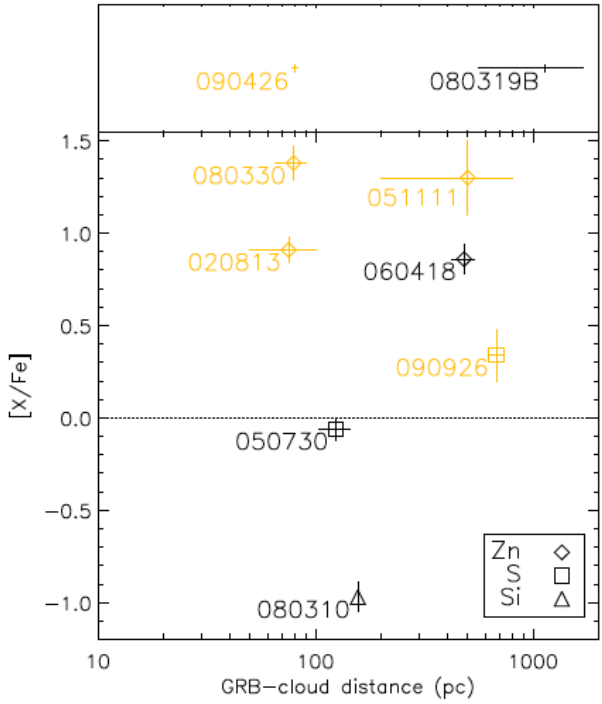}
\caption{The abundance ratio of zinc, sulphur or
silicon over iron, compared to the GRB-absorber
distance inferred from absorption-line modelling (black points come from high-resolution spectra and are more reliable). From \cite{Vreeswijk2012}.}
\label{fig:vreeswijk}
\end{figure}

To date, this analysis has been possible for $\sim10$ GRBs. It demonstrated that the gas is at least at $\sim100$\,pc from the GRB explosion site (see Fig.\,\ref{fig:vreeswijk}), and that the GRB ionizes its surrounding gas at least up to 20\,pc \citep{Vreeswijk2008,Prochaska2008,Vreeswijk2012,DElia+2014}. Therefore, through optical spectroscopy, we are not probing the close environment of the GRB progenitor, but possibly its star-forming region, the inner region of its host galaxy and its circumgalactic gas.

Many attempts to find signatures of the close (pc-scale) GRB environment have been carried out.  Unfortunately, there are no robust identifications to date, and possible evidences have been found only in a couple of cases (\citealt{Fox2008,Castro-Tirado2010}, but see \citealt{Chen2007}).  On the other hand, by studying the profile of high ionization lines, some evidence of outflows has been found \citep{Fox2008}.


\subsection{Comparison between ISM and ionized gas}
\label{sec:ismvsionized}

Once the afterglow vanishes, it is possible to carry out photometric and spectroscopic campaigns to study the properties of the GRB host galaxy.  Most exciting is the combination of the gas properties obtained by the afterglow spectroscopy and those retrieved by the host observations. A systematic study in this sense can provide important information on the physical processes of galaxy evolution and cosmological star-formation. The kinematics and geometry of the gas can be assessed as well, looking for inflow/outflow signatures. The physical state of the gas and its connection with the star-formation activity in the host galaxy can also be modelled. It is possible to obtain unique information to understand the link between the properties of the ISM and star formation activity at any redshift.

\begin{figure}[!h]
\includegraphics[width=0.85\linewidth]{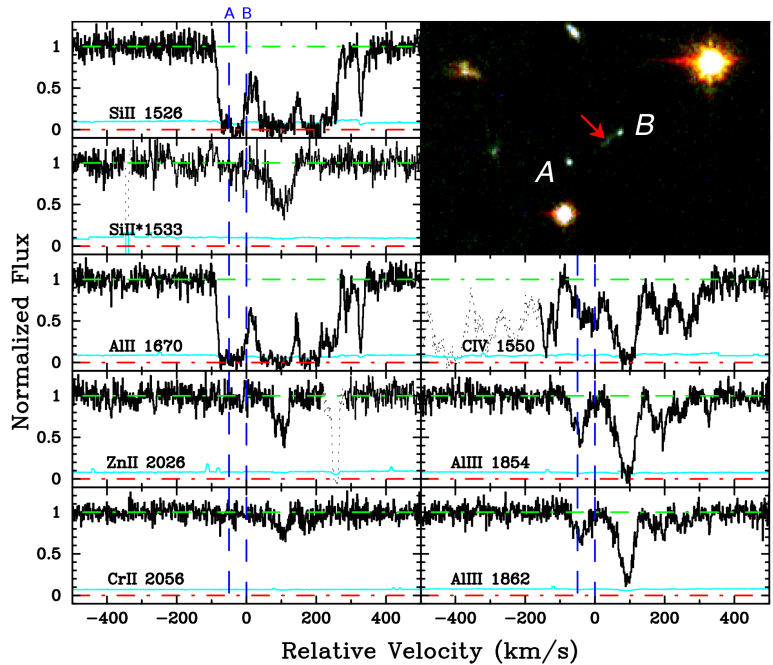}
\caption{
HST image of the field of GRB050820 at $z=2.6$, superimposed on some absorption lines from the afterglow spectrum.
The arrow indicates the afterglow position. The comparison between the velocities of the afterglow absorption lines and the
emission lines of interacting galaxies A and B (represented by blue-dashed lines) indicates that the GRB occurred in an extension of
galaxy B. Adapted from \cite{Chen2012}.}
\label{fig:chenfig}
\end{figure}

First attempts in this direction have been performed by \cite{Vergani2011a} and \cite{Chen2012} (see Fig.\,\ref{fig:chenfig}). With the larger samples of afterglow and host galaxy spectra now available, it will be possible to perform more systematic studies on this topic.


\subsection{Extinction}
\label{sec:extinction}
The issue of dust extinction in GRB afterglows has been central in the study of GRBs, their afterglows, host galaxies and birth sites since the first afterglows were detected in 1997 \citep[e.g.,][]{Groot+1998,Djorgovski+2001}.  The advantage of GRB afterglows for the study of extinction is (at least) threefold: {\it i)} afterglows can be extremely bright, allowing for actual detection of the spectra even when affected by substantial dust extinction; {\it ii)} afterglows have intrinsic relatively simple spectra consisting of power-law segments; {\it iii)} afterglows cover a very broad range of frequencies from radio to gamma-rays allowing for a study of the effects of dust over this broad range of frequencies.

A rich literature on the issue of extinction in GRB afterglows has developed over the past 20 years. Early studies focused on the issue of ``dark'' bursts \citep{Groot+1998,Djorgovski+2001,Fynbo+2001,Lazzati2002}, the name given to bursts that evaded detection in the optical.  Evidence then already suggested that these events were fairly common (perhaps as much as 50\% of the sample) and that a majority of these are partially or entirely the result of dust extinction. A study of dust-extinction in pre-Swift afterglows can be found in \citet{Kann2006}.   However, because follow-up of pre-Swift GRBs was usually delayed until the afterglow faded significantly and because observations at dust-unobscured (e.g. X-ray) wavelengths could not always be obtained, large and systematically-complete analyses of dust extinction were not possible in the early years.

Swift has allowed much larger and more complete studies to be built. First of all the concept of {\it dark burst} became more precisely defined based on the expected power-law slopes of GRB afterglows between the optical and X-ray bands \citep{Jakobsson+2004,Rol+2005,vanderHorst+2009}. Examples of single-event studies of well observed and dust extinguished dark bursts are  in \citet{Watson2006,Kruehler+2008,Eliasdottir+2009,Prochaska+2009}.  Efforts to build complete samples in particular allowed for more robust statistics on the amount of extinction towards GRBs: in particular the fraction of dark bursts (as defined, e.g., by \citealt{Jakobsson+2004}) was found to be about $\sim40\%$, and about half of these (i.e., $\sim20\%$ of GRBs) are \emph{very} obscured \citep[e.g.,][]{Cenko+2009,Perley+2009a,Fynbo+2009,Kann+2010,Greiner+2011,Hjorth+2012,Melandri+2012,Perley+2013a,Covino+2013,Littlejohns2015}.

An important, still open issue is that of photoelectric absorption and dust destruction.  \citet{GalamaandWijers01} first noted that there were cases of afterglows with very strong photoelectric absorption in the X-rays and yet apparently rather  limited extinction in the optical. This raised the issue of dust destruction \citep[e.g.,][]{Fruchter01}. The issue of photoelectric absorption in the X-rays is further discussed in several dedicated papers \citep{Behar2011,Watson12,Watson13,Starling2013,Krongold2013}, but without concensus.  A possibly related phenomenon is bursts with high dust depletion of refractory elements, but limited extinction in the optical \citep{Savaglio2003,Savaglio04,Perley+2008,Friis+2015}. 

More detailed studies of extinction curves derived from afterglow spectra are presented in several sample papers:  \citet{Starling07,Zafar2011,Schady2010,Schady2012,Covino+2013,Japelj+2015}.  Extinction curves similar to those found towards stars in the  Small Magellanic Cloud give the best fit for most GRB sightlines, but the 2175\AA \ extinction bump is detected towards a handful of bursts.



\begin{figure*}
\includegraphics[width=0.95\textwidth]{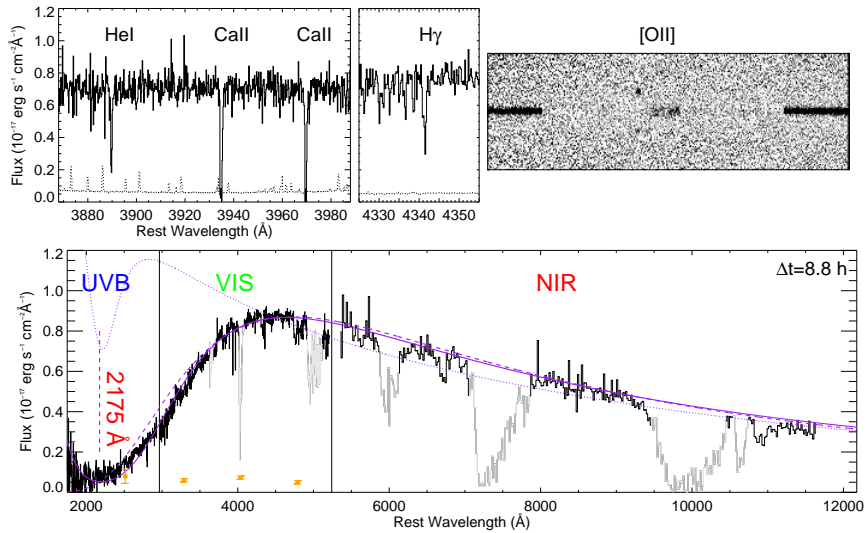}
\caption{
Spectroscopy of GRB\,140506A \citep{Fynbo+2014} and its unusual sightline.  {\bf
Bottom panel}: An X-shooter spectrum taken 8.8 hr after the burst covering the
range from about 2000 to 12000 \AA \ in the rest frame of the $z=0.889$ GRB is
plotted in black and late time host photometry is overplotted in orange. The
spectrum shows a very steep drop in the blue, which is interpreted as the result
of unusual extinction, either an extremely strong 2175~{\AA} extinction bump (the dotted
purple line shows a fit with the MW extinction curve and the full drawn line an
extreme 2175~{\AA} extinction bump) or more likely extinction similar to what
is also seen towards some type Ia supernovae and AGN (the purple dashed line,
see \citealt{Fynbo+2014} for a full discussion).  {\bf Top left:} The spectrum
also shows very unusual absorption lines including absorption from excited
helium and strong Calcium absorption as well as hydrogen Balmer lines and
molecular CH$^+$ lines (the latter not shown here, but in \citealt{Fynbo+2014})
-- never seen before in any afterglow spectra. {\bf Top right:}
Emission lines from the underlying host galaxy visible under the light of the
afterglow; here [OII].}
\label{fig:140506a}
\end{figure*}

Some GRB sightlines show evidence for extinction laws with no local analog: for example, the apparently-grey extinction towards GRB 061126 \citep{Perley+2008} or the unusual extinction towards the high-redshift GRB 071025 \citep{Perley+2010}.  Perhaps the most exotic system yet is that of GRB\,140506A, whose extinction curve has been measured using X-shooter spectroscopy \citep{Fynbo+2014}.  This sightline shows several peculiarities: {\it i)} absorption lines from excited hydrogen and helium, {\it ii)} molecular absorption from CH$^+$, and most importantly for this discussion, {\it iii)} very strong dust reddening bluewards of about 4000~\AA \ in the rest frame (see Fig.~\ref{fig:140506a}). Furthermore, the afterglow was observed on two consecutive nights and significant change in the shape of the reddening was found.  

This strong reddening cannot be fitted by extinction curves known from the Local Group and its nature remains somewhat mysterious. It has some similarity to the sightline to SN\,2014J in the starburst galaxy M82 \citep{Amanullah14}.  Along this sightline similar absorption line properties are also seen with strong Calcium absorption and absorption from CH$^+$ \citep{Ritchey14}.  Another place where similar reddening has been observed is towards some QSOs \citep[e.g.,][]{Hall2002,Karen14}: the extinction curves for some reddened QSOs appear to be very steep \citep[e.g.,][]{Fynbo2013} and one of the most remarkable QSOs in this regard, CQ0127+0114 \citep{Hall2002}, can be fitted with a similar prescription as the supernova in M82.  There is also evidence for different extinction towards the central regions of the Galaxy \citep[e.g.,][]{Sumi2004,Nishiyama2008,Nishiyama2009,Gosling2009,Nataf2015} and taken all together there seems to be a hint that extinction towards the central regions of galaxies is different and in particular steeper.

GRBs are strongly centred on the light of their host galaxies \citep{Fruchter+2006} and hence often probe the central regions of their hosts. Furthermore, there are still a significant number of bursts every year where we do not detect any afterglow light in the optical or near-IR so it appears quite plausible that this kind of unusual extinction could be more frequent for GRB sightlines than this (so far) only detection suggests. In fact, there is evidence that similar extinction was observed towards GRB\,070318 \citep{Watson2009}.  Future GRBs may provide additional constraints on unusual extinction of this type and others.

\section{Conclusions}

GRBs, their afterglows, and their host galaxies provide powerful means of studying both the nearby Universe and the most distant galaxies.  Studies of the population in different ways using different techniques (in emission and absorption, photometrically and spectroscopically, and at a wide range of wavelengths) provide unique insight into the factors governing GRB production and the nature and evolution of galaxies in the Universe.

Although the subset of GRBs with bright optical afterglows probes a low-metallicity, low-luminosity, low-mass host population compared to the population of galaxies that provide the bulk of the Universe's star-formation over most of its history, optically-obscured dark GRBs probe a redder and more massive galaxy population and are essential to provide a complete view of the GRB host population.  Even with these included, GRBs are not perfect tracers of all of cosmic star-formation: their lower abundance in high-mass and metal-rich galaxies and a tendency towards physically smaller systems point to the existence of physical factors that operate to favor the production of GRBs (relative to other end-states of massive stellar evolution) in some environments over others. 
Metallicity is a leading candidate, and a model in which the GRB rate relative to star-formation is heavily suppressed above approximately solar metallicity but largely constant at lower metallicity appears to provide good consistency with photometric and spectroscopic observations of the GRB host population.  On the other hand, other aspects of the host population are harder to explain by this model alone: the small sizes of GRB hosts, the concentration of GRB sites on the brightest star-forming regions, and the low-mass nature of the most luminous host galaxies may indicate that stellar density may play a supporting role in encouraging GRB production.   These possibilities will soon be tested by detailed analysis of the large and uniform samples that have recently been acquired, and by more sophisticated modeling and analysis techniques (including the effects of chemical inhomogeneity within galaxies).


In either case, these environmental influences are expected, and observed, to become less significant at higher redshifts, where galaxies are uniformly small, star-forming, and metal-poor.  GRBs are already being used in this way to place independent constraints on the sites of cosmic star-formation and sources of reionization.  The nondetection of most $z>5$ GRB hosts in very deep HST observations is fully consistent with the leading view that reionization was driven by galaxies too faint to be currently observed---as is the recent detection of two $z\sim6$ GRB hosts.


GRB host studies using afterglow spectroscopy provide information independent of that deriving from the host light.  Very detailed information about metal abundances, molecular content, escape of ionizing radiation, dust depletion patterns, and dust extinction curves can be inferred for host galaxies over a wide range of redshifts. Variability of fine structure lines provides unique information about the distance between the burst site and the absorbing material. In addition, information about the ionization state of the IGM can be inferred from the shape of the red damping wing of the Lyman-$\alpha$ line.  However, absorption studies still provide only an incomplete view of the overall host population because optical afterglow spectroscopy generally requires a relatively unobscured afterglow.  The few exceptions to this limitation (such as GRB 080607) suggest that sight lines towards GRBs in more mature and massive hosts are extremely rich in metals and molecules and there may be much to learn in the future when these sight lines can be studied with the next generation of extremely large telescopes.

For the past decade, Swift (supported by a wide range of ground-based observatories) has been the predominant facility for detecting and locating new GRBs, as well as a source of critical afterglow observations needed to characterize their sightlines and build complete samples.  In spite of its achievements, Swift does have some limitations---in particular, difficulty in rapidly confirming the rare but valuable high-redshift ($z>6$) GRBs.  For these purposes, a satellite a satellite capable of triggering on a larger number of high-redshift GRBs and earlier identification of the afterglow counterparts of these events through optical and NIR photometry are essential. The SVOM mission\footnote{http://smsc.cnes.fr/SVOM/} will be particularly useful in this sense, thanks to the sensitive Visible Telescope (VT) onboard and to the network of optical-NIR robotic telescopes dedicated to the photometric follow-up of SVOM GRBs. A further step forward towards a large sample of high-redshift GRB DLAs could be made with the proposed space mission THESEUS, specifically developed to detect high-redshift GRBs and observe their afterglows via on-board NIR photometry and spectroscopy.

\begin{acknowledgements}
The authors would like to thank the organizers of the ``Gamma-Ray Bursts---A Tool to Explore the Young Universe'' workshop (in particular Diego G\"otz and Maurizio Falanga) for a stimulating and productive conference and for their excellent hospitality.

Research leading to these results has received funding from the European Research
Council under the European Union's Seventh Framework Program
(FP7/2007-2013)/ERC Grant agreement no. EGGS-278202.

SV would like to thank Patrick Petitjean for accepting the duty of presenting her contribution at the ISSI-BJ meeting. 

\end{acknowledgements}

\bibliographystyle{aps-nameyear}  
\bibliographystyle{apj}

\begin{thebibliography}{195}
\ifx \bisbn   \undefined \def \bisbn  #1{ISBN #1}\fi
\ifx \binits  \undefined \def \binits#1{#1} \fi
\ifx \bauthor  \undefined \def \bauthor#1{#1} \fi
\ifx \bjtitle  \undefined \def \bjtitle#1{\textrm{#1}}\fi
\ifx \batitle  \undefined \def \batitle#1{#1} \fi
\ifx \bctitle  \undefined \def \bctitle#1{#1} \fi
\ifx \bvolume  \undefined \def \bvolume#1{\textbf{#1}}\fi
\ifx \byear  \undefined \def \byear#1{#1} \fi
\ifx \bissue  \undefined \def \bissue#1{#1} \fi
\ifx \bfpage  \undefined \def \bfpage#1{#1} \fi
\ifx \blpage  \undefined \def \blpage #1{#1} \fi
\ifx \burl  \undefined \def \burl#1{#1} \fi
\ifx \doiurl  \undefined \def \doiurl#1{#1} \fi
\ifx \betal  \undefined \def \betal{et al.} \fi
\ifx \binstitute  \undefined \def \binstitute#1{#1} \fi
\ifx \beditor  \undefined \def \beditor#1{#1} \fi
\ifx \bpublisher  \undefined \def \bpublisher#1{#1} \fi
\ifx \bbtitle  \undefined \def \bbtitle#1{\textit{#1}} \fi
\ifx \bedition  \undefined \def \bedition#1{#1} \fi
\ifx \bseriesno  \undefined \def \bseriesno#1{#1} \fi
\ifx \blocation  \undefined \def \blocation#1{#1} \fi
\ifx \bsertitle  \undefined \def \bsertitle#1{#1} \fi
\ifx \bsnm \undefined \def \bsnm#1{#1} \fi
\ifx \bsuffix \undefined \def \bsuffix#1{#1} \fi
\ifx \bparticle \undefined \def \bparticle#1{#1} \fi
\ifx \barticle \undefined \def \barticle#1{#1} \fi
\ifx \botherref \undefined \def \botherref #1{#1} \fi
\ifx \url \undefined \def \url#1{#1} \fi
\ifx \bchapter \undefined \def \bchapter#1{#1} \fi
\ifx \bbook \undefined \def \bbook#1{#1} \fi
\ifx \bcomment \undefined \def \bcomment#1{#1} \fi
\ifx \oauthor \undefined \def \oauthor#1{#1} \fi
\ifx \citeauthoryear \undefined \def \citeauthoryear#1{#1} \fi
\ifx \texttildelow  \undefined \def \texttildelow{\symbol{126}} \fi
\def \endbibitem {}
\ifx \bconflocation  \undefined \def \bconflocation#1{#1} \fi

\bibitem[\protect\citeauthoryear{{Afflerbach} et~al.}{1997}]{Afflerbach+1997a}
\begin{barticle}
\bauthor{\binits{A.} \bsnm{{Afflerbach}}},
\bauthor{\binits{E.} \bsnm{{Churchwell}}},
\bauthor{\binits{M.W.} \bsnm{{Werner}}},
\batitle{{Galactic Abundance Gradients from Infrared Fine-Structure Lines in
  Compact H II Regions}}.
\bjtitle{\apj}
\bvolume{478},
\bfpage{190}
(\byear{1997}).
doi:\doiurl{10.1086/303771}
\end{barticle}
\endbibitem

\bibitem[\protect\citeauthoryear{{Amanullah} et~al.}{2014}]{Amanullah14}
\begin{barticle}
\bauthor{\binits{R.} \bsnm{{Amanullah}}},
\bauthor{\binits{A.} \bsnm{{Goobar}}},
\bauthor{\binits{J.} \bsnm{{Johansson}}},
\bauthor{\binits{D.P.K.} \bsnm{{Banerjee}}},
\bauthor{\binits{V.} \bsnm{{Venkataraman}}},
\bauthor{\binits{V.} \bsnm{{Joshi}}},
\bauthor{\binits{N.M.} \bsnm{{Ashok}}},
\bauthor{\binits{Y.} \bsnm{{Cao}}},
\bauthor{\binits{M.M.} \bsnm{{Kasliwal}}},
\bauthor{\binits{S.R.} \bsnm{{Kulkarni}}},
\bauthor{\binits{P.E.} \bsnm{{Nugent}}},
\bauthor{\binits{T.} \bsnm{{Petrushevska}}},
\bauthor{\binits{V.} \bsnm{{Stanishev}}},
\batitle{{The Peculiar Extinction Law of SN 2014J Measured with the Hubble
  Space Telescope}}.
\bjtitle{\apjl}
\bvolume{788},
\bfpage{21}
(\byear{2014}).
doi:\doiurl{10.1088/2041-8205/788/2/L21}
\end{barticle}
\endbibitem

\bibitem[\protect\citeauthoryear{{Behar} et~al.}{2011}]{Behar2011}
\begin{barticle}
\bauthor{\binits{E.} \bsnm{{Behar}}},
\bauthor{\binits{S.} \bsnm{{Dado}}},
\bauthor{\binits{A.} \bsnm{{Dar}}},
\bauthor{\binits{A.} \bsnm{{Laor}}},
\batitle{{Can the Soft X-Ray Opacity Toward High-redshift Sources Probe the
  Missing Baryons?}}
\bjtitle{\apj}
\bvolume{734},
\bfpage{26}
(\byear{2011}).
doi:\doiurl{10.1088/0004-637X/734/1/26}
\end{barticle}
\endbibitem

\bibitem[\protect\citeauthoryear{{Berger}}{2014}]{Berger+2014}
\begin{barticle}
\bauthor{\binits{E.} \bsnm{{Berger}}},
\batitle{{Short-Duration Gamma-Ray Bursts}}.
\bjtitle{\araa}
\bvolume{52},
\bfpage{43}--\blpage{105}
(\byear{2014}).
doi:\doiurl{10.1146/annurev-astro-081913-035926}
\end{barticle}
\endbibitem

\bibitem[\protect\citeauthoryear{{Berger} et~al.}{2003}]{Berger+2003}
\begin{barticle}
\bauthor{\binits{E.} \bsnm{{Berger}}},
\bauthor{\binits{L.L.} \bsnm{{Cowie}}},
\bauthor{\binits{S.R.} \bsnm{{Kulkarni}}},
\bauthor{\binits{D.A.} \bsnm{{Frail}}},
\bauthor{\binits{H.} \bsnm{{Aussel}}},
\bauthor{\binits{A.J.} \bsnm{{Barger}}},
\batitle{{A Submillimeter and Radio Survey of Gamma-Ray Burst Host Galaxies: A
  Glimpse into the Future of Star Formation Studies}}.
\bjtitle{\apj}
\bvolume{588},
\bfpage{99}--\blpage{112}
(\byear{2003}).
doi:\doiurl{10.1086/373991}
\end{barticle}
\endbibitem

\bibitem[\protect\citeauthoryear{{Berger} et~al.}{2007}]{Berger+2007}
\begin{barticle}
\bauthor{\binits{E.} \bsnm{{Berger}}},
\bauthor{\binits{D.B.} \bsnm{{Fox}}},
\bauthor{\binits{S.R.} \bsnm{{Kulkarni}}},
\bauthor{\binits{D.A.} \bsnm{{Frail}}},
\bauthor{\binits{S.G.} \bsnm{{Djorgovski}}},
\batitle{{The ERO Host Galaxy of GRB 020127: Implications for the Metallicity
  of GRB Progenitors}}.
\bjtitle{\apj}
\bvolume{660},
\bfpage{504}--\blpage{508}
(\byear{2007}).
doi:\doiurl{10.1086/513007}
\end{barticle}
\endbibitem

\bibitem[\protect\citeauthoryear{{Blanchard} et~al.}{2015}]{Blanchard+2015}
\begin{botherref}
\oauthor{\binits{P.K.} \bsnm{{Blanchard}}},
\oauthor{\binits{E.} \bsnm{{Berger}}},
\oauthor{\binits{W.-f.} \bsnm{{Fong}}},
{The Offset and Host Light Distributions of Long Gamma-Ray Bursts: A New View
  from HST Observations of Swift Bursts}.
ArXiv e-prints
(2015)
\end{botherref}
\endbibitem

\bibitem[\protect\citeauthoryear{{Bloom} et~al.}{2002}]{Bloom+2002}
\begin{barticle}
\bauthor{\binits{J.S.} \bsnm{{Bloom}}},
\bauthor{\binits{S.R.} \bsnm{{Kulkarni}}},
\bauthor{\binits{S.G.} \bsnm{{Djorgovski}}},
\batitle{{The Observed Offset Distribution of Gamma-Ray Bursts from Their Host
  Galaxies: A Robust Clue to the Nature of the Progenitors}}.
\bjtitle{\aj}
\bvolume{123},
\bfpage{1111}--\blpage{1148}
(\byear{2002}).
doi:\doiurl{10.1086/338893}
\end{barticle}
\endbibitem

\bibitem[\protect\citeauthoryear{{Bloom} et~al.}{2009}]{Bloom+2009}
\begin{barticle}
\bauthor{\binits{J.S.} \bsnm{{Bloom}}},
\bauthor{\binits{D.A.} \bsnm{{Perley}}},
\bauthor{\binits{W.} \bsnm{{Li}}},
\bauthor{\binits{N.R.} \bsnm{{Butler}}},
\bauthor{\binits{A.A.} \bsnm{{Miller}}},
\bauthor{\binits{D.} \bsnm{{Kocevski}}},
\bauthor{\binits{D.A.} \bsnm{{Kann}}},
\bauthor{\binits{R.J.} \bsnm{{Foley}}},
\bauthor{\binits{H.-W.} \bsnm{{Chen}}},
\bauthor{\binits{A.V.} \bsnm{{Filippenko}}},
\bauthor{\binits{D.L.} \bsnm{{Starr}}},
\bauthor{\binits{B.} \bsnm{{Macomber}}},
\bauthor{\binits{J.X.} \bsnm{{Prochaska}}},
\bauthor{\binits{R.} \bsnm{{Chornock}}},
\bauthor{\binits{D.} \bsnm{{Poznanski}}},
\bauthor{\binits{S.} \bsnm{{Klose}}},
\bauthor{\binits{M.F.} \bsnm{{Skrutskie}}},
\bauthor{\binits{S.} \bsnm{{Lopez}}},
\bauthor{\binits{P.} \bsnm{{Hall}}},
\bauthor{\binits{K.} \bsnm{{Glazebrook}}},
\bauthor{\binits{C.H.} \bsnm{{Blake}}},
\batitle{{Observations of the Naked-Eye GRB 080319B: Implications of Nature's
  Brightest Explosion}}.
\bjtitle{\apj}
\bvolume{691},
\bfpage{723}--\blpage{737}
(\byear{2009}).
doi:\doiurl{10.1088/0004-637X/691/1/723}
\end{barticle}
\endbibitem

\bibitem[\protect\citeauthoryear{{Bohlin} et~al.}{1978}]{Bohlin1978}
\begin{barticle}
\bauthor{\binits{R.C.} \bsnm{{Bohlin}}},
\bauthor{\binits{B.D.} \bsnm{{Savage}}},
\bauthor{\binits{J.F.} \bsnm{{Drake}}},
\batitle{{A survey of interstellar H I from L-alpha absorption measurements.
  II}}.
\bjtitle{\apj}
\bvolume{224},
\bfpage{132}--\blpage{142}
(\byear{1978}).
doi:\doiurl{10.1086/156357}
\end{barticle}
\endbibitem

\bibitem[\protect\citeauthoryear{{Bouwens} et~al.}{2015}]{Bouwens+2015}
\begin{barticle}
\bauthor{\binits{R.J.} \bsnm{{Bouwens}}},
\bauthor{\binits{G.D.} \bsnm{{Illingworth}}},
\bauthor{\binits{P.A.} \bsnm{{Oesch}}},
\bauthor{\binits{M.} \bsnm{{Trenti}}},
\bauthor{\binits{I.} \bsnm{{Labb{\'e}}}},
\bauthor{\binits{L.} \bsnm{{Bradley}}},
\bauthor{\binits{M.} \bsnm{{Carollo}}},
\bauthor{\binits{P.G.} \bsnm{{van Dokkum}}},
\bauthor{\binits{V.} \bsnm{{Gonzalez}}},
\bauthor{\binits{B.} \bsnm{{Holwerda}}},
\bauthor{\binits{M.} \bsnm{{Franx}}},
\bauthor{\binits{L.} \bsnm{{Spitler}}},
\bauthor{\binits{R.} \bsnm{{Smit}}},
\bauthor{\binits{D.} \bsnm{{Magee}}},
\batitle{{UV Luminosity Functions at Redshifts z=4-10: 10,000 Galaxies from HST
  Legacy Fields}}.
\bjtitle{\apj}
\bvolume{803},
\bfpage{34}
(\byear{2015}).
doi:\doiurl{10.1088/0004-637X/803/1/34}
\end{barticle}
\endbibitem

\bibitem[\protect\citeauthoryear{{Butler}}{2007}]{Butler2007}
\begin{barticle}
\bauthor{\binits{N.R.} \bsnm{{Butler}}},
\batitle{{Refined Astrometry and Positions for 179 Swift X-Ray Afterglows}}.
\bjtitle{\aj}
\bvolume{133},
\bfpage{1027}--\blpage{1033}
(\byear{2007}).
doi:\doiurl{10.1086/510836}
\end{barticle}
\endbibitem

\bibitem[\protect\citeauthoryear{{Campisi} et~al.}{2009}]{Campisi+2009a}
\begin{barticle}
\bauthor{\binits{M.A.} \bsnm{{Campisi}}},
\bauthor{\binits{G.} \bsnm{{De Lucia}}},
\bauthor{\binits{L.} \bsnm{{Li}}},
\bauthor{\binits{S.} \bsnm{{Mao}}},
\bauthor{\binits{X.} \bsnm{{Kang}}},
\batitle{{Properties of long gamma-ray burst host galaxies in cosmological
  simulations}}.
\bjtitle{\mnras}
\bvolume{400},
\bfpage{1613}--\blpage{1624}
(\byear{2009}).
doi:\doiurl{10.1111/j.1365-2966.2009.15568.x}
\end{barticle}
\endbibitem

\bibitem[\protect\citeauthoryear{{Casey} et~al.}{2014}]{Casey+2014}
\begin{barticle}
\bauthor{\binits{C.M.} \bsnm{{Casey}}},
\bauthor{\binits{D.} \bsnm{{Narayanan}}},
\bauthor{\binits{A.} \bsnm{{Cooray}}},
\batitle{{Dusty star-forming galaxies at high redshift}}.
\bjtitle{\physrep}
\bvolume{541},
\bfpage{45}--\blpage{161}
(\byear{2014}).
doi:\doiurl{10.1016/j.physrep.2014.02.009}
\end{barticle}
\endbibitem

\bibitem[\protect\citeauthoryear{{Castro Cer{\'o}n}
  et~al.}{2006}]{CastroCeron+2006}
\begin{barticle}
\bauthor{\binits{J.M.} \bsnm{{Castro Cer{\'o}n}}},
\bauthor{\binits{M.J.} \bsnm{{Micha{\l}owski}}},
\bauthor{\binits{J.} \bsnm{{Hjorth}}},
\bauthor{\binits{D.} \bsnm{{Watson}}},
\bauthor{\binits{J.P.U.} \bsnm{{Fynbo}}},
\bauthor{\binits{J.} \bsnm{{Gorosabel}}},
\batitle{{Star Formation Rates and Stellar Masses in z \~{} 1 Gamma-Ray Burst
  Hosts}}.
\bjtitle{\apjl}
\bvolume{653},
\bfpage{85}--\blpage{88}
(\byear{2006}).
doi:\doiurl{10.1086/510618}
\end{barticle}
\endbibitem

\bibitem[\protect\citeauthoryear{{Castro Cer{\'o}n}
  et~al.}{2010}]{CastroCeron+2010}
\begin{barticle}
\bauthor{\binits{J.M.} \bsnm{{Castro Cer{\'o}n}}},
\bauthor{\binits{M.J.} \bsnm{{Micha{\l}owski}}},
\bauthor{\binits{J.} \bsnm{{Hjorth}}},
\bauthor{\binits{D.} \bsnm{{Malesani}}},
\bauthor{\binits{J.} \bsnm{{Gorosabel}}},
\bauthor{\binits{D.} \bsnm{{Watson}}},
\bauthor{\binits{J.P.U.} \bsnm{{Fynbo}}},
\bauthor{\binits{M.} \bsnm{{Morales Calder{\'o}n}}},
\batitle{{On the Distribution of Stellar Masses in Gamma-ray Burst Host
  Galaxies}}.
\bjtitle{\apj}
\bvolume{721},
\bfpage{1919}--\blpage{1927}
(\byear{2010}).
doi:\doiurl{10.1088/0004-637X/721/2/1919}
\end{barticle}
\endbibitem

\bibitem[\protect\citeauthoryear{{Castro-Tirado}
  et~al.}{2007}]{CastroTirado+2007}
\begin{barticle}
\bauthor{\binits{A.J.} \bsnm{{Castro-Tirado}}},
\bauthor{\binits{M.} \bsnm{{Bremer}}},
\bauthor{\binits{S.} \bsnm{{McBreen}}},
\bauthor{\binits{J.} \bsnm{{Gorosabel}}},
\bauthor{\binits{S.} \bsnm{{Guziy}}},
\bauthor{\binits{T.A.} \bsnm{{Fakthullin}}},
\bauthor{\binits{V.V.} \bsnm{{Sokolov}}},
\bauthor{\binits{R.M.} \bsnm{{Gonz{\'a}lez Delgado}}},
\bauthor{\binits{G.} \bsnm{{Bihain}}},
\bauthor{\binits{S.B.} \bsnm{{Pandey}}},
\bauthor{\binits{M.} \bsnm{{Jel{\'{\i}}nek}}},
\bauthor{\binits{A.} \bsnm{{de Ugarte Postigo}}},
\bauthor{\binits{K.} \bsnm{{Misra}}},
\bauthor{\binits{R.} \bsnm{{Sagar}}},
\bauthor{\binits{P.} \bsnm{{Bama}}},
\bauthor{\binits{A.P.} \bsnm{{Kamble}}},
\bauthor{\binits{G.C.} \bsnm{{Anupama}}},
\bauthor{\binits{J.} \bsnm{{Licandro}}},
\bauthor{\binits{D.} \bsnm{{P{\'e}rez-Ram{\'{\i}}rez}}},
\bauthor{\binits{D.} \bsnm{{Bhattacharya}}},
\bauthor{\binits{F.J.} \bsnm{{Aceituno}}},
\bauthor{\binits{R.} \bsnm{{Neri}}},
\batitle{{The dark nature of GRB 051022 and its host galaxy}}.
\bjtitle{\aap}
\bvolume{475},
\bfpage{101}--\blpage{107}
(\byear{2007}).
doi:\doiurl{10.1051/0004-6361+20066748}
\end{barticle}
\endbibitem

\bibitem[\protect\citeauthoryear{{Castro-Tirado}
  et~al.}{2010}]{Castro-Tirado2010}
\begin{barticle}
\bauthor{\binits{A.J.} \bsnm{{Castro-Tirado}}},
\bauthor{\binits{P.} \bsnm{{M{\o}ller}}},
\bauthor{\binits{G.} \bsnm{{Garc{\'{\i}}a-Segura}}},
\bauthor{\binits{J.} \bsnm{{Gorosabel}}},
\bauthor{\binits{E.} \bsnm{{P{\'e}rez}}},
\bauthor{\binits{A.} \bsnm{{de Ugarte Postigo}}},
\bauthor{\binits{E.} \bsnm{{Solano}}},
\bauthor{\binits{D.} \bsnm{{Barrado}}},
\bauthor{\binits{S.} \bsnm{{Klose}}},
\bauthor{\binits{D.A.} \bsnm{{Kann}}},
\bauthor{\binits{J.M.} \bsnm{{Castro Cer{\'o}n}}},
\bauthor{\binits{C.} \bsnm{{Kouveliotou}}},
\bauthor{\binits{J.P.U.} \bsnm{{Fynbo}}},
\bauthor{\binits{J.} \bsnm{{Hjorth}}},
\bauthor{\binits{H.} \bsnm{{Pedersen}}},
\bauthor{\binits{E.} \bsnm{{Pian}}},
\bauthor{\binits{E.} \bsnm{{Rol}}},
\bauthor{\binits{E.} \bsnm{{Palazzi}}},
\bauthor{\binits{N.} \bsnm{{Masetti}}},
\bauthor{\binits{N.R.} \bsnm{{Tanvir}}},
\bauthor{\binits{P.M.} \bsnm{{Vreeswijk}}},
\bauthor{\binits{M.I.} \bsnm{{Andersen}}},
\bauthor{\binits{A.S.} \bsnm{{Fruchter}}},
\bauthor{\binits{J.} \bsnm{{Greiner}}},
\bauthor{\binits{R.A.M.J.} \bsnm{{Wijers}}},
\bauthor{\binits{E.P.J.} \bsnm{{van den Heuvel}}},
\batitle{{GRB 021004: Tomography of a gamma-ray burst progenitor and its host
  galaxy}}.
\bjtitle{\aap}
\bvolume{517},
\bfpage{61}
(\byear{2010}).
doi:\doiurl{10.1051/0004-6361/200913966}
\end{barticle}
\endbibitem

\bibitem[\protect\citeauthoryear{{Cenko} et~al.}{2009}]{Cenko+2009}
\begin{barticle}
\bauthor{\binits{S.B.} \bsnm{{Cenko}}},
\bauthor{\binits{J.} \bsnm{{Kelemen}}},
\bauthor{\binits{F.A.} \bsnm{{Harrison}}},
\bauthor{\binits{D.B.} \bsnm{{Fox}}},
\bauthor{\binits{S.R.} \bsnm{{Kulkarni}}},
\bauthor{\binits{M.M.} \bsnm{{Kasliwal}}},
\bauthor{\binits{E.O.} \bsnm{{Ofek}}},
\bauthor{\binits{A.} \bsnm{{Rau}}},
\bauthor{\binits{A.} \bsnm{{Gal-Yam}}},
\bauthor{\binits{D.A.} \bsnm{{Frail}}},
\bauthor{\binits{D.-S.} \bsnm{{Moon}}},
\batitle{{Dark Bursts in the Swift Era: The Palomar 60 Inch-Swift Early Optical
  Afterglow Catalog}}.
\bjtitle{\apj}
\bvolume{693},
\bfpage{1484}--\blpage{1493}
(\byear{2009}).
doi:\doiurl{10.1088/0004-637X/693/2/1484}
\end{barticle}
\endbibitem

\bibitem[\protect\citeauthoryear{{Chary} et~al.}{2002}]{Chary+2002}
\begin{barticle}
\bauthor{\binits{R.} \bsnm{{Chary}}},
\bauthor{\binits{E.E.} \bsnm{{Becklin}}},
\bauthor{\binits{L.} \bsnm{{Armus}}},
\batitle{{Are Starburst Galaxies the Hosts of Gamma-Ray Bursts?}}
\bjtitle{\apj}
\bvolume{566},
\bfpage{229}--\blpage{238}
(\byear{2002}).
doi:\doiurl{10.1086/337964}
\end{barticle}
\endbibitem

\bibitem[\protect\citeauthoryear{{Chen}}{2012}]{Chen2012}
\begin{barticle}
\bauthor{\binits{H.-W.} \bsnm{{Chen}}},
\batitle{{Near-infrared spectroscopy of gamma-ray burst host galaxies at z>1.5:
  insights into host galaxy dynamics and interpretations of afterglow
  absorption spectra}}.
\bjtitle{\mnras}
\bvolume{419},
\bfpage{3039}--\blpage{3047}
(\byear{2012}).
doi:\doiurl{10.1111/j.1365-2966.2011.19944.x}
\end{barticle}
\endbibitem

\bibitem[\protect\citeauthoryear{{Chen} et~al.}{2007}]{Chen2007}
\begin{barticle}
\bauthor{\binits{H.-W.} \bsnm{{Chen}}},
\bauthor{\binits{J.X.} \bsnm{{Prochaska}}},
\bauthor{\binits{E.} \bsnm{{Ramirez-Ruiz}}},
\bauthor{\binits{J.S.} \bsnm{{Bloom}}},
\bauthor{\binits{M.} \bsnm{{Dessauges-Zavadsky}}},
\bauthor{\binits{R.J.} \bsnm{{Foley}}},
\batitle{{On the Absence of Wind Signatures in GRB Afterglow Spectra:
  Constraints on the Wolf-Rayet Winds of GRB Progenitors}}.
\bjtitle{\apj}
\bvolume{663},
\bfpage{420}--\blpage{436}
(\byear{2007}).
doi:\doiurl{10.1086/518111}
\end{barticle}
\endbibitem

\bibitem[\protect\citeauthoryear{{Chisari} et~al.}{2010}]{Chisari+2010a}
\begin{barticle}
\bauthor{\binits{N.E.} \bsnm{{Chisari}}},
\bauthor{\binits{P.B.} \bsnm{{Tissera}}},
\bauthor{\binits{L.J.} \bsnm{{Pellizza}}},
\batitle{{Host galaxies of long gamma-ray bursts in the Millennium
  Simulation}}.
\bjtitle{\mnras}
\bvolume{408},
\bfpage{647}--\blpage{656}
(\byear{2010}).
doi:\doiurl{10.1111/j.1365-2966.2010.17169.x}
\end{barticle}
\endbibitem

\bibitem[\protect\citeauthoryear{{Chornock} et~al.}{2013}]{Chornock+2013}
\begin{barticle}
\bauthor{\binits{R.} \bsnm{{Chornock}}},
\bauthor{\binits{E.} \bsnm{{Berger}}},
\bauthor{\binits{D.B.} \bsnm{{Fox}}},
\bauthor{\binits{R.} \bsnm{{Lunnan}}},
\bauthor{\binits{M.R.} \bsnm{{Drout}}},
\bauthor{\binits{W.-f.} \bsnm{{Fong}}},
\bauthor{\binits{T.} \bsnm{{Laskar}}},
\bauthor{\binits{K.C.} \bsnm{{Roth}}},
\batitle{{GRB 130606A as a Probe of the Intergalactic Medium and the
  Interstellar Medium in a Star-forming Galaxy in the First Gyr after the Big
  Bang}}.
\bjtitle{\apj}
\bvolume{774},
\bfpage{26}
(\byear{2013}).
doi:\doiurl{10.1088/0004-637X/774/1/26}
\end{barticle}
\endbibitem

\bibitem[\protect\citeauthoryear{{Christensen}
  et~al.}{2008}]{Christensen+2008a}
\begin{barticle}
\bauthor{\binits{L.} \bsnm{{Christensen}}},
\bauthor{\binits{P.M.} \bsnm{{Vreeswijk}}},
\bauthor{\binits{J.} \bsnm{{Sollerman}}},
\bauthor{\binits{C.C.} \bsnm{{Th{\"o}ne}}},
\bauthor{\binits{E.} \bsnm{{Le Floc'h}}},
\bauthor{\binits{K.} \bsnm{{Wiersema}}},
\batitle{{IFU observations of the GRB 980425/SN 1998bw host galaxy: emission
  line ratios in GRB regions}}.
\bjtitle{\aap}
\bvolume{490},
\bfpage{45}--\blpage{59}
(\byear{2008}).
doi:\doiurl{10.1051/0004-6361+200809896}
\end{barticle}
\endbibitem

\bibitem[\protect\citeauthoryear{{Cioni}}{2009}]{Cioni+2009a}
\begin{barticle}
\bauthor{\binits{M.} \bsnm{{Cioni}}},
\batitle{{The metallicity gradient as a tracer of history and structure: the
  Magellanic Clouds and M33 galaxies}}.
\bjtitle{\aap}
\bvolume{506},
\bfpage{1137}--\blpage{1146}
(\byear{2009}).
doi:\doiurl{10.1051/0004-6361/200912138}
\end{barticle}
\endbibitem

\bibitem[\protect\citeauthoryear{{Covino} et~al.}{2013}]{Covino+2013}
\begin{barticle}
\bauthor{\binits{S.} \bsnm{{Covino}}},
\bauthor{\binits{A.} \bsnm{{Melandri}}},
\bauthor{\binits{R.} \bsnm{{Salvaterra}}},
\bauthor{\binits{S.} \bsnm{{Campana}}},
\bauthor{\binits{S.D.} \bsnm{{Vergani}}},
\bauthor{\binits{M.G.} \bsnm{{Bernardini}}},
\bauthor{\binits{P.} \bsnm{{D'Avanzo}}},
\bauthor{\binits{V.} \bsnm{{D'Elia}}},
\bauthor{\binits{D.} \bsnm{{Fugazza}}},
\bauthor{\binits{G.} \bsnm{{Ghirlanda}}},
\bauthor{\binits{G.} \bsnm{{Ghisellini}}},
\bauthor{\binits{A.} \bsnm{{Gomboc}}},
\bauthor{\binits{Z.P.} \bsnm{{Jin}}},
\bauthor{\binits{T.} \bsnm{{Kr{\"u}hler}}},
\bauthor{\binits{D.} \bsnm{{Malesani}}},
\bauthor{\binits{L.} \bsnm{{Nava}}},
\bauthor{\binits{B.} \bsnm{{Sbarufatti}}},
\bauthor{\binits{G.} \bsnm{{Tagliaferri}}},
\batitle{{Dust extinctions for an unbiased sample of gamma-ray burst
  afterglows}}.
\bjtitle{\mnras}
\bvolume{432},
\bfpage{1231}--\blpage{1244}
(\byear{2013}).
doi:\doiurl{10.1093/mnras/stt540}
\end{barticle}
\endbibitem

\bibitem[\protect\citeauthoryear{{Crowther} et~al.}{2002}]{Crowther+2002}
\begin{barticle}
\bauthor{\binits{P.A.} \bsnm{{Crowther}}},
\bauthor{\binits{L.} \bsnm{{Dessart}}},
\bauthor{\binits{D.J.} \bsnm{{Hillier}}},
\bauthor{\binits{J.B.} \bsnm{{Abbott}}},
\bauthor{\binits{A.W.} \bsnm{{Fullerton}}},
\batitle{{Stellar and wind properties of LMC WC4 stars. A metallicity
  dependence for Wolf-Rayet mass-loss rates}}.
\bjtitle{\aap}
\bvolume{392},
\bfpage{653}--\blpage{669}
(\byear{2002}).
doi:\doiurl{10.1051/0004-6361:20020941}
\end{barticle}
\endbibitem

\bibitem[\protect\citeauthoryear{{Cucchiara} et~al.}{2011}]{Cucchiara+2011}
\begin{barticle}
\bauthor{\binits{A.} \bsnm{{Cucchiara}}},
\bauthor{\binits{A.J.} \bsnm{{Levan}}},
\bauthor{\binits{D.B.} \bsnm{{Fox}}},
\bauthor{\binits{N.R.} \bsnm{{Tanvir}}},
\bauthor{\binits{T.N.} \bsnm{{Ukwatta}}},
\bauthor{\binits{E.} \bsnm{{Berger}}},
\bauthor{\binits{T.} \bsnm{{Kr{\"u}hler}}},
\bauthor{\binits{A.} \bsnm{{K{\"u}pc{\"u} Yolda{\c s}}}},
\bauthor{\binits{X.F.} \bsnm{{Wu}}},
\bauthor{\binits{K.} \bsnm{{Toma}}},
\bauthor{\binits{J.} \bsnm{{Greiner}}},
\bauthor{\binits{F.E.} \bsnm{{Olivares}}},
\bauthor{\binits{A.} \bsnm{{Rowlinson}}},
\bauthor{\binits{L.} \bsnm{{Amati}}},
\bauthor{\binits{T.} \bsnm{{Sakamoto}}},
\bauthor{\binits{K.} \bsnm{{Roth}}},
\bauthor{\binits{A.} \bsnm{{Stephens}}},
\bauthor{\binits{A.} \bsnm{{Fritz}}},
\bauthor{\binits{J.P.U.} \bsnm{{Fynbo}}},
\bauthor{\binits{J.} \bsnm{{Hjorth}}},
\bauthor{\binits{D.} \bsnm{{Malesani}}},
\bauthor{\binits{P.} \bsnm{{Jakobsson}}},
\bauthor{\binits{K.} \bsnm{{Wiersema}}},
\bauthor{\binits{P.T.} \bsnm{{O'Brien}}},
\bauthor{\binits{A.M.} \bsnm{{Soderberg}}},
\bauthor{\binits{R.J.} \bsnm{{Foley}}},
\bauthor{\binits{A.S.} \bsnm{{Fruchter}}},
\bauthor{\binits{J.} \bsnm{{Rhoads}}},
\bauthor{\binits{R.E.} \bsnm{{Rutledge}}},
\bauthor{\binits{B.P.} \bsnm{{Schmidt}}},
\bauthor{\binits{M.A.} \bsnm{{Dopita}}},
\bauthor{\binits{P.} \bsnm{{Podsiadlowski}}},
\bauthor{\binits{R.} \bsnm{{Willingale}}},
\bauthor{\binits{C.} \bsnm{{Wolf}}},
\bauthor{\binits{S.R.} \bsnm{{Kulkarni}}},
\bauthor{\binits{P.} \bsnm{{D'Avanzo}}},
\batitle{{A Photometric Redshift of z \~{} 9.4 for GRB 090429B}}.
\bjtitle{\apj}
\bvolume{736},
\bfpage{7}
(\byear{2011}).
doi:\doiurl{10.1088/0004-637X/736/1/7}
\end{barticle}
\endbibitem

\bibitem[\protect\citeauthoryear{{Dav{\'e}}}{2008}]{Dave2008}
\begin{barticle}
\bauthor{\binits{R.} \bsnm{{Dav{\'e}}}},
\batitle{{The galaxy stellar mass-star formation rate relation: evidence for an
  evolving stellar initial mass function?}}
\bjtitle{\mnras}
\bvolume{385},
\bfpage{147}--\blpage{160}
(\byear{2008}).
doi:\doiurl{10.1111/j.1365-2966.2008.12866.x}
\end{barticle}
\endbibitem

\bibitem[\protect\citeauthoryear{{De Cia} et~al.}{2013}]{De-Cia2013}
\begin{barticle}
\bauthor{\binits{A.} \bsnm{{De Cia}}},
\bauthor{\binits{C.} \bsnm{{Ledoux}}},
\bauthor{\binits{S.} \bsnm{{Savaglio}}},
\bauthor{\binits{P.} \bsnm{{Schady}}},
\bauthor{\binits{P.M.} \bsnm{{Vreeswijk}}},
\batitle{{Dust-to-metal ratios in damped Lyman-{$\alpha$} absorbers. Fresh
  clues to the origins of dust and optical extinction towards {$\gamma$}-ray
  bursts}}.
\bjtitle{\aap}
\bvolume{560},
\bfpage{88}
(\byear{2013}).
doi:\doiurl{10.1051/0004-6361/201321834}
\end{barticle}
\endbibitem

\bibitem[\protect\citeauthoryear{{D'Elia} et~al.}{2014}]{DElia+2014}
\begin{barticle}
\bauthor{\binits{V.} \bsnm{{D'Elia}}},
\bauthor{\binits{J.P.U.} \bsnm{{Fynbo}}},
\bauthor{\binits{P.} \bsnm{{Goldoni}}},
\bauthor{\binits{S.} \bsnm{{Covino}}},
\bauthor{\binits{A.} \bsnm{{de Ugarte Postigo}}},
\bauthor{\binits{C.} \bsnm{{Ledoux}}},
\bauthor{\binits{F.} \bsnm{{Calura}}},
\bauthor{\binits{J.} \bsnm{{Gorosabel}}},
\bauthor{\binits{D.} \bsnm{{Malesani}}},
\bauthor{\binits{F.} \bsnm{{Matteucci}}},
\bauthor{\binits{R.} \bsnm{{S{\'a}nchez-Ram{\'{\i}}rez}}},
\bauthor{\binits{S.} \bsnm{{Savaglio}}},
\bauthor{\binits{A.J.} \bsnm{{Castro-Tirado}}},
\bauthor{\binits{O.E.} \bsnm{{Hartoog}}},
\bauthor{\binits{L.} \bsnm{{Kaper}}},
\bauthor{\binits{T.} \bsnm{{Mu{\~n}oz-Darias}}},
\bauthor{\binits{E.} \bsnm{{Pian}}},
\bauthor{\binits{S.} \bsnm{{Piranomonte}}},
\bauthor{\binits{G.} \bsnm{{Tagliaferri}}},
\bauthor{\binits{N.} \bsnm{{Tanvir}}},
\bauthor{\binits{S.D.} \bsnm{{Vergani}}},
\bauthor{\binits{D.J.} \bsnm{{Watson}}},
\bauthor{\binits{D.} \bsnm{{Xu}}},
\batitle{{VLT/X-shooter spectroscopy of the GRB 120327A afterglow}}.
\bjtitle{\aap}
\bvolume{564},
\bfpage{38}
(\byear{2014}).
doi:\doiurl{10.1051/0004-6361/201323057}
\end{barticle}
\endbibitem

\bibitem[\protect\citeauthoryear{{Djorgovski} et~al.}{2001}]{Djorgovski+2001}
\begin{barticle}
\bauthor{\binits{S.G.} \bsnm{{Djorgovski}}},
\bauthor{\binits{D.A.} \bsnm{{Frail}}},
\bauthor{\binits{S.R.} \bsnm{{Kulkarni}}},
\bauthor{\binits{J.S.} \bsnm{{Bloom}}},
\bauthor{\binits{S.C.} \bsnm{{Odewahn}}},
\bauthor{\binits{A.} \bsnm{{Diercks}}},
\batitle{{The Afterglow and the Host Galaxy of the Dark Burst GRB 970828}}.
\bjtitle{\apj}
\bvolume{562},
\bfpage{654}--\blpage{663}
(\byear{2001}).
doi:\doiurl{10.1086/323845}
\end{barticle}
\endbibitem

\bibitem[\protect\citeauthoryear{{Draine} and {Hao}}{2002}]{Draine2002}
\begin{barticle}
\bauthor{\binits{B.T.} \bsnm{{Draine}}},
\bauthor{\binits{L.} \bsnm{{Hao}}},
\batitle{{Gamma-Ray Burst in a Molecular Cloud: Destruction of Dust and H$_{2}$
  and the Emergent Spectrum}}.
\bjtitle{\apj}
\bvolume{569},
\bfpage{780}--\blpage{791}
(\byear{2002}).
doi:\doiurl{10.1086/339394}
\end{barticle}
\endbibitem

\bibitem[\protect\citeauthoryear{{El{\'{\i}}asd{\'o}ttir}
  et~al.}{2009}]{Eliasdottir+2009}
\begin{barticle}
\bauthor{\binits{{\'A}.} \bsnm{{El{\'{\i}}asd{\'o}ttir}}},
\bauthor{\binits{J.P.U.} \bsnm{{Fynbo}}},
\bauthor{\binits{J.} \bsnm{{Hjorth}}},
\bauthor{\binits{C.} \bsnm{{Ledoux}}},
\bauthor{\binits{D.J.} \bsnm{{Watson}}},
\bauthor{\binits{A.C.} \bsnm{{Andersen}}},
\bauthor{\binits{D.} \bsnm{{Malesani}}},
\bauthor{\binits{P.M.} \bsnm{{Vreeswijk}}},
\bauthor{\binits{J.X.} \bsnm{{Prochaska}}},
\bauthor{\binits{J.} \bsnm{{Sollerman}}},
\bauthor{\binits{A.O.} \bsnm{{Jaunsen}}},
\batitle{{Dust Extinction in High-z Galaxies with Gamma-Ray Burst Afterglow
  Spectroscopy: The 2175 {\AA} Feature at z = 2.45}}.
\bjtitle{\apj}
\bvolume{697},
\bfpage{1725}--\blpage{1740}
(\byear{2009}).
doi:\doiurl{10.1088/0004-637X/697/2/1725}
\end{barticle}
\endbibitem

\bibitem[\protect\citeauthoryear{{Elliott} et~al.}{2013}]{Elliott+2013}
\begin{barticle}
\bauthor{\binits{J.} \bsnm{{Elliott}}},
\bauthor{\binits{T.} \bsnm{{Kr{\"u}hler}}},
\bauthor{\binits{J.} \bsnm{{Greiner}}},
\bauthor{\binits{S.} \bsnm{{Savaglio}}},
\bauthor{\binits{F.} \bsnm{{Olivares}}},
\bauthor{\binits{E.A.} \bsnm{{Rau}}},
\bauthor{\binits{A.} \bsnm{{de Ugarte Postigo}}},
\bauthor{\binits{R.} \bsnm{{S{\'a}nchez-Ram{\'{\i}}rez}}},
\bauthor{\binits{K.} \bsnm{{Wiersema}}},
\bauthor{\binits{P.} \bsnm{{Schady}}},
\bauthor{\binits{D.A.} \bsnm{{Kann}}},
\bauthor{\binits{R.} \bsnm{{Filgas}}},
\bauthor{\binits{M.} \bsnm{{Nardini}}},
\bauthor{\binits{E.} \bsnm{{Berger}}},
\bauthor{\binits{D.} \bsnm{{Fox}}},
\bauthor{\binits{J.} \bsnm{{Gorosabel}}},
\bauthor{\binits{S.} \bsnm{{Klose}}},
\bauthor{\binits{A.} \bsnm{{Levan}}},
\bauthor{\binits{A.} \bsnm{{Nicuesa Guelbenzu}}},
\bauthor{\binits{A.} \bsnm{{Rossi}}},
\bauthor{\binits{S.} \bsnm{{Schmidl}}},
\bauthor{\binits{V.} \bsnm{{Sudilovsky}}},
\bauthor{\binits{N.R.} \bsnm{{Tanvir}}},
\bauthor{\binits{C.C.} \bsnm{{Th{\"o}ne}}},
\batitle{{The low-extinction afterglow in the solar-metallicity host galaxy of
  {$\gamma$}-ray burst 110918A}}.
\bjtitle{\aap}
\bvolume{556},
\bfpage{23}
(\byear{2013}).
doi:\doiurl{10.1051/0004-6361/201220968}
\end{barticle}
\endbibitem

\bibitem[\protect\citeauthoryear{{Elliott} et~al.}{2015}]{Elliott+2015a}
\begin{barticle}
\bauthor{\binits{J.} \bsnm{{Elliott}}},
\bauthor{\binits{S.} \bsnm{{Khochfar}}},
\bauthor{\binits{J.} \bsnm{{Greiner}}},
\bauthor{\binits{C.} \bsnm{{Dalla Vecchia}}},
\batitle{{The First Billion Years project: gamma-ray bursts at z $>$ 5}}.
\bjtitle{\mnras}
\bvolume{446},
\bfpage{4239}--\blpage{4249}
(\byear{2015}).
doi:\doiurl{10.1093/mnras/stu2417}
\end{barticle}
\endbibitem

\bibitem[\protect\citeauthoryear{{Erb} et~al.}{2006}]{Erb+2006}
\begin{barticle}
\bauthor{\binits{D.K.} \bsnm{{Erb}}},
\bauthor{\binits{A.E.} \bsnm{{Shapley}}},
\bauthor{\binits{M.} \bsnm{{Pettini}}},
\bauthor{\binits{C.C.} \bsnm{{Steidel}}},
\bauthor{\binits{N.A.} \bsnm{{Reddy}}},
\bauthor{\binits{K.L.} \bsnm{{Adelberger}}},
\batitle{{The Mass-Metallicity Relation at z{$>$}\~{}2}}.
\bjtitle{\apj}
\bvolume{644},
\bfpage{813}--\blpage{828}
(\byear{2006}).
doi:\doiurl{10.1086/503623}
\end{barticle}
\endbibitem

\bibitem[\protect\citeauthoryear{{Evans} et~al.}{2009}]{Evans+2009}
\begin{barticle}
\bauthor{\binits{P.A.} \bsnm{{Evans}}},
\bauthor{\binits{A.P.} \bsnm{{Beardmore}}},
\bauthor{\binits{K.L.} \bsnm{{Page}}},
\bauthor{\binits{J.P.} \bsnm{{Osborne}}},
\bauthor{\binits{P.T.} \bsnm{{O'Brien}}},
\bauthor{\binits{R.} \bsnm{{Willingale}}},
\bauthor{\binits{R.L.C.} \bsnm{{Starling}}},
\bauthor{\binits{D.N.} \bsnm{{Burrows}}},
\bauthor{\binits{O.} \bsnm{{Godet}}},
\bauthor{\binits{L.} \bsnm{{Vetere}}},
\bauthor{\binits{J.} \bsnm{{Racusin}}},
\bauthor{\binits{M.R.} \bsnm{{Goad}}},
\bauthor{\binits{K.} \bsnm{{Wiersema}}},
\bauthor{\binits{L.} \bsnm{{Angelini}}},
\bauthor{\binits{M.} \bsnm{{Capalbi}}},
\bauthor{\binits{G.} \bsnm{{Chincarini}}},
\bauthor{\binits{N.} \bsnm{{Gehrels}}},
\bauthor{\binits{J.A.} \bsnm{{Kennea}}},
\bauthor{\binits{R.} \bsnm{{Margutti}}},
\bauthor{\binits{D.C.} \bsnm{{Morris}}},
\bauthor{\binits{C.J.} \bsnm{{Mountford}}},
\bauthor{\binits{C.} \bsnm{{Pagani}}},
\bauthor{\binits{M.} \bsnm{{Perri}}},
\bauthor{\binits{P.} \bsnm{{Romano}}},
\bauthor{\binits{N.} \bsnm{{Tanvir}}},
\batitle{{Methods and results of an automatic analysis of a complete sample of
  Swift-XRT observations of GRBs}}.
\bjtitle{\mnras}
\bvolume{397},
\bfpage{1177}--\blpage{1201}
(\byear{2009}).
doi:\doiurl{10.1111/j.1365-2966.2009.14913.x}
\end{barticle}
\endbibitem

\bibitem[\protect\citeauthoryear{{Fox} et~al.}{2008}]{Fox2008}
\begin{barticle}
\bauthor{\binits{A.J.} \bsnm{{Fox}}},
\bauthor{\binits{C.} \bsnm{{Ledoux}}},
\bauthor{\binits{P.M.} \bsnm{{Vreeswijk}}},
\bauthor{\binits{A.} \bsnm{{Smette}}},
\bauthor{\binits{A.O.} \bsnm{{Jaunsen}}},
\batitle{{High-ion absorption in seven GRB host galaxies at z = 2-4. Evidence
  for both circumburst plasma and outflowing interstellar gas}}.
\bjtitle{\aap}
\bvolume{491},
\bfpage{189}--\blpage{207}
(\byear{2008}).
doi:\doiurl{10.1051/0004-6361+200810286}
\end{barticle}
\endbibitem

\bibitem[\protect\citeauthoryear{{Friis} et~al.}{2015}]{Friis+2015}
\begin{barticle}
\bauthor{\binits{M.} \bsnm{{Friis}}},
\bauthor{\binits{A.} \bsnm{{De Cia}}},
\bauthor{\binits{T.} \bsnm{{Kr{\"u}hler}}},
\bauthor{\binits{J.P.U.} \bsnm{{Fynbo}}},
\bauthor{\binits{C.} \bsnm{{Ledoux}}},
\bauthor{\binits{P.M.} \bsnm{{Vreeswijk}}},
\bauthor{\binits{D.J.} \bsnm{{Watson}}},
\bauthor{\binits{D.} \bsnm{{Malesani}}},
\bauthor{\binits{J.} \bsnm{{Gorosabel}}},
\bauthor{\binits{R.L.C.} \bsnm{{Starling}}},
\bauthor{\binits{P.} \bsnm{{Jakobsson}}},
\bauthor{\binits{K.} \bsnm{{Varela}}},
\bauthor{\binits{K.} \bsnm{{Wiersema}}},
\bauthor{\binits{A.P.} \bsnm{{Drachmann}}},
\bauthor{\binits{A.} \bsnm{{Trotter}}},
\bauthor{\binits{C.C.} \bsnm{{Th{\"o}ne}}},
\bauthor{\binits{A.} \bsnm{{de Ugarte Postigo}}},
\bauthor{\binits{V.} \bsnm{{D'Elia}}},
\bauthor{\binits{J.} \bsnm{{Elliott}}},
\bauthor{\binits{M.} \bsnm{{Maturi}}},
\bauthor{\binits{P.} \bsnm{{Goldoni}}},
\bauthor{\binits{J.} \bsnm{{Greiner}}},
\bauthor{\binits{J.} \bsnm{{Haislip}}},
\bauthor{\binits{L.} \bsnm{{Kaper}}},
\bauthor{\binits{F.} \bsnm{{Knust}}},
\bauthor{\binits{A.} \bsnm{{LaCluyze}}},
\bauthor{\binits{B.} \bsnm{{Milvang-Jensen}}},
\bauthor{\binits{D.} \bsnm{{Reichart}}},
\bauthor{\binits{S.} \bsnm{{Schulze}}},
\bauthor{\binits{V.} \bsnm{{Sudilovsky}}},
\bauthor{\binits{N.} \bsnm{{Tanvir}}},
\bauthor{\binits{S.D.} \bsnm{{Vergani}}},
\batitle{{The warm, the excited, and the molecular gas: GRB 121024A shining
  through its star-forming galaxy}}.
\bjtitle{\mnras}
\bvolume{451},
\bfpage{167}--\blpage{183}
(\byear{2015}).
doi:\doiurl{10.1093/mnras/stv960}
\end{barticle}
\endbibitem

\bibitem[\protect\citeauthoryear{{Fruchter} et~al.}{2001}]{Fruchter01}
\begin{barticle}
\bauthor{\binits{A.} \bsnm{{Fruchter}}},
\bauthor{\binits{J.H.} \bsnm{{Krolik}}},
\bauthor{\binits{J.E.} \bsnm{{Rhoads}}},
\batitle{{X-Ray Destruction of Dust along the Line of Sight to {$\gamma$}-Ray
  Bursts}}.
\bjtitle{\apj}
\bvolume{563},
\bfpage{597}--\blpage{610}
(\byear{2001}).
doi:\doiurl{10.1086/323520}
\end{barticle}
\endbibitem

\bibitem[\protect\citeauthoryear{{Fruchter} et~al.}{2006}]{Fruchter+2006}
\begin{barticle}
\bauthor{\binits{A.S.} \bsnm{{Fruchter}}},
\bauthor{\binits{A.J.} \bsnm{{Levan}}},
\bauthor{\binits{L.} \bsnm{{Strolger}}},
\bauthor{\binits{P.M.} \bsnm{{Vreeswijk}}},
\bauthor{\binits{S.E.} \bsnm{{Thorsett}}},
\bauthor{\binits{D.} \bsnm{{Bersier}}},
\bauthor{\binits{I.} \bsnm{{Burud}}},
\bauthor{\binits{J.M.} \bsnm{{Castro Cer{\'o}n}}},
\bauthor{\binits{A.J.} \bsnm{{Castro-Tirado}}},
\bauthor{\binits{C.} \bsnm{{Conselice}}},
\bauthor{\binits{T.} \bsnm{{Dahlen}}},
\bauthor{\binits{H.C.} \bsnm{{Ferguson}}},
\bauthor{\binits{J.P.U.} \bsnm{{Fynbo}}},
\bauthor{\binits{P.M.} \bsnm{{Garnavich}}},
\bauthor{\binits{R.A.} \bsnm{{Gibbons}}},
\bauthor{\binits{J.} \bsnm{{Gorosabel}}},
\bauthor{\binits{T.R.} \bsnm{{Gull}}},
\bauthor{\binits{J.} \bsnm{{Hjorth}}},
\bauthor{\binits{S.T.} \bsnm{{Holland}}},
\bauthor{\binits{C.} \bsnm{{Kouveliotou}}},
\bauthor{\binits{Z.} \bsnm{{Levay}}},
\bauthor{\binits{M.} \bsnm{{Livio}}},
\bauthor{\binits{M.R.} \bsnm{{Metzger}}},
\bauthor{\binits{P.E.} \bsnm{{Nugent}}},
\bauthor{\binits{L.} \bsnm{{Petro}}},
\bauthor{\binits{E.} \bsnm{{Pian}}},
\bauthor{\binits{J.E.} \bsnm{{Rhoads}}},
\bauthor{\binits{A.G.} \bsnm{{Riess}}},
\bauthor{\binits{K.C.} \bsnm{{Sahu}}},
\bauthor{\binits{A.} \bsnm{{Smette}}},
\bauthor{\binits{N.R.} \bsnm{{Tanvir}}},
\bauthor{\binits{R.A.M.J.} \bsnm{{Wijers}}},
\bauthor{\binits{S.E.} \bsnm{{Woosley}}},
\batitle{{Long {$\gamma$}-ray bursts and core-collapse supernovae have
  different environments}}.
\bjtitle{\nat}
\bvolume{441},
\bfpage{463}--\blpage{468}
(\byear{2006}).
doi:\doiurl{10.1038/nature04787}
\end{barticle}
\endbibitem

\bibitem[\protect\citeauthoryear{{Fryer} and {Heger}}{2005}]{Fryer+2005}
\begin{barticle}
\bauthor{\binits{C.L.} \bsnm{{Fryer}}},
\bauthor{\binits{A.} \bsnm{{Heger}}},
\batitle{{Binary Merger Progenitors for Gamma-Ray Bursts and Hypernovae}}.
\bjtitle{\apj}
\bvolume{623},
\bfpage{302}--\blpage{313}
(\byear{2005}).
doi:\doiurl{10.1086/428379}
\end{barticle}
\endbibitem

\bibitem[\protect\citeauthoryear{{Fynbo} et~al.}{2003}]{Fynbo+2003}
\begin{barticle}
\bauthor{\binits{J.P.U.} \bsnm{{Fynbo}}},
\bauthor{\binits{P.} \bsnm{{Jakobsson}}},
\bauthor{\binits{P.} \bsnm{{M{\"o}ller}}},
\bauthor{\binits{J.} \bsnm{{Hjorth}}},
\bauthor{\binits{B.} \bsnm{{Thomsen}}},
\bauthor{\binits{M.I.} \bsnm{{Andersen}}},
\bauthor{\binits{A.S.} \bsnm{{Fruchter}}},
\bauthor{\binits{J.} \bsnm{{Gorosabel}}},
\bauthor{\binits{S.T.} \bsnm{{Holland}}},
\bauthor{\binits{C.} \bsnm{{Ledoux}}},
\bauthor{\binits{H.} \bsnm{{Pedersen}}},
\bauthor{\binits{J.} \bsnm{{Rhoads}}},
\bauthor{\binits{M.} \bsnm{{Weidinger}}},
\bauthor{\binits{R.A.M.J.} \bsnm{{Wijers}}},
\batitle{{On the Lyalpha emission from gamma-ray burst host galaxies: Evidence
  for low metallicities}}.
\bjtitle{\aap}
\bvolume{406},
\bfpage{63}--\blpage{66}
(\byear{2003}).
doi:\doiurl{10.1051/0004-6361+20030931}
\end{barticle}
\endbibitem

\bibitem[\protect\citeauthoryear{{Fynbo} et~al.}{2006}]{Fynbo+2006a}
\begin{barticle}
\bauthor{\binits{J.P.U.} \bsnm{{Fynbo}}},
\bauthor{\binits{R.L.C.} \bsnm{{Starling}}},
\bauthor{\binits{C.} \bsnm{{Ledoux}}},
\bauthor{\binits{K.} \bsnm{{Wiersema}}},
\bauthor{\binits{C.C.} \bsnm{{Th{\"o}ne}}},
\bauthor{\binits{J.} \bsnm{{Sollerman}}},
\bauthor{\binits{P.} \bsnm{{Jakobsson}}},
\bauthor{\binits{J.} \bsnm{{Hjorth}}},
\bauthor{\binits{D.} \bsnm{{Watson}}},
\bauthor{\binits{P.M.} \bsnm{{Vreeswijk}}},
\bauthor{\binits{P.} \bsnm{{M{\o}ller}}},
\bauthor{\binits{E.} \bsnm{{Rol}}},
\bauthor{\binits{J.} \bsnm{{Gorosabel}}},
\bauthor{\binits{J.} \bsnm{{N{\"a}r{\"a}nen}}},
\bauthor{\binits{R.A.M.J.} \bsnm{{Wijers}}},
\bauthor{\binits{G.} \bsnm{{Bj{\"o}rnsson}}},
\bauthor{\binits{J.M.} \bsnm{{Castro Cer{\'o}n}}},
\bauthor{\binits{P.} \bsnm{{Curran}}},
\bauthor{\binits{D.H.} \bsnm{{Hartmann}}},
\bauthor{\binits{S.T.} \bsnm{{Holland}}},
\bauthor{\binits{B.L.} \bsnm{{Jensen}}},
\bauthor{\binits{A.J.} \bsnm{{Levan}}},
\bauthor{\binits{M.} \bsnm{{Limousin}}},
\bauthor{\binits{C.} \bsnm{{Kouveliotou}}},
\bauthor{\binits{G.} \bsnm{{Nelemans}}},
\bauthor{\binits{K.} \bsnm{{Pedersen}}},
\bauthor{\binits{R.S.} \bsnm{{Priddey}}},
\bauthor{\binits{N.R.} \bsnm{{Tanvir}}},
\batitle{{Probing cosmic chemical evolution with gamma-ray bursts: GRB 060206
  at z = 4.048}}.
\bjtitle{\aap}
\bvolume{451},
\bfpage{47}--\blpage{50}
(\byear{2006}).
doi:\doiurl{10.1051/0004-6361+20065056}
\end{barticle}
\endbibitem

\bibitem[\protect\citeauthoryear{{Fynbo} et~al.}{2009}]{Fynbo+2009}
\begin{barticle}
\bauthor{\binits{J.P.U.} \bsnm{{Fynbo}}},
\bauthor{\binits{P.} \bsnm{{Jakobsson}}},
\bauthor{\binits{J.X.} \bsnm{{Prochaska}}},
\bauthor{\binits{D.} \bsnm{{Malesani}}},
\bauthor{\binits{C.} \bsnm{{Ledoux}}},
\bauthor{\binits{A.} \bsnm{{de Ugarte Postigo}}},
\bauthor{\binits{M.} \bsnm{{Nardini}}},
\bauthor{\binits{P.M.} \bsnm{{Vreeswijk}}},
\bauthor{\binits{K.} \bsnm{{Wiersema}}},
\bauthor{\binits{J.} \bsnm{{Hjorth}}},
\bauthor{\binits{J.} \bsnm{{Sollerman}}},
\bauthor{\binits{H.-W.} \bsnm{{Chen}}},
\bauthor{\binits{C.C.} \bsnm{{Th{\"o}ne}}},
\bauthor{\binits{G.} \bsnm{{Bj{\"o}rnsson}}},
\bauthor{\binits{J.S.} \bsnm{{Bloom}}},
\bauthor{\binits{A.J.} \bsnm{{Castro-Tirado}}},
\bauthor{\binits{L.} \bsnm{{Christensen}}},
\bauthor{\binits{A.} \bsnm{{De Cia}}},
\bauthor{\binits{A.S.} \bsnm{{Fruchter}}},
\bauthor{\binits{J.} \bsnm{{Gorosabel}}},
\bauthor{\binits{J.F.} \bsnm{{Graham}}},
\bauthor{\binits{A.O.} \bsnm{{Jaunsen}}},
\bauthor{\binits{B.L.} \bsnm{{Jensen}}},
\bauthor{\binits{D.A.} \bsnm{{Kann}}},
\bauthor{\binits{C.} \bsnm{{Kouveliotou}}},
\bauthor{\binits{A.J.} \bsnm{{Levan}}},
\bauthor{\binits{J.} \bsnm{{Maund}}},
\bauthor{\binits{N.} \bsnm{{Masetti}}},
\bauthor{\binits{B.} \bsnm{{Milvang-Jensen}}},
\bauthor{\binits{E.} \bsnm{{Palazzi}}},
\bauthor{\binits{D.A.} \bsnm{{Perley}}},
\bauthor{\binits{E.} \bsnm{{Pian}}},
\bauthor{\binits{E.} \bsnm{{Rol}}},
\bauthor{\binits{P.} \bsnm{{Schady}}},
\bauthor{\binits{R.L.C.} \bsnm{{Starling}}},
\bauthor{\binits{N.R.} \bsnm{{Tanvir}}},
\bauthor{\binits{D.J.} \bsnm{{Watson}}},
\bauthor{\binits{D.} \bsnm{{Xu}}},
\bauthor{\binits{T.} \bsnm{{Augusteijn}}},
\bauthor{\binits{F.} \bsnm{{Grundahl}}},
\bauthor{\binits{J.} \bsnm{{Telting}}},
\bauthor{\binits{P.-O.} \bsnm{{Quirion}}},
\batitle{{Low-resolution Spectroscopy of Gamma-ray Burst Optical Afterglows:
  Biases in the Swift Sample and Characterization of the Absorbers}}.
\bjtitle{\apjs}
\bvolume{185},
\bfpage{526}--\blpage{573}
(\byear{2009}).
doi:\doiurl{10.1088/0067-0049/185/2/526}
\end{barticle}
\endbibitem

\bibitem[\protect\citeauthoryear{{Fynbo} et~al.}{2013}]{Fynbo2013}
\begin{barticle}
\bauthor{\binits{J.P.U.} \bsnm{{Fynbo}}},
\bauthor{\binits{J.-K.} \bsnm{{Krogager}}},
\bauthor{\binits{B.} \bsnm{{Venemans}}},
\bauthor{\binits{P.} \bsnm{{Noterdaeme}}},
\bauthor{\binits{M.} \bsnm{{Vestergaard}}},
\bauthor{\binits{P.} \bsnm{{M{\o}ller}}},
\bauthor{\binits{C.} \bsnm{{Ledoux}}},
\bauthor{\binits{S.} \bsnm{{Geier}}},
\batitle{{Optical/Near-infrared Selection of Red Quasi-stellar Objects:
  Evidence for Steep Extinction Curves toward Galactic Centers?}}
\bjtitle{\apjs}
\bvolume{204},
\bfpage{6}
(\byear{2013}).
doi:\doiurl{10.1088/0067-0049/204/1/6}
\end{barticle}
\endbibitem

\bibitem[\protect\citeauthoryear{{Fynbo} et~al.}{2014}]{Fynbo+2014}
\begin{barticle}
\bauthor{\binits{J.P.U.} \bsnm{{Fynbo}}},
\bauthor{\binits{T.} \bsnm{{Kr{\"u}hler}}},
\bauthor{\binits{K.} \bsnm{{Leighly}}},
\bauthor{\binits{C.} \bsnm{{Ledoux}}},
\bauthor{\binits{P.M.} \bsnm{{Vreeswijk}}},
\bauthor{\binits{S.} \bsnm{{Schulze}}},
\bauthor{\binits{P.} \bsnm{{Noterdaeme}}},
\bauthor{\binits{D.} \bsnm{{Watson}}},
\bauthor{\binits{R.A.M.J.} \bsnm{{Wijers}}},
\bauthor{\binits{J.} \bsnm{{Bolmer}}},
\bauthor{\binits{Z.} \bsnm{{Cano}}},
\bauthor{\binits{L.} \bsnm{{Christensen}}},
\bauthor{\binits{S.} \bsnm{{Covino}}},
\bauthor{\binits{V.} \bsnm{{D'Elia}}},
\bauthor{\binits{H.} \bsnm{{Flores}}},
\bauthor{\binits{M.} \bsnm{{Friis}}},
\bauthor{\binits{P.} \bsnm{{Goldoni}}},
\bauthor{\binits{J.} \bsnm{{Greiner}}},
\bauthor{\binits{F.} \bsnm{{Hammer}}},
\bauthor{\binits{J.} \bsnm{{Hjorth}}},
\bauthor{\binits{P.} \bsnm{{Jakobsson}}},
\bauthor{\binits{J.} \bsnm{{Japelj}}},
\bauthor{\binits{L.} \bsnm{{Kaper}}},
\bauthor{\binits{S.} \bsnm{{Klose}}},
\bauthor{\binits{F.} \bsnm{{Knust}}},
\bauthor{\binits{G.} \bsnm{{Leloudas}}},
\bauthor{\binits{A.} \bsnm{{Levan}}},
\bauthor{\binits{D.} \bsnm{{Malesani}}},
\bauthor{\binits{B.} \bsnm{{Milvang-Jensen}}},
\bauthor{\binits{P.} \bsnm{{M{\o}ller}}},
\bauthor{\binits{A.} \bsnm{{Nicuesa Guelbenzu}}},
\bauthor{\binits{S.} \bsnm{{Oates}}},
\bauthor{\binits{E.} \bsnm{{Pian}}},
\bauthor{\binits{P.} \bsnm{{Schady}}},
\bauthor{\binits{M.} \bsnm{{Sparre}}},
\bauthor{\binits{G.} \bsnm{{Tagliaferri}}},
\bauthor{\binits{N.} \bsnm{{Tanvir}}},
\bauthor{\binits{C.C.} \bsnm{{Th{\"o}ne}}},
\bauthor{\binits{A.} \bsnm{{de Ugarte Postigo}}},
\bauthor{\binits{S.} \bsnm{{Vergani}}},
\bauthor{\binits{K.} \bsnm{{Wiersema}}},
\bauthor{\binits{D.} \bsnm{{Xu}}},
\bauthor{\binits{T.} \bsnm{{Zafar}}},
\batitle{{The mysterious optical afterglow spectrum of GRB 140506A at z =
  0.889}}.
\bjtitle{\aap}
\bvolume{572},
\bfpage{12}
(\byear{2014}).
doi:\doiurl{10.1051/0004-6361/201424726}
\end{barticle}
\endbibitem

\bibitem[\protect\citeauthoryear{{Fynbo} et~al.}{2001}]{Fynbo+2001}
\begin{barticle}
\bauthor{\binits{J.U.} \bsnm{{Fynbo}}},
\bauthor{\binits{B.L.} \bsnm{{Jensen}}},
\bauthor{\binits{J.} \bsnm{{Gorosabel}}},
\bauthor{\binits{J.} \bsnm{{Hjorth}}},
\bauthor{\binits{H.} \bsnm{{Pedersen}}},
\bauthor{\binits{P.} \bsnm{{M{\o}ller}}},
\bauthor{\binits{T.} \bsnm{{Abbott}}},
\bauthor{\binits{A.J.} \bsnm{{Castro-Tirado}}},
\bauthor{\binits{D.} \bsnm{{Delgado}}},
\bauthor{\binits{J.} \bsnm{{Greiner}}},
\bauthor{\binits{A.} \bsnm{{Henden}}},
\bauthor{\binits{A.} \bsnm{{Magazz{\`u}}}},
\bauthor{\binits{N.} \bsnm{{Masetti}}},
\bauthor{\binits{S.} \bsnm{{Merlino}}},
\bauthor{\binits{J.} \bsnm{{Masegosa}}},
\bauthor{\binits{R.} \bsnm{{{\O}stensen}}},
\bauthor{\binits{E.} \bsnm{{Palazzi}}},
\bauthor{\binits{E.} \bsnm{{Pian}}},
\bauthor{\binits{H.E.} \bsnm{{Schwarz}}},
\bauthor{\binits{T.} \bsnm{{Cline}}},
\bauthor{\binits{C.} \bsnm{{Guidorzi}}},
\bauthor{\binits{J.} \bsnm{{Goldsten}}},
\bauthor{\binits{K.} \bsnm{{Hurley}}},
\bauthor{\binits{E.} \bsnm{{Mazets}}},
\bauthor{\binits{T.} \bsnm{{McClanahan}}},
\bauthor{\binits{E.} \bsnm{{Montanari}}},
\bauthor{\binits{R.} \bsnm{{Starr}}},
\bauthor{\binits{J.} \bsnm{{Trombka}}},
\batitle{{Detection of the optical afterglow of GRB 000630: Implications for
  dark bursts}}.
\bjtitle{\aap}
\bvolume{369},
\bfpage{373}--\blpage{379}
(\byear{2001}).
doi:\doiurl{10.1051/0004-6361+20010112}
\end{barticle}
\endbibitem

\bibitem[\protect\citeauthoryear{{Galama} and
  {Wijers}}{2001}]{GalamaandWijers01}
\begin{barticle}
\bauthor{\binits{T.J.} \bsnm{{Galama}}},
\bauthor{\binits{R.A.M.J.} \bsnm{{Wijers}}},
\batitle{{High Column Densities and Low Extinctions of Gamma-Ray Bursts:
  Evidence for Hypernovae and Dust Destruction}}.
\bjtitle{\apjl}
\bvolume{549},
\bfpage{209}--\blpage{213}
(\byear{2001}).
doi:\doiurl{10.1086/319162}
\end{barticle}
\endbibitem

\bibitem[\protect\citeauthoryear{{Galama} et~al.}{1998}]{Galama+1998}
\begin{barticle}
\bauthor{\binits{T.J.} \bsnm{{Galama}}},
\bauthor{\binits{P.M.} \bsnm{{Vreeswijk}}},
\bauthor{\binits{J.} \bsnm{{van Paradijs}}},
\bauthor{\binits{C.} \bsnm{{Kouveliotou}}},
\bauthor{\binits{T.} \bsnm{{Augusteijn}}},
\bauthor{\binits{H.} \bsnm{{B{\"o}hnhardt}}},
\bauthor{\binits{J.P.} \bsnm{{Brewer}}},
\bauthor{\binits{V.} \bsnm{{Doublier}}},
\bauthor{\binits{J.-F.} \bsnm{{Gonzalez}}},
\bauthor{\binits{B.} \bsnm{{Leibundgut}}},
\bauthor{\binits{C.} \bsnm{{Lidman}}},
\bauthor{\binits{O.R.} \bsnm{{Hainaut}}},
\bauthor{\binits{F.} \bsnm{{Patat}}},
\bauthor{\binits{J.} \bsnm{{Heise}}},
\bauthor{\binits{J.} \bsnm{{in't Zand}}},
\bauthor{\binits{K.} \bsnm{{Hurley}}},
\bauthor{\binits{P.J.} \bsnm{{Groot}}},
\bauthor{\binits{R.G.} \bsnm{{Strom}}},
\bauthor{\binits{P.A.} \bsnm{{Mazzali}}},
\bauthor{\binits{K.} \bsnm{{Iwamoto}}},
\bauthor{\binits{K.} \bsnm{{Nomoto}}},
\bauthor{\binits{H.} \bsnm{{Umeda}}},
\bauthor{\binits{T.} \bsnm{{Nakamura}}},
\bauthor{\binits{T.R.} \bsnm{{Young}}},
\bauthor{\binits{T.} \bsnm{{Suzuki}}},
\bauthor{\binits{T.} \bsnm{{Shigeyama}}},
\bauthor{\binits{T.} \bsnm{{Koshut}}},
\bauthor{\binits{M.} \bsnm{{Kippen}}},
\bauthor{\binits{C.} \bsnm{{Robinson}}},
\bauthor{\binits{P.} \bsnm{{de Wildt}}},
\bauthor{\binits{R.A.M.J.} \bsnm{{Wijers}}},
\bauthor{\binits{N.} \bsnm{{Tanvir}}},
\bauthor{\binits{J.} \bsnm{{Greiner}}},
\bauthor{\binits{E.} \bsnm{{Pian}}},
\bauthor{\binits{E.} \bsnm{{Palazzi}}},
\bauthor{\binits{F.} \bsnm{{Frontera}}},
\bauthor{\binits{N.} \bsnm{{Masetti}}},
\bauthor{\binits{L.} \bsnm{{Nicastro}}},
\bauthor{\binits{M.} \bsnm{{Feroci}}},
\bauthor{\binits{E.} \bsnm{{Costa}}},
\bauthor{\binits{L.} \bsnm{{Piro}}},
\bauthor{\binits{B.A.} \bsnm{{Peterson}}},
\bauthor{\binits{C.} \bsnm{{Tinney}}},
\bauthor{\binits{B.} \bsnm{{Boyle}}},
\bauthor{\binits{R.} \bsnm{{Cannon}}},
\bauthor{\binits{R.} \bsnm{{Stathakis}}},
\bauthor{\binits{E.} \bsnm{{Sadler}}},
\bauthor{\binits{M.C.} \bsnm{{Begam}}},
\bauthor{\binits{P.} \bsnm{{Ianna}}},
\batitle{{An unusual supernova in the error box of the {$\gamma$}-ray burst of
  25 April 1998}}.
\bjtitle{\nat}
\bvolume{395},
\bfpage{670}--\blpage{672}
(\byear{1998}).
doi:\doiurl{10.1038/27150}
\end{barticle}
\endbibitem

\bibitem[\protect\citeauthoryear{{Gehrels} et~al.}{2004}]{Gehrels+2004}
\begin{barticle}
\bauthor{\binits{N.} \bsnm{{Gehrels}}},
\bauthor{\binits{G.} \bsnm{{Chincarini}}},
\bauthor{\binits{P.} \bsnm{{Giommi}}},
\bauthor{\binits{K.O.} \bsnm{{Mason}}},
\bauthor{\binits{J.A.} \bsnm{{Nousek}}},
\bauthor{\binits{A.A.} \bsnm{{Wells}}},
\bauthor{\binits{N.E.} \bsnm{{White}}},
\bauthor{\binits{S.D.} \bsnm{{Barthelmy}}},
\bauthor{\binits{D.N.} \bsnm{{Burrows}}},
\bauthor{\binits{L.R.} \bsnm{{Cominsky}}},
\bauthor{\binits{K.C.} \bsnm{{Hurley}}},
\bauthor{\binits{F.E.} \bsnm{{Marshall}}},
\bauthor{\binits{P.} \bsnm{{M{\'e}sz{\'a}ros}}},
\bauthor{\binits{P.W.A.} \bsnm{{Roming}}},
\bauthor{\binits{L.} \bsnm{{Angelini}}},
\bauthor{\binits{L.M.} \bsnm{{Barbier}}},
\bauthor{\binits{T.} \bsnm{{Belloni}}},
\bauthor{\binits{S.} \bsnm{{Campana}}},
\bauthor{\binits{P.A.} \bsnm{{Caraveo}}},
\bauthor{\binits{M.M.} \bsnm{{Chester}}},
\bauthor{\binits{O.} \bsnm{{Citterio}}},
\bauthor{\binits{T.L.} \bsnm{{Cline}}},
\bauthor{\binits{M.S.} \bsnm{{Cropper}}},
\bauthor{\binits{J.R.} \bsnm{{Cummings}}},
\bauthor{\binits{A.J.} \bsnm{{Dean}}},
\bauthor{\binits{E.D.} \bsnm{{Feigelson}}},
\bauthor{\binits{E.E.} \bsnm{{Fenimore}}},
\bauthor{\binits{D.A.} \bsnm{{Frail}}},
\bauthor{\binits{A.S.} \bsnm{{Fruchter}}},
\bauthor{\binits{G.P.} \bsnm{{Garmire}}},
\bauthor{\binits{K.} \bsnm{{Gendreau}}},
\bauthor{\binits{G.} \bsnm{{Ghisellini}}},
\bauthor{\binits{J.} \bsnm{{Greiner}}},
\bauthor{\binits{J.E.} \bsnm{{Hill}}},
\bauthor{\binits{S.D.} \bsnm{{Hunsberger}}},
\bauthor{\binits{H.A.} \bsnm{{Krimm}}},
\bauthor{\binits{S.R.} \bsnm{{Kulkarni}}},
\bauthor{\binits{P.} \bsnm{{Kumar}}},
\bauthor{\binits{F.} \bsnm{{Lebrun}}},
\bauthor{\binits{N.M.} \bsnm{{Lloyd-Ronning}}},
\bauthor{\binits{C.B.} \bsnm{{Markwardt}}},
\bauthor{\binits{B.J.} \bsnm{{Mattson}}},
\bauthor{\binits{R.F.} \bsnm{{Mushotzky}}},
\bauthor{\binits{J.P.} \bsnm{{Norris}}},
\bauthor{\binits{J.} \bsnm{{Osborne}}},
\bauthor{\binits{B.} \bsnm{{Paczynski}}},
\bauthor{\binits{D.M.} \bsnm{{Palmer}}},
\bauthor{\binits{H.-S.} \bsnm{{Park}}},
\bauthor{\binits{A.M.} \bsnm{{Parsons}}},
\bauthor{\binits{J.} \bsnm{{Paul}}},
\bauthor{\binits{M.J.} \bsnm{{Rees}}},
\bauthor{\binits{C.S.} \bsnm{{Reynolds}}},
\bauthor{\binits{J.E.} \bsnm{{Rhoads}}},
\bauthor{\binits{T.P.} \bsnm{{Sasseen}}},
\bauthor{\binits{B.E.} \bsnm{{Schaefer}}},
\bauthor{\binits{A.T.} \bsnm{{Short}}},
\bauthor{\binits{A.P.} \bsnm{{Smale}}},
\bauthor{\binits{I.A.} \bsnm{{Smith}}},
\bauthor{\binits{L.} \bsnm{{Stella}}},
\bauthor{\binits{G.} \bsnm{{Tagliaferri}}},
\bauthor{\binits{T.} \bsnm{{Takahashi}}},
\bauthor{\binits{M.} \bsnm{{Tashiro}}},
\bauthor{\binits{L.K.} \bsnm{{Townsley}}},
\bauthor{\binits{J.} \bsnm{{Tueller}}},
\bauthor{\binits{M.J.L.} \bsnm{{Turner}}},
\bauthor{\binits{M.} \bsnm{{Vietri}}},
\bauthor{\binits{W.} \bsnm{{Voges}}},
\bauthor{\binits{M.J.} \bsnm{{Ward}}},
\bauthor{\binits{R.} \bsnm{{Willingale}}},
\bauthor{\binits{F.M.} \bsnm{{Zerbi}}},
\bauthor{\binits{W.W.} \bsnm{{Zhang}}},
\batitle{{The Swift Gamma-Ray Burst Mission}}.
\bjtitle{\apj}
\bvolume{611},
\bfpage{1005}--\blpage{1020}
(\byear{2004}).
doi:\doiurl{10.1086/422091}
\end{barticle}
\endbibitem

\bibitem[\protect\citeauthoryear{{Gosling} et~al.}{2009}]{Gosling2009}
\begin{barticle}
\bauthor{\binits{A.J.} \bsnm{{Gosling}}},
\bauthor{\binits{R.M.} \bsnm{{Bandyopadhyay}}},
\bauthor{\binits{K.M.} \bsnm{{Blundell}}},
\batitle{{The complex, variable near-infrared extinction towards the Nuclear
  Bulge}}.
\bjtitle{\mnras}
\bvolume{394},
\bfpage{2247}--\blpage{2254}
(\byear{2009}).
doi:\doiurl{10.1111/j.1365-2966.2009.14493.x}
\end{barticle}
\endbibitem

\bibitem[\protect\citeauthoryear{{Graham} and {Fruchter}}{2013}]{Graham+2013}
\begin{barticle}
\bauthor{\binits{J.F.} \bsnm{{Graham}}},
\bauthor{\binits{A.S.} \bsnm{{Fruchter}}},
\batitle{{The Metal Aversion of Long-duration Gamma-Ray Bursts}}.
\bjtitle{\apj}
\bvolume{774},
\bfpage{119}
(\byear{2013}).
doi:\doiurl{10.1088/0004-637X/774/2/119}
\end{barticle}
\endbibitem

\bibitem[\protect\citeauthoryear{{Greiner} et~al.}{2011}]{Greiner+2011}
\begin{barticle}
\bauthor{\binits{J.} \bsnm{{Greiner}}},
\bauthor{\binits{T.} \bsnm{{Kr{\"u}hler}}},
\bauthor{\binits{S.} \bsnm{{Klose}}},
\bauthor{\binits{P.} \bsnm{{Afonso}}},
\bauthor{\binits{C.} \bsnm{{Clemens}}},
\bauthor{\binits{R.} \bsnm{{Filgas}}},
\bauthor{\binits{D.H.} \bsnm{{Hartmann}}},
\bauthor{\binits{A.} \bsnm{{K{\"u}pc{\"u} Yolda{\c s}}}},
\bauthor{\binits{M.} \bsnm{{Nardini}}},
\bauthor{\binits{F.} \bsnm{{Olivares E.}}},
\bauthor{\binits{A.} \bsnm{{Rau}}},
\bauthor{\binits{A.} \bsnm{{Rossi}}},
\bauthor{\binits{P.} \bsnm{{Schady}}},
\bauthor{\binits{A.} \bsnm{{Updike}}},
\batitle{{The nature of ``dark'' gamma-ray bursts}}.
\bjtitle{\aap}
\bvolume{526},
\bfpage{30}
(\byear{2011}).
doi:\doiurl{10.1051/0004-6361/201015458}
\end{barticle}
\endbibitem

\bibitem[\protect\citeauthoryear{{Greiner} et~al.}{2015}]{Greiner+2015}
\begin{barticle}
\bauthor{\binits{J.} \bsnm{{Greiner}}},
\bauthor{\binits{D.B.} \bsnm{{Fox}}},
\bauthor{\binits{P.} \bsnm{{Schady}}},
\bauthor{\binits{T.} \bsnm{{Kr{\"u}hler}}},
\bauthor{\binits{M.} \bsnm{{Trenti}}},
\bauthor{\binits{A.} \bsnm{{Cikota}}},
\bauthor{\binits{J.} \bsnm{{Bolmer}}},
\bauthor{\binits{J.} \bsnm{{Elliott}}},
\bauthor{\binits{C.} \bsnm{{Delvaux}}},
\bauthor{\binits{R.} \bsnm{{Perna}}},
\bauthor{\binits{P.} \bsnm{{Afonso}}},
\bauthor{\binits{D.A.} \bsnm{{Kann}}},
\bauthor{\binits{S.} \bsnm{{Klose}}},
\bauthor{\binits{S.} \bsnm{{Savaglio}}},
\bauthor{\binits{S.} \bsnm{{Schmidl}}},
\bauthor{\binits{T.} \bsnm{{Schweyer}}},
\bauthor{\binits{M.} \bsnm{{Tanga}}},
\bauthor{\binits{K.} \bsnm{{Varela}}},
\batitle{{Gamma-Ray Bursts Trace UV Metrics of Star Formation over 3 {$<$} z
  {$<$} 5}}.
\bjtitle{\apj}
\bvolume{809},
\bfpage{76}
(\byear{2015}).
doi:\doiurl{10.1088/0004-637X/809/1/76}
\end{barticle}
\endbibitem

\bibitem[\protect\citeauthoryear{{Groot} et~al.}{1998}]{Groot+1998}
\begin{barticle}
\bauthor{\binits{P.J.} \bsnm{{Groot}}},
\bauthor{\binits{T.J.} \bsnm{{Galama}}},
\bauthor{\binits{J.} \bsnm{{van Paradijs}}},
\bauthor{\binits{C.} \bsnm{{Kouveliotou}}},
\bauthor{\binits{R.A.M.J.} \bsnm{{Wijers}}},
\bauthor{\binits{J.} \bsnm{{Bloom}}},
\bauthor{\binits{N.} \bsnm{{Tanvir}}},
\bauthor{\binits{R.} \bsnm{{Vanderspek}}},
\bauthor{\binits{J.} \bsnm{{Greiner}}},
\bauthor{\binits{A.J.} \bsnm{{Castro-Tirado}}},
\bauthor{\binits{J.} \bsnm{{Gorosabel}}},
\bauthor{\binits{T.} \bsnm{{von Hippel}}},
\bauthor{\binits{M.} \bsnm{{Lehnert}}},
\bauthor{\binits{K.} \bsnm{{Kuijken}}},
\bauthor{\binits{H.} \bsnm{{Hoekstra}}},
\bauthor{\binits{N.} \bsnm{{Metcalfe}}},
\bauthor{\binits{C.} \bsnm{{Howk}}},
\bauthor{\binits{C.} \bsnm{{Conselice}}},
\bauthor{\binits{J.} \bsnm{{Telting}}},
\bauthor{\binits{R.G.M.} \bsnm{{Rutten}}},
\bauthor{\binits{J.} \bsnm{{Rhoads}}},
\bauthor{\binits{A.} \bsnm{{Cole}}},
\bauthor{\binits{D.J.} \bsnm{{Pisano}}},
\bauthor{\binits{R.} \bsnm{{Naber}}},
\bauthor{\binits{R.} \bsnm{{Schwarz}}},
\batitle{{A Search for Optical Afterglow from GRB 970828}}.
\bjtitle{\apjl}
\bvolume{493},
\bfpage{27}--\blpage{30}
(\byear{1998}).
doi:\doiurl{10.1086/311125}
\end{barticle}
\endbibitem

\bibitem[\protect\citeauthoryear{{Hall} et~al.}{2002}]{Hall2002}
\begin{barticle}
\bauthor{\binits{P.B.} \bsnm{{Hall}}},
\bauthor{\binits{S.F.} \bsnm{{Anderson}}},
\bauthor{\binits{M.A.} \bsnm{{Strauss}}},
\bauthor{\binits{D.G.} \bsnm{{York}}},
\bauthor{\binits{G.T.} \bsnm{{Richards}}},
\bauthor{\binits{X.} \bsnm{{Fan}}},
\bauthor{\binits{G.R.} \bsnm{{Knapp}}},
\bauthor{\binits{D.P.} \bsnm{{Schneider}}},
\bauthor{\binits{D.E.} \bsnm{{Vanden Berk}}},
\bauthor{\binits{T.R.} \bsnm{{Geballe}}},
\bauthor{\binits{A.E.} \bsnm{{Bauer}}},
\bauthor{\binits{R.H.} \bsnm{{Becker}}},
\bauthor{\binits{M.} \bsnm{{Davis}}},
\bauthor{\binits{H.-W.} \bsnm{{Rix}}},
\bauthor{\binits{R.C.} \bsnm{{Nichol}}},
\bauthor{\binits{N.A.} \bsnm{{Bahcall}}},
\bauthor{\binits{J.} \bsnm{{Brinkmann}}},
\bauthor{\binits{R.} \bsnm{{Brunner}}},
\bauthor{\binits{A.J.} \bsnm{{Connolly}}},
\bauthor{\binits{I.} \bsnm{{Csabai}}},
\bauthor{\binits{M.} \bsnm{{Doi}}},
\bauthor{\binits{M.} \bsnm{{Fukugita}}},
\bauthor{\binits{J.E.} \bsnm{{Gunn}}},
\bauthor{\binits{Z.} \bsnm{{Haiman}}},
\bauthor{\binits{M.} \bsnm{{Harvanek}}},
\bauthor{\binits{T.M.} \bsnm{{Heckman}}},
\bauthor{\binits{G.S.} \bsnm{{Hennessy}}},
\bauthor{\binits{N.} \bsnm{{Inada}}},
\bauthor{\binits{{\v Z}.} \bsnm{{Ivezi{\'c}}}},
\bauthor{\binits{D.} \bsnm{{Johnston}}},
\bauthor{\binits{S.} \bsnm{{Kleinman}}},
\bauthor{\binits{J.H.} \bsnm{{Krolik}}},
\bauthor{\binits{J.} \bsnm{{Krzesinski}}},
\bauthor{\binits{P.Z.} \bsnm{{Kunszt}}},
\bauthor{\binits{D.Q.} \bsnm{{Lamb}}},
\bauthor{\binits{D.C.} \bsnm{{Long}}},
\bauthor{\binits{R.H.} \bsnm{{Lupton}}},
\bauthor{\binits{G.} \bsnm{{Miknaitis}}},
\bauthor{\binits{J.A.} \bsnm{{Munn}}},
\bauthor{\binits{V.K.} \bsnm{{Narayanan}}},
\bauthor{\binits{E.} \bsnm{{Neilsen}}},
\bauthor{\binits{P.R.} \bsnm{{Newman}}},
\bauthor{\binits{A.} \bsnm{{Nitta}}},
\bauthor{\binits{S.} \bsnm{{Okamura}}},
\bauthor{\binits{L.} \bsnm{{Pentericci}}},
\bauthor{\binits{J.R.} \bsnm{{Pier}}},
\bauthor{\binits{D.J.} \bsnm{{Schlegel}}},
\bauthor{\binits{S.} \bsnm{{Snedden}}},
\bauthor{\binits{A.S.} \bsnm{{Szalay}}},
\bauthor{\binits{A.R.} \bsnm{{Thakar}}},
\bauthor{\binits{Z.} \bsnm{{Tsvetanov}}},
\bauthor{\binits{R.L.} \bsnm{{White}}},
\bauthor{\binits{W.} \bsnm{{Zheng}}},
\batitle{{Unusual Broad Absorption Line Quasars from the Sloan Digital Sky
  Survey}}.
\bjtitle{\apjs}
\bvolume{141},
\bfpage{267}--\blpage{309}
(\byear{2002}).
doi:\doiurl{10.1086/340546}
\end{barticle}
\endbibitem

\bibitem[\protect\citeauthoryear{{Han} et~al.}{2010}]{Han+2010}
\begin{barticle}
\bauthor{\binits{X.H.} \bsnm{{Han}}},
\bauthor{\binits{F.} \bsnm{{Hammer}}},
\bauthor{\binits{Y.C.} \bsnm{{Liang}}},
\bauthor{\binits{H.} \bsnm{{Flores}}},
\bauthor{\binits{M.} \bsnm{{Rodrigues}}},
\bauthor{\binits{J.L.} \bsnm{{Hou}}},
\bauthor{\binits{J.Y.} \bsnm{{Wei}}},
\batitle{{The Wolf-Rayet features and mass-metallicity relation of
  long-duration gamma-ray burst host galaxies}}.
\bjtitle{\aap}
\bvolume{514},
\bfpage{24}
(\byear{2010}).
doi:\doiurl{10.1051/0004-6361/200912475}
\end{barticle}
\endbibitem

\bibitem[\protect\citeauthoryear{{Hartoog} et~al.}{2015}]{Hartoog+2015}
\begin{barticle}
\bauthor{\binits{O.E.} \bsnm{{Hartoog}}},
\bauthor{\binits{D.} \bsnm{{Malesani}}},
\bauthor{\binits{J.P.U.} \bsnm{{Fynbo}}},
\bauthor{\binits{T.} \bsnm{{Goto}}},
\bauthor{\binits{T.} \bsnm{{Kr{\"u}hler}}},
\bauthor{\binits{P.M.} \bsnm{{Vreeswijk}}},
\bauthor{\binits{A.} \bsnm{{De Cia}}},
\bauthor{\binits{D.} \bsnm{{Xu}}},
\bauthor{\binits{P.} \bsnm{{M{\o}ller}}},
\bauthor{\binits{S.} \bsnm{{Covino}}},
\bauthor{\binits{V.} \bsnm{{D'Elia}}},
\bauthor{\binits{H.} \bsnm{{Flores}}},
\bauthor{\binits{P.} \bsnm{{Goldoni}}},
\bauthor{\binits{J.} \bsnm{{Hjorth}}},
\bauthor{\binits{P.} \bsnm{{Jakobsson}}},
\bauthor{\binits{J.-K.} \bsnm{{Krogager}}},
\bauthor{\binits{L.} \bsnm{{Kaper}}},
\bauthor{\binits{C.} \bsnm{{Ledoux}}},
\bauthor{\binits{A.J.} \bsnm{{Levan}}},
\bauthor{\binits{B.} \bsnm{{Milvang-Jensen}}},
\bauthor{\binits{J.} \bsnm{{Sollerman}}},
\bauthor{\binits{M.} \bsnm{{Sparre}}},
\bauthor{\binits{G.} \bsnm{{Tagliaferri}}},
\bauthor{\binits{N.R.} \bsnm{{Tanvir}}},
\bauthor{\binits{A.} \bsnm{{de Ugarte Postigo}}},
\bauthor{\binits{S.D.} \bsnm{{Vergani}}},
\bauthor{\binits{K.} \bsnm{{Wiersema}}},
\bauthor{\binits{J.} \bsnm{{Datson}}},
\bauthor{\binits{R.} \bsnm{{Salinas}}},
\bauthor{\binits{K.} \bsnm{{Mikkelsen}}},
\bauthor{\binits{N.} \bsnm{{Aghanim}}},
\batitle{{VLT/X-Shooter spectroscopy of the afterglow of the Swift GRB 130606A.
  Chemical abundances and reionisation at z \~{} 6}}.
\bjtitle{\aap}
\bvolume{580},
\bfpage{139}
(\byear{2015}).
doi:\doiurl{10.1051/0004-6361/201425001}
\end{barticle}
\endbibitem

\bibitem[\protect\citeauthoryear{{Hashimoto} et~al.}{2010}]{Hashimoto+2010}
\begin{barticle}
\bauthor{\binits{T.} \bsnm{{Hashimoto}}},
\bauthor{\binits{K.} \bsnm{{Ohta}}},
\bauthor{\binits{K.} \bsnm{{Aoki}}},
\bauthor{\binits{I.} \bsnm{{Tanaka}}},
\bauthor{\binits{K.} \bsnm{{Yabe}}},
\bauthor{\binits{N.} \bsnm{{Kawai}}},
\bauthor{\binits{W.} \bsnm{{Aoki}}},
\bauthor{\binits{H.} \bsnm{{Furusawa}}},
\bauthor{\binits{T.} \bsnm{{Hattori}}},
\bauthor{\binits{M.} \bsnm{{Iye}}},
\bauthor{\binits{K.S.} \bsnm{{Kawabata}}},
\bauthor{\binits{N.} \bsnm{{Kobayashi}}},
\bauthor{\binits{Y.} \bsnm{{Komiyama}}},
\bauthor{\binits{G.} \bsnm{{Kosugi}}},
\bauthor{\binits{Y.} \bsnm{{Minowa}}},
\bauthor{\binits{Y.} \bsnm{{Mizumoto}}},
\bauthor{\binits{Y.} \bsnm{{Niino}}},
\bauthor{\binits{K.} \bsnm{{Nomoto}}},
\bauthor{\binits{J.} \bsnm{{Noumaru}}},
\bauthor{\binits{R.} \bsnm{{Ogasawara}}},
\bauthor{\binits{T.-S.} \bsnm{{Pyo}}},
\bauthor{\binits{T.} \bsnm{{Sakamoto}}},
\bauthor{\binits{K.} \bsnm{{Sekiguchi}}},
\bauthor{\binits{Y.} \bsnm{{Shirasaki}}},
\bauthor{\binits{M.} \bsnm{{Suzuki}}},
\bauthor{\binits{A.} \bsnm{{Tajitsu}}},
\bauthor{\binits{T.} \bsnm{{Takata}}},
\bauthor{\binits{T.} \bsnm{{Tamagawa}}},
\bauthor{\binits{H.} \bsnm{{Terada}}},
\bauthor{\binits{T.} \bsnm{{Totani}}},
\bauthor{\binits{J.} \bsnm{{Watanabe}}},
\bauthor{\binits{T.} \bsnm{{Yamada}}},
\bauthor{\binits{A.} \bsnm{{Yoshida}}},
\batitle{{''Dark'' GRB 080325 in a Dusty Massive Galaxy at z \~{} 2}}.
\bjtitle{\apj}
\bvolume{719},
\bfpage{378}--\blpage{384}
(\byear{2010}).
doi:\doiurl{10.1088/0004-637X/719/1/378}
\end{barticle}
\endbibitem

\bibitem[\protect\citeauthoryear{{Hashimoto} et~al.}{2015}]{Hashimoto+2015}
\begin{barticle}
\bauthor{\binits{T.} \bsnm{{Hashimoto}}},
\bauthor{\binits{D.A.} \bsnm{{Perley}}},
\bauthor{\binits{K.} \bsnm{{Ohta}}},
\bauthor{\binits{K.} \bsnm{{Aoki}}},
\bauthor{\binits{I.} \bsnm{{Tanaka}}},
\bauthor{\binits{Y.} \bsnm{{Niino}}},
\bauthor{\binits{K.} \bsnm{{Yabe}}},
\bauthor{\binits{N.} \bsnm{{Kawai}}},
\batitle{{The Star Formation Rate and Metallicity of the Host Galaxy of the
  Dark GRB 080325 at z=1.78}}.
\bjtitle{\apj}
\bvolume{806},
\bfpage{250}
(\byear{2015}).
doi:\doiurl{10.1088/0004-637X/806/2/250}
\end{barticle}
\endbibitem

\bibitem[\protect\citeauthoryear{{Hatsukade} et~al.}{2014}]{Hatsukade+2014}
\begin{barticle}
\bauthor{\binits{B.} \bsnm{{Hatsukade}}},
\bauthor{\binits{K.} \bsnm{{Ohta}}},
\bauthor{\binits{A.} \bsnm{{Endo}}},
\bauthor{\binits{K.} \bsnm{{Nakanishi}}},
\bauthor{\binits{Y.} \bsnm{{Tamura}}},
\bauthor{\binits{T.} \bsnm{{Hashimoto}}},
\bauthor{\binits{K.} \bsnm{{Kohno}}},
\batitle{{Two {$\gamma$}-ray bursts from dusty regions with little molecular
  gas}}.
\bjtitle{\nat}
\bvolume{510},
\bfpage{247}--\blpage{249}
(\byear{2014}).
doi:\doiurl{10.1038/nature13325}
\end{barticle}
\endbibitem

\bibitem[\protect\citeauthoryear{{Heger} et~al.}{2003}]{Heger+2003}
\begin{barticle}
\bauthor{\binits{A.} \bsnm{{Heger}}},
\bauthor{\binits{C.L.} \bsnm{{Fryer}}},
\bauthor{\binits{S.E.} \bsnm{{Woosley}}},
\bauthor{\binits{N.} \bsnm{{Langer}}},
\bauthor{\binits{D.H.} \bsnm{{Hartmann}}},
\batitle{{How Massive Single Stars End Their Life}}.
\bjtitle{\apj}
\bvolume{591},
\bfpage{288}--\blpage{300}
(\byear{2003}).
doi:\doiurl{10.1086/375341}
\end{barticle}
\endbibitem

\bibitem[\protect\citeauthoryear{{Hirschi} et~al.}{2005}]{Hirschi+2005}
\begin{barticle}
\bauthor{\binits{R.} \bsnm{{Hirschi}}},
\bauthor{\binits{G.} \bsnm{{Meynet}}},
\bauthor{\binits{A.} \bsnm{{Maeder}}},
\batitle{{Stellar evolution with rotation. XIII. Predicted GRB rates at various
  Z}}.
\bjtitle{\aap}
\bvolume{443},
\bfpage{581}--\blpage{591}
(\byear{2005}).
doi:\doiurl{10.1051/0004-6361+20053329}
\end{barticle}
\endbibitem

\bibitem[\protect\citeauthoryear{{Hirschi} et~al.}{2008}]{Hirschi+2008}
\begin{bchapter}
\bauthor{\binits{R.} \bsnm{{Hirschi}}},
\bauthor{\binits{C.} \bsnm{{Chiappini}}},
\bauthor{\binits{G.} \bsnm{{Meynet}}},
\bauthor{\binits{A.} \bsnm{{Maeder}}},
\bauthor{\binits{S.} \bsnm{{Ekstr{\"o}m}}},
\bctitle{{Stellar Evolution at Low Metallicity}},
in \bbtitle{IAU Symposium},
ed. by \beditor{\binits{F.} \bsnm{{Bresolin}}},
\beditor{\binits{P.A.} \bsnm{{Crowther}}},
\beditor{\binits{J.} \bsnm{{Puls}}}
\bsertitle{IAU Symposium},
vol. \bseriesno{250},
\byear{2008},
pp. \bfpage{217}--\blpage{230}.
doi:\doiurl{10.1017/S1743921308020528}
\end{bchapter}
\endbibitem

\bibitem[\protect\citeauthoryear{{Hjorth} and {Bloom}}{2012}]{Hjorth+2006}
\begin{botherref}
\oauthor{\binits{J.} \bsnm{{Hjorth}}},
\oauthor{\binits{J.S.} \bsnm{{Bloom}}},
{The Gamma-Ray Burst - Supernova Connection}
2012,
pp. 169--190
\end{botherref}
\endbibitem

\bibitem[\protect\citeauthoryear{{Hjorth} et~al.}{2012}]{Hjorth+2012}
\begin{barticle}
\bauthor{\binits{J.} \bsnm{{Hjorth}}},
\bauthor{\binits{D.} \bsnm{{Malesani}}},
\bauthor{\binits{P.} \bsnm{{Jakobsson}}},
\bauthor{\binits{A.O.} \bsnm{{Jaunsen}}},
\bauthor{\binits{J.P.U.} \bsnm{{Fynbo}}},
\bauthor{\binits{J.} \bsnm{{Gorosabel}}},
\bauthor{\binits{T.} \bsnm{{Kr{\"u}hler}}},
\bauthor{\binits{A.J.} \bsnm{{Levan}}},
\bauthor{\binits{M.J.} \bsnm{{Micha{\l}owski}}},
\bauthor{\binits{B.} \bsnm{{Milvang-Jensen}}},
\bauthor{\binits{P.} \bsnm{{M{\o}ller}}},
\bauthor{\binits{S.} \bsnm{{Schulze}}},
\bauthor{\binits{N.R.} \bsnm{{Tanvir}}},
\bauthor{\binits{D.} \bsnm{{Watson}}},
\batitle{{The Optically Unbiased Gamma-Ray Burst Host (TOUGH) Survey. I. Survey
  Design and Catalogs}}.
\bjtitle{\apj}
\bvolume{756},
\bfpage{187}
(\byear{2012}).
doi:\doiurl{10.1088/0004-637X/756/2/187}
\end{barticle}
\endbibitem

\bibitem[\protect\citeauthoryear{{Hunt} et~al.}{2014}]{Hunt+2014}
\begin{barticle}
\bauthor{\binits{L.K.} \bsnm{{Hunt}}},
\bauthor{\binits{E.} \bsnm{{Palazzi}}},
\bauthor{\binits{M.J.} \bsnm{{Micha{\l}owski}}},
\bauthor{\binits{A.} \bsnm{{Rossi}}},
\bauthor{\binits{S.} \bsnm{{Savaglio}}},
\bauthor{\binits{S.} \bsnm{{Basa}}},
\bauthor{\binits{S.} \bsnm{{Berta}}},
\bauthor{\binits{S.} \bsnm{{Bianchi}}},
\bauthor{\binits{S.} \bsnm{{Covino}}},
\bauthor{\binits{V.} \bsnm{{D'Elia}}},
\bauthor{\binits{P.} \bsnm{{Ferrero}}},
\bauthor{\binits{D.} \bsnm{{G{\"o}tz}}},
\bauthor{\binits{J.} \bsnm{{Greiner}}},
\bauthor{\binits{S.} \bsnm{{Klose}}},
\bauthor{\binits{D.} \bsnm{{Le Borgne}}},
\bauthor{\binits{E.} \bsnm{{Le Floc'h}}},
\bauthor{\binits{E.} \bsnm{{Pian}}},
\bauthor{\binits{S.} \bsnm{{Piranomonte}}},
\bauthor{\binits{P.} \bsnm{{Schady}}},
\bauthor{\binits{S.D.} \bsnm{{Vergani}}},
\batitle{{New light on gamma-ray burst host galaxies with Herschel}}.
\bjtitle{\aap}
\bvolume{565},
\bfpage{112}
(\byear{2014}).
doi:\doiurl{10.1051/0004-6361/201323340}
\end{barticle}
\endbibitem

\bibitem[\protect\citeauthoryear{{Hunt} et~al.}{2011}]{Hunt+2011}
\begin{barticle}
\bauthor{\binits{L.} \bsnm{{Hunt}}},
\bauthor{\binits{E.} \bsnm{{Palazzi}}},
\bauthor{\binits{A.} \bsnm{{Rossi}}},
\bauthor{\binits{S.} \bsnm{{Savaglio}}},
\bauthor{\binits{G.} \bsnm{{Cresci}}},
\bauthor{\binits{S.} \bsnm{{Klose}}},
\bauthor{\binits{M.} \bsnm{{Micha{\l}owski}}},
\bauthor{\binits{E.} \bsnm{{Pian}}},
\batitle{{The Extremely Red Host Galaxy of GRB 080207}}.
\bjtitle{\apjl}
\bvolume{736},
\bfpage{36}
(\byear{2011}).
doi:\doiurl{10.1088/2041-8205/736/2/L36}
\end{barticle}
\endbibitem

\bibitem[\protect\citeauthoryear{{Izzard} et~al.}{2004}]{Izzard+2004}
\begin{barticle}
\bauthor{\binits{R.G.} \bsnm{{Izzard}}},
\bauthor{\binits{E.} \bsnm{{Ramirez-Ruiz}}},
\bauthor{\binits{C.A.} \bsnm{{Tout}}},
\batitle{{Formation rates of core-collapse supernovae and gamma-ray bursts}}.
\bjtitle{\mnras}
\bvolume{348},
\bfpage{1215}--\blpage{1228}
(\byear{2004}).
doi:\doiurl{10.1111/j.1365-2966.2004.07436.x}
\end{barticle}
\endbibitem

\bibitem[\protect\citeauthoryear{{Jakobsson} et~al.}{2004}]{Jakobsson+2004}
\begin{barticle}
\bauthor{\binits{P.} \bsnm{{Jakobsson}}},
\bauthor{\binits{J.} \bsnm{{Hjorth}}},
\bauthor{\binits{J.P.U.} \bsnm{{Fynbo}}},
\bauthor{\binits{D.} \bsnm{{Watson}}},
\bauthor{\binits{K.} \bsnm{{Pedersen}}},
\bauthor{\binits{G.} \bsnm{{Bj{\"o}rnsson}}},
\bauthor{\binits{J.} \bsnm{{Gorosabel}}},
\batitle{{Swift Identification of Dark Gamma-Ray Bursts}}.
\bjtitle{\apjl}
\bvolume{617},
\bfpage{21}--\blpage{24}
(\byear{2004}).
doi:\doiurl{10.1086/427089}
\end{barticle}
\endbibitem

\bibitem[\protect\citeauthoryear{{Jakobsson} et~al.}{2005}]{Jakobsson+2005}
\begin{barticle}
\bauthor{\binits{P.} \bsnm{{Jakobsson}}},
\bauthor{\binits{G.} \bsnm{{Bj{\"o}rnsson}}},
\bauthor{\binits{J.P.U.} \bsnm{{Fynbo}}},
\bauthor{\binits{G.} \bsnm{{J{\'o}hannesson}}},
\bauthor{\binits{J.} \bsnm{{Hjorth}}},
\bauthor{\binits{B.} \bsnm{{Thomsen}}},
\bauthor{\binits{P.} \bsnm{{M{\o}ller}}},
\bauthor{\binits{D.} \bsnm{{Watson}}},
\bauthor{\binits{B.L.} \bsnm{{Jensen}}},
\bauthor{\binits{G.} \bsnm{{{\"O}stlin}}},
\bauthor{\binits{J.} \bsnm{{Gorosabel}}},
\bauthor{\binits{E.H.} \bsnm{{Gudmundsson}}},
\batitle{{Ly-{$\alpha$} and ultraviolet emission from high-redshift gamma-ray
  burst hosts: to what extent do gamma-ray bursts trace star formation?}}
\bjtitle{\mnras}
\bvolume{362},
\bfpage{245}--\blpage{251}
(\byear{2005}).
doi:\doiurl{10.1111/j.1365-2966.2005.09289.x}
\end{barticle}
\endbibitem

\bibitem[\protect\citeauthoryear{{Japelj} et~al.}{2015}]{Japelj+2015}
\begin{barticle}
\bauthor{\binits{J.} \bsnm{{Japelj}}},
\bauthor{\binits{S.} \bsnm{{Covino}}},
\bauthor{\binits{A.} \bsnm{{Gomboc}}},
\bauthor{\binits{S.D.} \bsnm{{Vergani}}},
\bauthor{\binits{P.} \bsnm{{Goldoni}}},
\bauthor{\binits{J.} \bsnm{{Selsing}}},
\bauthor{\binits{Z.} \bsnm{{Cano}}},
\bauthor{\binits{V.} \bsnm{{D'Elia}}},
\bauthor{\binits{H.} \bsnm{{Flores}}},
\bauthor{\binits{J.P.U.} \bsnm{{Fynbo}}},
\bauthor{\binits{F.} \bsnm{{Hammer}}},
\bauthor{\binits{J.} \bsnm{{Hjorth}}},
\bauthor{\binits{P.} \bsnm{{Jakobsson}}},
\bauthor{\binits{L.} \bsnm{{Kaper}}},
\bauthor{\binits{D.} \bsnm{{Kopa{\v c}}}},
\bauthor{\binits{T.} \bsnm{{Kr{\"u}hler}}},
\bauthor{\binits{A.} \bsnm{{Melandri}}},
\bauthor{\binits{S.} \bsnm{{Piranomonte}}},
\bauthor{\binits{R.} \bsnm{{S{\'a}nchez-Ram{\'{\i}}rez}}},
\bauthor{\binits{G.} \bsnm{{Tagliaferri}}},
\bauthor{\binits{N.R.} \bsnm{{Tanvir}}},
\bauthor{\binits{A.} \bsnm{{de Ugarte Postigo}}},
\bauthor{\binits{D.} \bsnm{{Watson}}},
\bauthor{\binits{R.A.M.J.} \bsnm{{Wijers}}},
\batitle{{Spectrophotometric analysis of gamma-ray burst afterglow extinction
  curves with X-Shooter}}.
\bjtitle{\aap}
\bvolume{579},
\bfpage{74}
(\byear{2015}).
doi:\doiurl{10.1051/0004-6361/201525665}
\end{barticle}
\endbibitem

\bibitem[\protect\citeauthoryear{{Kajisawa} et~al.}{2011}]{Kajisawa+2011}
\begin{barticle}
\bauthor{\binits{M.} \bsnm{{Kajisawa}}},
\bauthor{\binits{T.} \bsnm{{Ichikawa}}},
\bauthor{\binits{I.} \bsnm{{Tanaka}}},
\bauthor{\binits{T.} \bsnm{{Yamada}}},
\bauthor{\binits{M.} \bsnm{{Akiyama}}},
\bauthor{\binits{R.} \bsnm{{Suzuki}}},
\bauthor{\binits{C.} \bsnm{{Tokoku}}},
\bauthor{\binits{Y.} \bsnm{{Katsuno Uchimoto}}},
\bauthor{\binits{M.} \bsnm{{Konishi}}},
\bauthor{\binits{T.} \bsnm{{Yoshikawa}}},
\bauthor{\binits{T.} \bsnm{{Nishimura}}},
\bauthor{\binits{K.} \bsnm{{Omata}}},
\bauthor{\binits{M.} \bsnm{{Ouchi}}},
\bauthor{\binits{I.} \bsnm{{Iwata}}},
\bauthor{\binits{T.} \bsnm{{Hamana}}},
\bauthor{\binits{M.} \bsnm{{Onodera}}},
\batitle{{MOIRCS Deep Survey. IX. Deep Near-Infrared Imaging Data and Source
  Catalog}}.
\bjtitle{\pasj}
\bvolume{63},
\bfpage{379}
(\byear{2011}).
doi:\doiurl{10.1093/pasj/63.sp2.S379}
\end{barticle}
\endbibitem

\bibitem[\protect\citeauthoryear{{Kann} et~al.}{2006}]{Kann2006}
\begin{barticle}
\bauthor{\binits{D.A.} \bsnm{{Kann}}},
\bauthor{\binits{S.} \bsnm{{Klose}}},
\bauthor{\binits{A.} \bsnm{{Zeh}}},
\batitle{{Signatures of Extragalactic Dust in Pre-Swift GRB Afterglows}}.
\bjtitle{\apj}
\bvolume{641},
\bfpage{993}--\blpage{1009}
(\byear{2006}).
doi:\doiurl{10.1086/500652}
\end{barticle}
\endbibitem

\bibitem[\protect\citeauthoryear{{Kann} et~al.}{2010}]{Kann+2010}
\begin{barticle}
\bauthor{\binits{D.A.} \bsnm{{Kann}}},
\bauthor{\binits{S.} \bsnm{{Klose}}},
\bauthor{\binits{B.} \bsnm{{Zhang}}},
\bauthor{\binits{D.} \bsnm{{Malesani}}},
\bauthor{\binits{E.} \bsnm{{Nakar}}},
\bauthor{\binits{A.} \bsnm{{Pozanenko}}},
\bauthor{\binits{A.C.} \bsnm{{Wilson}}},
\bauthor{\binits{N.R.} \bsnm{{Butler}}},
\bauthor{\binits{P.} \bsnm{{Jakobsson}}},
\bauthor{\binits{S.} \bsnm{{Schulze}}},
\bauthor{\binits{M.} \bsnm{{Andreev}}},
\bauthor{\binits{L.A.} \bsnm{{Antonelli}}},
\bauthor{\binits{I.F.} \bsnm{{Bikmaev}}},
\bauthor{\binits{V.} \bsnm{{Biryukov}}},
\bauthor{\binits{M.} \bsnm{{B{\"o}ttcher}}},
\bauthor{\binits{R.A.} \bsnm{{Burenin}}},
\bauthor{\binits{J.M.} \bsnm{{Castro Cer{\'o}n}}},
\bauthor{\binits{A.J.} \bsnm{{Castro-Tirado}}},
\bauthor{\binits{G.} \bsnm{{Chincarini}}},
\bauthor{\binits{B.E.} \bsnm{{Cobb}}},
\bauthor{\binits{S.} \bsnm{{Covino}}},
\bauthor{\binits{P.} \bsnm{{D'Avanzo}}},
\bauthor{\binits{V.} \bsnm{{D'Elia}}},
\bauthor{\binits{M.} \bsnm{{Della Valle}}},
\bauthor{\binits{A.} \bsnm{{de Ugarte Postigo}}},
\bauthor{\binits{Y.} \bsnm{{Efimov}}},
\bauthor{\binits{P.} \bsnm{{Ferrero}}},
\bauthor{\binits{D.} \bsnm{{Fugazza}}},
\bauthor{\binits{J.P.U.} \bsnm{{Fynbo}}},
\bauthor{\binits{M.} \bsnm{{G{\aa}lfalk}}},
\bauthor{\binits{F.} \bsnm{{Grundahl}}},
\bauthor{\binits{J.} \bsnm{{Gorosabel}}},
\bauthor{\binits{S.} \bsnm{{Gupta}}},
\bauthor{\binits{S.} \bsnm{{Guziy}}},
\bauthor{\binits{B.} \bsnm{{Hafizov}}},
\bauthor{\binits{J.} \bsnm{{Hjorth}}},
\bauthor{\binits{K.} \bsnm{{Holhjem}}},
\bauthor{\binits{M.} \bsnm{{Ibrahimov}}},
\bauthor{\binits{M.} \bsnm{{Im}}},
\bauthor{\binits{G.L.} \bsnm{{Israel}}},
\bauthor{\binits{M.} \bsnm{{Je{\'l}inek}}},
\bauthor{\binits{B.L.} \bsnm{{Jensen}}},
\bauthor{\binits{R.} \bsnm{{Karimov}}},
\bauthor{\binits{I.M.} \bsnm{{Khamitov}}},
\bauthor{\binits{{\"U}.} \bsnm{{Kizilo{\v g}lu}}},
\bauthor{\binits{E.} \bsnm{{Klunko}}},
\bauthor{\binits{P.} \bsnm{{Kub{\'a}nek}}},
\bauthor{\binits{A.S.} \bsnm{{Kutyrev}}},
\bauthor{\binits{P.} \bsnm{{Laursen}}},
\bauthor{\binits{A.J.} \bsnm{{Levan}}},
\bauthor{\binits{F.} \bsnm{{Mannucci}}},
\bauthor{\binits{C.M.} \bsnm{{Martin}}},
\bauthor{\binits{A.} \bsnm{{Mescheryakov}}},
\bauthor{\binits{N.} \bsnm{{Mirabal}}},
\bauthor{\binits{J.P.} \bsnm{{Norris}}},
\bauthor{\binits{J.-E.} \bsnm{{Ovaldsen}}},
\bauthor{\binits{D.} \bsnm{{Paraficz}}},
\bauthor{\binits{E.} \bsnm{{Pavlenko}}},
\bauthor{\binits{S.} \bsnm{{Piranomonte}}},
\bauthor{\binits{A.} \bsnm{{Rossi}}},
\bauthor{\binits{V.} \bsnm{{Rumyantsev}}},
\bauthor{\binits{R.} \bsnm{{Salinas}}},
\bauthor{\binits{A.} \bsnm{{Sergeev}}},
\bauthor{\binits{D.} \bsnm{{Sharapov}}},
\bauthor{\binits{J.} \bsnm{{Sollerman}}},
\bauthor{\binits{B.} \bsnm{{Stecklum}}},
\bauthor{\binits{L.} \bsnm{{Stella}}},
\bauthor{\binits{G.} \bsnm{{Tagliaferri}}},
\bauthor{\binits{N.R.} \bsnm{{Tanvir}}},
\bauthor{\binits{J.} \bsnm{{Telting}}},
\bauthor{\binits{V.} \bsnm{{Testa}}},
\bauthor{\binits{A.C.} \bsnm{{Updike}}},
\bauthor{\binits{A.} \bsnm{{Volnova}}},
\bauthor{\binits{D.} \bsnm{{Watson}}},
\bauthor{\binits{K.} \bsnm{{Wiersema}}},
\bauthor{\binits{D.} \bsnm{{Xu}}},
\batitle{{The Afterglows of Swift-era Gamma-ray Bursts. I. Comparing pre-Swift
  and Swift-era Long/Soft (Type II) GRB Optical Afterglows}}.
\bjtitle{\apj}
\bvolume{720},
\bfpage{1513}--\blpage{1558}
(\byear{2010}).
doi:\doiurl{10.1088/0004-637X/720/2/1513}
\end{barticle}
\endbibitem

\bibitem[\protect\citeauthoryear{{Kelly} et~al.}{2008}]{Kelly+2008}
\begin{barticle}
\bauthor{\binits{P.L.} \bsnm{{Kelly}}},
\bauthor{\binits{R.P.} \bsnm{{Kirshner}}},
\bauthor{\binits{M.} \bsnm{{Pahre}}},
\batitle{{Long {$\gamma$}-Ray Bursts and Type Ic Core-Collapse Supernovae Have
  Similar Locations in Hosts}}.
\bjtitle{\apj}
\bvolume{687},
\bfpage{1201}--\blpage{1207}
(\byear{2008}).
doi:\doiurl{10.1086/591925}
\end{barticle}
\endbibitem

\bibitem[\protect\citeauthoryear{{Kelly} et~al.}{2014}]{Kelly+2014}
\begin{barticle}
\bauthor{\binits{P.L.} \bsnm{{Kelly}}},
\bauthor{\binits{A.V.} \bsnm{{Filippenko}}},
\bauthor{\binits{M.} \bsnm{{Modjaz}}},
\bauthor{\binits{D.} \bsnm{{Kocevski}}},
\batitle{{The Host Galaxies of Fast-ejecta Core-collapse Supernovae}}.
\bjtitle{\apj}
\bvolume{789},
\bfpage{23}
(\byear{2014}).
doi:\doiurl{10.1088/0004-637X/789/1/23}
\end{barticle}
\endbibitem

\bibitem[\protect\citeauthoryear{{Kewley} and {Dopita}}{2002}]{Kewley+2002}
\begin{barticle}
\bauthor{\binits{L.J.} \bsnm{{Kewley}}},
\bauthor{\binits{M.A.} \bsnm{{Dopita}}},
\batitle{{Using Strong Lines to Estimate Abundances in Extragalactic H II
  Regions and Starburst Galaxies}}.
\bjtitle{\apjs}
\bvolume{142},
\bfpage{35}--\blpage{52}
(\byear{2002}).
doi:\doiurl{10.1086/341326}
\end{barticle}
\endbibitem

\bibitem[\protect\citeauthoryear{{Kobulnicky} and
  {Kewley}}{2004}]{Kobulnicky+2004}
\begin{barticle}
\bauthor{\binits{H.A.} \bsnm{{Kobulnicky}}},
\bauthor{\binits{L.J.} \bsnm{{Kewley}}},
\batitle{{Metallicities of z=0.3-1.0 Galaxies in the GOODS-North Field}}.
\bjtitle{\apj}
\bvolume{617},
\bfpage{240}--\blpage{261}
(\byear{2004}).
doi:\doiurl{10.1086/425299}
\end{barticle}
\endbibitem

\bibitem[\protect\citeauthoryear{{Kocevski} et~al.}{2009}]{Kocevski+2009}
\begin{barticle}
\bauthor{\binits{D.} \bsnm{{Kocevski}}},
\bauthor{\binits{A.A.} \bsnm{{West}}},
\bauthor{\binits{M.} \bsnm{{Modjaz}}},
\batitle{{Modeling the GRB Host Galaxy Mass Distribution: Are GRBs Unbiased
  Tracers of Star Formation?}}
\bjtitle{\apj}
\bvolume{702},
\bfpage{377}--\blpage{385}
(\byear{2009}).
doi:\doiurl{10.1088/0004-637X/702/1/377}
\end{barticle}
\endbibitem

\bibitem[\protect\citeauthoryear{{Kohn} et~al.}{2015}]{Kohn+2015}
\begin{barticle}
\bauthor{\binits{S.A.} \bsnm{{Kohn}}},
\bauthor{\binits{M.J.} \bsnm{{Micha{\l}owski}}},
\bauthor{\binits{N.} \bsnm{{Bourne}}},
\bauthor{\binits{M.} \bsnm{{Baes}}},
\bauthor{\binits{J.} \bsnm{{Fritz}}},
\bauthor{\binits{A.} \bsnm{{Cooray}}},
\bauthor{\binits{I.} \bsnm{{De Looze}}},
\bauthor{\binits{G.} \bsnm{{De Zotti}}},
\bauthor{\binits{H.} \bsnm{{Dannerbauer}}},
\bauthor{\binits{L.} \bsnm{{Dunne}}},
\bauthor{\binits{S.} \bsnm{{Dye}}},
\bauthor{\binits{S.} \bsnm{{Eales}}},
\bauthor{\binits{C.} \bsnm{{Furlanetto}}},
\bauthor{\binits{J.} \bsnm{{Gonzalez-Nuevo}}},
\bauthor{\binits{E.} \bsnm{{Ibar}}},
\bauthor{\binits{R.J.} \bsnm{{Ivison}}},
\bauthor{\binits{S.J.} \bsnm{{Maddox}}},
\bauthor{\binits{D.} \bsnm{{Scott}}},
\bauthor{\binits{D.J.B.} \bsnm{{Smith}}},
\bauthor{\binits{M.W.L.} \bsnm{{Smith}}},
\bauthor{\binits{M.} \bsnm{{Symeonidis}}},
\bauthor{\binits{E.} \bsnm{{Valiante}}},
\batitle{{Far-infrared observations of an unbiased sample of gamma-ray burst
  host galaxies}}.
\bjtitle{\mnras}
\bvolume{448},
\bfpage{1494}--\blpage{1503}
(\byear{2015}).
doi:\doiurl{10.1093/mnras/stv088}
\end{barticle}
\endbibitem

\bibitem[\protect\citeauthoryear{{Krongold} and
  {Prochaska}}{2013}]{Krongold2013}
\begin{barticle}
\bauthor{\binits{Y.} \bsnm{{Krongold}}},
\bauthor{\binits{J.X.} \bsnm{{Prochaska}}},
\batitle{{An Explanation for the Different X-Ray to Optical Column Densities in
  the Environments of Gamma Ray Bursts: A Progenitor Embedded in a Dense
  Medium}}.
\bjtitle{\apj}
\bvolume{774},
\bfpage{115}
(\byear{2013}).
doi:\doiurl{10.1088/0004-637X/774/2/115}
\end{barticle}
\endbibitem

\bibitem[\protect\citeauthoryear{{Kr{\"u}hler} et~al.}{2008}]{Kruehler+2008}
\begin{barticle}
\bauthor{\binits{T.} \bsnm{{Kr{\"u}hler}}},
\bauthor{\binits{A.} \bsnm{{K{\"u}pc{\"u} Yolda{\c s}}}},
\bauthor{\binits{J.} \bsnm{{Greiner}}},
\bauthor{\binits{C.} \bsnm{{Clemens}}},
\bauthor{\binits{S.} \bsnm{{McBreen}}},
\bauthor{\binits{N.} \bsnm{{Primak}}},
\bauthor{\binits{S.} \bsnm{{Savaglio}}},
\bauthor{\binits{A.} \bsnm{{Yolda{\c s}}}},
\bauthor{\binits{G.P.} \bsnm{{Szokoly}}},
\bauthor{\binits{S.} \bsnm{{Klose}}},
\batitle{{The 2175 {\AA} Dust Feature in a Gamma-Ray Burst Afterglow at
  Redshift 2.45}}.
\bjtitle{\apj}
\bvolume{685},
\bfpage{376}--\blpage{383}
(\byear{2008}).
doi:\doiurl{10.1086/590240}
\end{barticle}
\endbibitem

\bibitem[\protect\citeauthoryear{{Kr{\"u}hler} et~al.}{2011}]{Kruehler+2011}
\begin{barticle}
\bauthor{\binits{T.} \bsnm{{Kr{\"u}hler}}},
\bauthor{\binits{J.} \bsnm{{Greiner}}},
\bauthor{\binits{P.} \bsnm{{Schady}}},
\bauthor{\binits{S.} \bsnm{{Savaglio}}},
\bauthor{\binits{P.M.J.} \bsnm{{Afonso}}},
\bauthor{\binits{C.} \bsnm{{Clemens}}},
\bauthor{\binits{J.} \bsnm{{Elliott}}},
\bauthor{\binits{R.} \bsnm{{Filgas}}},
\bauthor{\binits{D.} \bsnm{{Gruber}}},
\bauthor{\binits{D.A.} \bsnm{{Kann}}},
\bauthor{\binits{S.} \bsnm{{Klose}}},
\bauthor{\binits{A.} \bsnm{{K{\"u}pc{\"u}-Yolda{\c s}}}},
\bauthor{\binits{S.} \bsnm{{McBreen}}},
\bauthor{\binits{F.} \bsnm{{Olivares}}},
\bauthor{\binits{D.} \bsnm{{Pierini}}},
\bauthor{\binits{A.} \bsnm{{Rau}}},
\bauthor{\binits{A.} \bsnm{{Rossi}}},
\bauthor{\binits{M.} \bsnm{{Nardini}}},
\bauthor{\binits{A.} \bsnm{{Nicuesa Guelbenzu}}},
\bauthor{\binits{V.} \bsnm{{Sudilovsky}}},
\bauthor{\binits{A.C.} \bsnm{{Updike}}},
\batitle{{The SEDs and host galaxies of the dustiest GRB afterglows}}.
\bjtitle{\aap}
\bvolume{534},
\bfpage{108}
(\byear{2011}).
doi:\doiurl{10.1051/0004-6361/201117428}
\end{barticle}
\endbibitem

\bibitem[\protect\citeauthoryear{{Kr{\"u}hler} et~al.}{2013}]{Kruhler2013}
\begin{barticle}
\bauthor{\binits{T.} \bsnm{{Kr{\"u}hler}}},
\bauthor{\binits{C.} \bsnm{{Ledoux}}},
\bauthor{\binits{J.P.U.} \bsnm{{Fynbo}}},
\bauthor{\binits{P.M.} \bsnm{{Vreeswijk}}},
\bauthor{\binits{S.} \bsnm{{Schmidl}}},
\bauthor{\binits{D.} \bsnm{{Malesani}}},
\bauthor{\binits{L.} \bsnm{{Christensen}}},
\bauthor{\binits{A.} \bsnm{{De Cia}}},
\bauthor{\binits{J.} \bsnm{{Hjorth}}},
\bauthor{\binits{P.} \bsnm{{Jakobsson}}},
\bauthor{\binits{D.A.} \bsnm{{Kann}}},
\bauthor{\binits{L.} \bsnm{{Kaper}}},
\bauthor{\binits{S.D.} \bsnm{{Vergani}}},
\bauthor{\binits{P.M.J.} \bsnm{{Afonso}}},
\bauthor{\binits{S.} \bsnm{{Covino}}},
\bauthor{\binits{A.} \bsnm{{de Ugarte Postigo}}},
\bauthor{\binits{V.} \bsnm{{D'Elia}}},
\bauthor{\binits{R.} \bsnm{{Filgas}}},
\bauthor{\binits{P.} \bsnm{{Goldoni}}},
\bauthor{\binits{J.} \bsnm{{Greiner}}},
\bauthor{\binits{O.E.} \bsnm{{Hartoog}}},
\bauthor{\binits{B.} \bsnm{{Milvang-Jensen}}},
\bauthor{\binits{M.} \bsnm{{Nardini}}},
\bauthor{\binits{S.} \bsnm{{Piranomonte}}},
\bauthor{\binits{A.} \bsnm{{Rossi}}},
\bauthor{\binits{R.} \bsnm{{S{\'a}nchez-Ram{\'{\i}}rez}}},
\bauthor{\binits{P.} \bsnm{{Schady}}},
\bauthor{\binits{S.} \bsnm{{Schulze}}},
\bauthor{\binits{V.} \bsnm{{Sudilovsky}}},
\bauthor{\binits{N.R.} \bsnm{{Tanvir}}},
\bauthor{\binits{G.} \bsnm{{Tagliaferri}}},
\bauthor{\binits{D.J.} \bsnm{{Watson}}},
\bauthor{\binits{K.} \bsnm{{Wiersema}}},
\bauthor{\binits{R.A.M.J.} \bsnm{{Wijers}}},
\bauthor{\binits{D.} \bsnm{{Xu}}},
\batitle{{Molecular hydrogen in the damped Lyman {$\alpha$} system towards GRB
  120815A at z = 2.36}}.
\bjtitle{\aap}
\bvolume{557},
\bfpage{18}
(\byear{2013}).
doi:\doiurl{10.1051/0004-6361/201321772}
\end{barticle}
\endbibitem

\bibitem[\protect\citeauthoryear{{Kr{\"u}hler} et~al.}{2015}]{Kruehler+2015}
\begin{barticle}
\bauthor{\binits{T.} \bsnm{{Kr{\"u}hler}}},
\bauthor{\binits{D.} \bsnm{{Malesani}}},
\bauthor{\binits{J.P.U.} \bsnm{{Fynbo}}},
\bauthor{\binits{O.E.} \bsnm{{Hartoog}}},
\bauthor{\binits{J.} \bsnm{{Hjorth}}},
\bauthor{\binits{P.} \bsnm{{Jakobsson}}},
\bauthor{\binits{D.A.} \bsnm{{Perley}}},
\bauthor{\binits{A.} \bsnm{{Rossi}}},
\bauthor{\binits{P.} \bsnm{{Schady}}},
\bauthor{\binits{S.} \bsnm{{Schulze}}},
\bauthor{\binits{N.R.} \bsnm{{Tanvir}}},
\bauthor{\binits{S.D.} \bsnm{{Vergani}}},
\bauthor{\binits{K.} \bsnm{{Wiersema}}},
\bauthor{\binits{P.M.J.} \bsnm{{Afonso}}},
\bauthor{\binits{J.} \bsnm{{Bolmer}}},
\bauthor{\binits{Z.} \bsnm{{Cano}}},
\bauthor{\binits{S.} \bsnm{{Covino}}},
\bauthor{\binits{V.} \bsnm{{D'Elia}}},
\bauthor{\binits{A.} \bsnm{{de Ugarte Postigo}}},
\bauthor{\binits{R.} \bsnm{{Filgas}}},
\bauthor{\binits{M.} \bsnm{{Friis}}},
\bauthor{\binits{J.F.} \bsnm{{Graham}}},
\bauthor{\binits{J.} \bsnm{{Greiner}}},
\bauthor{\binits{P.} \bsnm{{Goldoni}}},
\bauthor{\binits{A.} \bsnm{{Gomboc}}},
\bauthor{\binits{F.} \bsnm{{Hammer}}},
\bauthor{\binits{J.} \bsnm{{Japelj}}},
\bauthor{\binits{D.A.} \bsnm{{Kann}}},
\bauthor{\binits{L.} \bsnm{{Kaper}}},
\bauthor{\binits{S.} \bsnm{{Klose}}},
\bauthor{\binits{A.J.} \bsnm{{Levan}}},
\bauthor{\binits{G.} \bsnm{{Leloudas}}},
\bauthor{\binits{B.} \bsnm{{Milvang-Jensen}}},
\bauthor{\binits{A.} \bsnm{{Nicuesa Guelbenzu}}},
\bauthor{\binits{E.} \bsnm{{Palazzi}}},
\bauthor{\binits{E.} \bsnm{{Pian}}},
\bauthor{\binits{S.} \bsnm{{Piranomonte}}},
\bauthor{\binits{R.} \bsnm{{S{\'a}nchez-Ram{\'{\i}}rez}}},
\bauthor{\binits{S.} \bsnm{{Savaglio}}},
\bauthor{\binits{J.} \bsnm{{Selsing}}},
\bauthor{\binits{G.} \bsnm{{Tagliaferri}}},
\bauthor{\binits{P.M.} \bsnm{{Vreeswijk}}},
\bauthor{\binits{D.J.} \bsnm{{Watson}}},
\bauthor{\binits{D.} \bsnm{{Xu}}},
\batitle{{GRB hosts through cosmic time. VLT/X-Shooter emission-line
  spectroscopy of 96 {$\gamma$}-ray-burst-selected galaxies at 0.1 {$<$}z {$<$}
  3.6}}.
\bjtitle{\aap}
\bvolume{581},
\bfpage{125}
(\byear{2015}).
doi:\doiurl{10.1051/0004-6361/201425561}
\end{barticle}
\endbibitem

\bibitem[\protect\citeauthoryear{{Lamb} and {Reichart}}{2000}]{Lamb+2000}
\begin{barticle}
\bauthor{\binits{D.Q.} \bsnm{{Lamb}}},
\bauthor{\binits{D.E.} \bsnm{{Reichart}}},
\batitle{{Gamma-Ray Bursts as a Probe of the Very High Redshift Universe}}.
\bjtitle{\apj}
\bvolume{536},
\bfpage{1}--\blpage{18}
(\byear{2000}).
doi:\doiurl{10.1086/308918}
\end{barticle}
\endbibitem

\bibitem[\protect\citeauthoryear{{Langer} and {Norman}}{2006}]{Langer+2006}
\begin{barticle}
\bauthor{\binits{N.} \bsnm{{Langer}}},
\bauthor{\binits{C.A.} \bsnm{{Norman}}},
\batitle{{On the Collapsar Model of Long Gamma-Ray Bursts:Constraints from
  Cosmic Metallicity Evolution}}.
\bjtitle{\apjl}
\bvolume{638},
\bfpage{63}--\blpage{66}
(\byear{2006}).
doi:\doiurl{10.1086/500363}
\end{barticle}
\endbibitem

\bibitem[\protect\citeauthoryear{{Lapi} et~al.}{2008}]{Lapi+2008a}
\begin{barticle}
\bauthor{\binits{A.} \bsnm{{Lapi}}},
\bauthor{\binits{N.} \bsnm{{Kawakatu}}},
\bauthor{\binits{Z.} \bsnm{{Bosnjak}}},
\bauthor{\binits{A.} \bsnm{{Celotti}}},
\bauthor{\binits{A.} \bsnm{{Bressan}}},
\bauthor{\binits{G.L.} \bsnm{{Granato}}},
\bauthor{\binits{L.} \bsnm{{Danese}}},
\batitle{{Long gamma-ray bursts and their host galaxies at high redshift}}.
\bjtitle{\mnras}
\bvolume{386},
\bfpage{608}--\blpage{618}
(\byear{2008}).
doi:\doiurl{10.1111/j.1365-2966.2008.13076.x}
\end{barticle}
\endbibitem

\bibitem[\protect\citeauthoryear{{Larsson} et~al.}{2007}]{Larsson+2007}
\begin{barticle}
\bauthor{\binits{J.} \bsnm{{Larsson}}},
\bauthor{\binits{A.J.} \bsnm{{Levan}}},
\bauthor{\binits{M.B.} \bsnm{{Davies}}},
\bauthor{\binits{A.S.} \bsnm{{Fruchter}}},
\batitle{{A new constraint for gamma-ray burst progenitor mass}}.
\bjtitle{\mnras}
\bvolume{376},
\bfpage{1285}--\blpage{1290}
(\byear{2007}).
doi:\doiurl{10.1111/j.1365-2966.2007.11523.x}
\end{barticle}
\endbibitem

\bibitem[\protect\citeauthoryear{{Laskar} et~al.}{2011}]{Laskar+2011}
\begin{barticle}
\bauthor{\binits{T.} \bsnm{{Laskar}}},
\bauthor{\binits{E.} \bsnm{{Berger}}},
\bauthor{\binits{R.-R.} \bsnm{{Chary}}},
\batitle{{Exploring the Galaxy Mass-metallicity Relation at z \~{} 3-5}}.
\bjtitle{\apj}
\bvolume{739},
\bfpage{1}
(\byear{2011}).
doi:\doiurl{10.1088/0004-637X/739/1/1}
\end{barticle}
\endbibitem

\bibitem[\protect\citeauthoryear{{Lazzati} et~al.}{2002}]{Lazzati2002}
\begin{barticle}
\bauthor{\binits{D.} \bsnm{{Lazzati}}},
\bauthor{\binits{S.} \bsnm{{Covino}}},
\bauthor{\binits{G.} \bsnm{{Ghisellini}}},
\batitle{{On the role of extinction in failed gamma-ray burst optical/infrared
  afterglows}}.
\bjtitle{\mnras}
\bvolume{330},
\bfpage{583}--\blpage{590}
(\byear{2002}).
doi:\doiurl{10.1046/j.1365-8711.2002.05076.x}
\end{barticle}
\endbibitem

\bibitem[\protect\citeauthoryear{{Le Floc'h} et~al.}{2003}]{LeFloch+2003}
\begin{barticle}
\bauthor{\binits{E.} \bsnm{{Le Floc'h}}},
\bauthor{\binits{P.-A.} \bsnm{{Duc}}},
\bauthor{\binits{I.F.} \bsnm{{Mirabel}}},
\bauthor{\binits{D.B.} \bsnm{{Sanders}}},
\bauthor{\binits{G.} \bsnm{{Bosch}}},
\bauthor{\binits{R.J.} \bsnm{{Diaz}}},
\bauthor{\binits{C.J.} \bsnm{{Donzelli}}},
\bauthor{\binits{I.} \bsnm{{Rodrigues}}},
\bauthor{\binits{T.J.-L.} \bsnm{{Courvoisier}}},
\bauthor{\binits{J.} \bsnm{{Greiner}}},
\bauthor{\binits{S.} \bsnm{{Mereghetti}}},
\bauthor{\binits{J.} \bsnm{{Melnick}}},
\bauthor{\binits{J.} \bsnm{{Maza}}},
\bauthor{\binits{D.} \bsnm{{Minniti}}},
\batitle{{Are the hosts of gamma-ray bursts sub-luminous and blue galaxies?}}
\bjtitle{\aap}
\bvolume{400},
\bfpage{499}--\blpage{510}
(\byear{2003}).
doi:\doiurl{10.1051/0004-6361+20030001}
\end{barticle}
\endbibitem

\bibitem[\protect\citeauthoryear{{Le Floc'h} et~al.}{2006}]{LeFloch+2006}
\begin{barticle}
\bauthor{\binits{E.} \bsnm{{Le Floc'h}}},
\bauthor{\binits{V.} \bsnm{{Charmandaris}}},
\bauthor{\binits{W.J.} \bsnm{{Forrest}}},
\bauthor{\binits{I.F.} \bsnm{{Mirabel}}},
\bauthor{\binits{L.} \bsnm{{Armus}}},
\bauthor{\binits{D.} \bsnm{{Devost}}},
\batitle{{Probing Cosmic Star Formation Using Long Gamma-Ray Bursts: New
  Constraints from the Spitzer Space Telescope}}.
\bjtitle{\apj}
\bvolume{642},
\bfpage{636}--\blpage{652}
(\byear{2006}).
doi:\doiurl{10.1086/501118}
\end{barticle}
\endbibitem

\bibitem[\protect\citeauthoryear{{Leighly} et~al.}{2014}]{Karen14}
\begin{barticle}
\bauthor{\binits{K.M.} \bsnm{{Leighly}}},
\bauthor{\binits{D.M.} \bsnm{{Terndrup}}},
\bauthor{\binits{E.} \bsnm{{Baron}}},
\bauthor{\binits{A.B.} \bsnm{{Lucy}}},
\bauthor{\binits{M.} \bsnm{{Dietrich}}},
\bauthor{\binits{S.C.} \bsnm{{Gallagher}}},
\batitle{{Evidence for Active Galactic Nucleus Feedback in the Broad Absorption
  Lines and Reddening of Mrk 231}}.
\bjtitle{\apj}
\bvolume{788},
\bfpage{123}
(\byear{2014}).
doi:\doiurl{10.1088/0004-637X/788/2/123}
\end{barticle}
\endbibitem

\bibitem[\protect\citeauthoryear{{Levan} et~al.}{2006}]{Levan+2006}
\begin{barticle}
\bauthor{\binits{A.} \bsnm{{Levan}}},
\bauthor{\binits{A.} \bsnm{{Fruchter}}},
\bauthor{\binits{J.} \bsnm{{Rhoads}}},
\bauthor{\binits{B.} \bsnm{{Mobasher}}},
\bauthor{\binits{N.} \bsnm{{Tanvir}}},
\bauthor{\binits{J.} \bsnm{{Gorosabel}}},
\bauthor{\binits{E.} \bsnm{{Rol}}},
\bauthor{\binits{C.} \bsnm{{Kouveliotou}}},
\bauthor{\binits{I.} \bsnm{{Dell'Antonio}}},
\bauthor{\binits{M.} \bsnm{{Merrill}}},
\bauthor{\binits{E.} \bsnm{{Bergeron}}},
\bauthor{\binits{J.M.} \bsnm{{Castro Cer{\'o}n}}},
\bauthor{\binits{N.} \bsnm{{Masetti}}},
\bauthor{\binits{P.} \bsnm{{Vreeswijk}}},
\bauthor{\binits{A.} \bsnm{{Antonelli}}},
\bauthor{\binits{D.} \bsnm{{Bersier}}},
\bauthor{\binits{A.} \bsnm{{Castro-Tirado}}},
\bauthor{\binits{J.} \bsnm{{Fynbo}}},
\bauthor{\binits{P.} \bsnm{{Garnavich}}},
\bauthor{\binits{S.} \bsnm{{Holland}}},
\bauthor{\binits{J.} \bsnm{{Hjorth}}},
\bauthor{\binits{P.} \bsnm{{Nugent}}},
\bauthor{\binits{E.} \bsnm{{Pian}}},
\bauthor{\binits{A.} \bsnm{{Smette}}},
\bauthor{\binits{B.} \bsnm{{Thomsen}}},
\bauthor{\binits{S.E.} \bsnm{{Thorsett}}},
\bauthor{\binits{R.} \bsnm{{Wijers}}},
\batitle{{Infrared and Optical Observations of GRB 030115 and its Extremely Red
  Host Galaxy: Implications for Dark Bursts}}.
\bjtitle{\apj}
\bvolume{647},
\bfpage{471}--\blpage{482}
(\byear{2006}).
doi:\doiurl{10.1086/503595}
\end{barticle}
\endbibitem

\bibitem[\protect\citeauthoryear{{Levesque} et~al.}{2010a}]{Levesque+2010zrich}
\begin{barticle}
\bauthor{\binits{E.M.} \bsnm{{Levesque}}},
\bauthor{\binits{L.J.} \bsnm{{Kewley}}},
\bauthor{\binits{J.F.} \bsnm{{Graham}}},
\bauthor{\binits{A.S.} \bsnm{{Fruchter}}},
\batitle{{A High-metallicity Host Environment for the Long-duration GRB
  020819}}.
\bjtitle{\apjl}
\bvolume{712},
\bfpage{26}--\blpage{30}
(\byear{2010}a).
doi:\doiurl{10.1088/2041-8205/712/1/L26}
\end{barticle}
\endbibitem

\bibitem[\protect\citeauthoryear{{Levesque} et~al.}{2010b}]{Levesque+2010eiso}
\begin{barticle}
\bauthor{\binits{E.M.} \bsnm{{Levesque}}},
\bauthor{\binits{A.M.} \bsnm{{Soderberg}}},
\bauthor{\binits{L.J.} \bsnm{{Kewley}}},
\bauthor{\binits{E.} \bsnm{{Berger}}},
\batitle{{No Correlation Between Host Galaxy Metallicity and Gamma-ray Energy
  Release for Long-duration Gamma-ray Bursts}}.
\bjtitle{\apj}
\bvolume{725},
\bfpage{1337}--\blpage{1341}
(\byear{2010}b).
doi:\doiurl{10.1088/0004-637X/725/1/1337}
\end{barticle}
\endbibitem

\bibitem[\protect\citeauthoryear{{Levesque} et~al.}{2010c}]{Levesque+2010mz}
\begin{barticle}
\bauthor{\binits{E.M.} \bsnm{{Levesque}}},
\bauthor{\binits{L.J.} \bsnm{{Kewley}}},
\bauthor{\binits{E.} \bsnm{{Berger}}},
\bauthor{\binits{H.J.} \bsnm{{Zahid}}},
\batitle{{The Host Galaxies of Gamma-ray Bursts. II. A Mass-metallicity
  Relation for Long-duration Gamma-ray Burst Host Galaxies}}.
\bjtitle{\aj}
\bvolume{140},
\bfpage{1557}--\blpage{1566}
(\byear{2010}c).
doi:\doiurl{10.1088/0004-6256/140/5/1557}
\end{barticle}
\endbibitem

\bibitem[\protect\citeauthoryear{{Levesque} et~al.}{2011}]{Levesque+2011}
\begin{barticle}
\bauthor{\binits{E.M.} \bsnm{{Levesque}}},
\bauthor{\binits{E.} \bsnm{{Berger}}},
\bauthor{\binits{A.M.} \bsnm{{Soderberg}}},
\bauthor{\binits{R.} \bsnm{{Chornock}}},
\batitle{{Metallicity in the GRB 100316D/SN 2010bh Host Complex}}.
\bjtitle{\apj}
\bvolume{739},
\bfpage{23}
(\byear{2011}).
doi:\doiurl{10.1088/0004-637X/739/1/23}
\end{barticle}
\endbibitem

\bibitem[\protect\citeauthoryear{{Littlejohns} et~al.}{2015}]{Littlejohns2015}
\begin{barticle}
\bauthor{\binits{O.M.} \bsnm{{Littlejohns}}},
\bauthor{\binits{N.R.} \bsnm{{Butler}}},
\bauthor{\binits{A.} \bsnm{{Cucchiara}}},
\bauthor{\binits{A.M.} \bsnm{{Watson}}},
\bauthor{\binits{O.D.} \bsnm{{Fox}}},
\bauthor{\binits{W.H.} \bsnm{{Lee}}},
\bauthor{\binits{A.S.} \bsnm{{Kutyrev}}},
\bauthor{\binits{M.G.} \bsnm{{Richer}}},
\bauthor{\binits{C.R.} \bsnm{{Klein}}},
\bauthor{\binits{J.X.} \bsnm{{Prochaska}}},
\bauthor{\binits{J.S.} \bsnm{{Bloom}}},
\bauthor{\binits{E.} \bsnm{{Troja}}},
\bauthor{\binits{E.} \bsnm{{Ramirez-Ruiz}}},
\bauthor{\binits{J.A.} \bsnm{{de Diego}}},
\bauthor{\binits{L.} \bsnm{{Georgiev}}},
\bauthor{\binits{J.} \bsnm{{Gonz{\'a}lez}}},
\bauthor{\binits{C.G.} \bsnm{{Rom{\'a}n-Z{\'u}{\~n}iga}}},
\bauthor{\binits{N.} \bsnm{{Gehrels}}},
\bauthor{\binits{H.} \bsnm{{Moseley}}},
\batitle{{A detailed study of the optical attenuation of gamma-ray bursts in
  the Swift era}}.
\bjtitle{\mnras}
\bvolume{449},
\bfpage{2919}--\blpage{2936}
(\byear{2015}).
doi:\doiurl{10.1093/mnras/stv479}
\end{barticle}
\endbibitem

\bibitem[\protect\citeauthoryear{{MacFadyen} and
  {Woosley}}{1999}]{MacFadyen+1999}
\begin{barticle}
\bauthor{\binits{A.I.} \bsnm{{MacFadyen}}},
\bauthor{\binits{S.E.} \bsnm{{Woosley}}},
\batitle{{Collapsars: Gamma-Ray Bursts and Explosions in ``Failed
  Supernovae''}}.
\bjtitle{\apj}
\bvolume{524},
\bfpage{262}--\blpage{289}
(\byear{1999}).
doi:\doiurl{10.1086/307790}
\end{barticle}
\endbibitem

\bibitem[\protect\citeauthoryear{{McGuire} et~al.}{2015}]{McGuire+2015}
\begin{botherref}
\oauthor{\binits{J.T.W.} \bsnm{{McGuire}}},
\oauthor{\binits{N.R.} \bsnm{{Tanvir}}},
\oauthor{\binits{A.J.} \bsnm{{Levan}}},
\oauthor{\binits{M.} \bsnm{{Trenti}}},
\oauthor{\binits{E.R.} \bsnm{{Stanway}}},
\oauthor{\binits{J.M.} \bsnm{{Shull}}},
\oauthor{\binits{K.} \bsnm{{Wiersema}}},
\oauthor{\binits{D.A.} \bsnm{{Perley}}},
\oauthor{\binits{R.L.C.} \bsnm{{Starling}}},
\oauthor{\binits{M.} \bsnm{{Bremer}}},
\oauthor{\binits{J.T.} \bsnm{{Stocke}}},
\oauthor{\binits{J.} \bsnm{{Hjorth}}},
\oauthor{\binits{J.E.} \bsnm{{Rhoads}}},
\oauthor{\binits{E.M.} \bsnm{{Levesque}}},
\oauthor{\binits{B.} \bsnm{{Robertson}}},
\oauthor{\binits{J.P.U.} \bsnm{{Fynbo}}},
\oauthor{\binits{R.S.} \bsnm{{Ellis}}},
\oauthor{\binits{A.S.} \bsnm{{Fruchter}}},
\oauthor{\binits{R.} \bsnm{{Perna}}},
{Detection of three Gamma-Ray Burst host galaxies at $z\sim6$}.
ArXiv e-prints
(2015)
\end{botherref}
\endbibitem

\bibitem[\protect\citeauthoryear{{Melandri} et~al.}{2012}]{Melandri+2012}
\begin{barticle}
\bauthor{\binits{A.} \bsnm{{Melandri}}},
\bauthor{\binits{B.} \bsnm{{Sbarufatti}}},
\bauthor{\binits{P.} \bsnm{{D'Avanzo}}},
\bauthor{\binits{R.} \bsnm{{Salvaterra}}},
\bauthor{\binits{S.} \bsnm{{Campana}}},
\bauthor{\binits{S.} \bsnm{{Covino}}},
\bauthor{\binits{S.D.} \bsnm{{Vergani}}},
\bauthor{\binits{L.} \bsnm{{Nava}}},
\bauthor{\binits{G.} \bsnm{{Ghisellini}}},
\bauthor{\binits{G.} \bsnm{{Ghirlanda}}},
\bauthor{\binits{D.} \bsnm{{Fugazza}}},
\bauthor{\binits{V.} \bsnm{{Mangano}}},
\bauthor{\binits{M.} \bsnm{{Capalbi}}},
\bauthor{\binits{G.} \bsnm{{Tagliaferri}}},
\batitle{{The dark bursts population in a complete sample of bright Swift long
  gamma-ray bursts}}.
\bjtitle{\mnras}
\bvolume{421},
\bfpage{1265}--\blpage{1272}
(\byear{2012}).
doi:\doiurl{10.1111/j.1365-2966.2011.20398.x}
\end{barticle}
\endbibitem

\bibitem[\protect\citeauthoryear{{Micha{\l}owski}
  et~al.}{2008}]{Michalowski+2008}
\begin{barticle}
\bauthor{\binits{M.J.} \bsnm{{Micha{\l}owski}}},
\bauthor{\binits{J.} \bsnm{{Hjorth}}},
\bauthor{\binits{J.M.} \bsnm{{Castro Cer{\'o}n}}},
\bauthor{\binits{D.} \bsnm{{Watson}}},
\batitle{{The Nature of GRB-selected Submillimeter Galaxies: Hot and Young}}.
\bjtitle{\apj}
\bvolume{672},
\bfpage{817}--\blpage{824}
(\byear{2008}).
doi:\doiurl{10.1086/523891}
\end{barticle}
\endbibitem

\bibitem[\protect\citeauthoryear{{Micha{\l}owski}
  et~al.}{2012}]{Michalowski+2012}
\begin{barticle}
\bauthor{\binits{M.J.} \bsnm{{Micha{\l}owski}}},
\bauthor{\binits{A.} \bsnm{{Kamble}}},
\bauthor{\binits{J.} \bsnm{{Hjorth}}},
\bauthor{\binits{D.} \bsnm{{Malesani}}},
\bauthor{\binits{R.F.} \bsnm{{Reinfrank}}},
\bauthor{\binits{L.} \bsnm{{Bonavera}}},
\bauthor{\binits{J.M.} \bsnm{{Castro Cer{\'o}n}}},
\bauthor{\binits{E.} \bsnm{{Ibar}}},
\bauthor{\binits{J.S.} \bsnm{{Dunlop}}},
\bauthor{\binits{J.P.U.} \bsnm{{Fynbo}}},
\bauthor{\binits{M.A.} \bsnm{{Garrett}}},
\bauthor{\binits{P.} \bsnm{{Jakobsson}}},
\bauthor{\binits{D.L.} \bsnm{{Kaplan}}},
\bauthor{\binits{T.} \bsnm{{Kr{\"u}hler}}},
\bauthor{\binits{A.J.} \bsnm{{Levan}}},
\bauthor{\binits{M.} \bsnm{{Massardi}}},
\bauthor{\binits{S.} \bsnm{{Pal}}},
\bauthor{\binits{J.} \bsnm{{Sollerman}}},
\bauthor{\binits{N.R.} \bsnm{{Tanvir}}},
\bauthor{\binits{A.J.} \bsnm{{van der Horst}}},
\bauthor{\binits{D.} \bsnm{{Watson}}},
\bauthor{\binits{K.} \bsnm{{Wiersema}}},
\batitle{{The Optically Unbiased GRB Host (TOUGH) Survey. VI. Radio
  Observations at z {$<$}\~{} 1 and Consistency with Typical Star-forming
  Galaxies}}.
\bjtitle{\apj}
\bvolume{755},
\bfpage{85}
(\byear{2012}).
doi:\doiurl{10.1088/0004-637X/755/2/85}
\end{barticle}
\endbibitem

\bibitem[\protect\citeauthoryear{{Micha{\l}owski}
  et~al.}{2015}]{Michalowski+2015}
\begin{botherref}
\oauthor{\binits{M.J.} \bsnm{{Micha{\l}owski}}},
\oauthor{\binits{G.} \bsnm{{Gentile}}},
\oauthor{\binits{J.} \bsnm{{Hjorth}}},
\oauthor{\binits{M.R.} \bsnm{{Krumholz}}},
\oauthor{\binits{N.R.} \bsnm{{Tanvir}}},
\oauthor{\binits{P.} \bsnm{{Kamphuis}}},
\oauthor{\binits{D.} \bsnm{{Burlon}}},
\oauthor{\binits{M.} \bsnm{{Baes}}},
\oauthor{\binits{S.} \bsnm{{Basa}}},
\oauthor{\binits{S.} \bsnm{{Berta}}},
\oauthor{\binits{J.M.} \bsnm{{Castro Ceron}}},
\oauthor{\binits{D.} \bsnm{{Crosby}}},
\oauthor{\binits{V.} \bsnm{{D'Elia}}},
\oauthor{\binits{J.} \bsnm{{Elliott}}},
\oauthor{\binits{J.} \bsnm{{Greiner}}},
\oauthor{\binits{L.K.} \bsnm{{Hunt}}},
\oauthor{\binits{S.} \bsnm{{Klose}}},
\oauthor{\binits{M.P.} \bsnm{{Koprowski}}},
\oauthor{\binits{E.} \bsnm{{Le Floc'h}}},
\oauthor{\binits{D.} \bsnm{{Malesani}}},
\oauthor{\binits{T.} \bsnm{{Murphy}}},
\oauthor{\binits{A.} \bsnm{{Nicuesa Guelbenzu}}},
\oauthor{\binits{E.} \bsnm{{Palazzi}}},
\oauthor{\binits{J.} \bsnm{{Rasmussen}}},
\oauthor{\binits{A.} \bsnm{{Rossi}}},
\oauthor{\binits{S.} \bsnm{{Savaglio}}},
\oauthor{\binits{P.} \bsnm{{Schady}}},
\oauthor{\binits{J.} \bsnm{{Sollerman}}},
\oauthor{\binits{A.} \bsnm{{de Ugarte Postigo}}},
\oauthor{\binits{D.} \bsnm{{Watson}}},
\oauthor{\binits{P.} \bsnm{{van der Werf}}},
\oauthor{\binits{S.D.} \bsnm{{Vergani}}},
\oauthor{\binits{D.} \bsnm{{Xu}}},
{Massive stars formed in atomic hydrogen reservoirs: HI observations of
  gamma-ray burst host galaxies}.
ArXiv e-prints
(2015)
\end{botherref}
\endbibitem

\bibitem[\protect\citeauthoryear{{Milvang-Jensen}
  et~al.}{2012}]{MilvangJensen+2012}
\begin{barticle}
\bauthor{\binits{B.} \bsnm{{Milvang-Jensen}}},
\bauthor{\binits{J.P.U.} \bsnm{{Fynbo}}},
\bauthor{\binits{D.} \bsnm{{Malesani}}},
\bauthor{\binits{J.} \bsnm{{Hjorth}}},
\bauthor{\binits{P.} \bsnm{{Jakobsson}}},
\bauthor{\binits{P.} \bsnm{{M{\o}ller}}},
\batitle{{The Optically Unbiased GRB Host (TOUGH) Survey. IV. Ly{$\alpha$}
  Emitters}}.
\bjtitle{\apj}
\bvolume{756},
\bfpage{25}
(\byear{2012}).
doi:\doiurl{10.1088/0004-637X/756/1/25}
\end{barticle}
\endbibitem

\bibitem[\protect\citeauthoryear{{Modjaz} et~al.}{2008}]{Modjaz+2008}
\begin{barticle}
\bauthor{\binits{M.} \bsnm{{Modjaz}}},
\bauthor{\binits{L.} \bsnm{{Kewley}}},
\bauthor{\binits{R.P.} \bsnm{{Kirshner}}},
\bauthor{\binits{K.Z.} \bsnm{{Stanek}}},
\bauthor{\binits{P.} \bsnm{{Challis}}},
\bauthor{\binits{P.M.} \bsnm{{Garnavich}}},
\bauthor{\binits{J.E.} \bsnm{{Greene}}},
\bauthor{\binits{P.L.} \bsnm{{Kelly}}},
\bauthor{\binits{J.L.} \bsnm{{Prieto}}},
\batitle{{Measured Metallicities at the Sites of Nearby Broad-Lined Type Ic
  Supernovae and Implications for the Supernovae Gamma-Ray Burst Connection}}.
\bjtitle{\aj}
\bvolume{135},
\bfpage{1136}--\blpage{1150}
(\byear{2008}).
doi:\doiurl{10.1088/0004-6256/135/4/1136}
\end{barticle}
\endbibitem

\bibitem[\protect\citeauthoryear{{Modjaz} et~al.}{2011}]{Modjaz+2011}
\begin{barticle}
\bauthor{\binits{M.} \bsnm{{Modjaz}}},
\bauthor{\binits{L.} \bsnm{{Kewley}}},
\bauthor{\binits{J.S.} \bsnm{{Bloom}}},
\bauthor{\binits{A.V.} \bsnm{{Filippenko}}},
\bauthor{\binits{D.} \bsnm{{Perley}}},
\bauthor{\binits{J.M.} \bsnm{{Silverman}}},
\batitle{{Progenitor Diagnostics for Stripped Core-collapse Supernovae:
  Measured Metallicities at Explosion Sites}}.
\bjtitle{\apjl}
\bvolume{731},
\bfpage{4}
(\byear{2011}).
doi:\doiurl{10.1088/2041-8205/731/1/L4}
\end{barticle}
\endbibitem

\bibitem[\protect\citeauthoryear{{Nataf} et~al.}{2015}]{Nataf2015}
\begin{botherref}
\oauthor{\binits{D.M.} \bsnm{{Nataf}}},
\oauthor{\binits{O.A.} \bsnm{{Gonzalez}}},
\oauthor{\binits{L.} \bsnm{{Casagrande}}},
\oauthor{\binits{G.} \bsnm{{Zasowski}}},
\oauthor{\binits{C.} \bsnm{{Wegg}}},
\oauthor{\binits{C.} \bsnm{{Wolf}}},
\oauthor{\binits{A.} \bsnm{{Kunder}}},
\oauthor{\binits{J.} \bsnm{{Alonso-Garcia}}},
\oauthor{\binits{D.} \bsnm{{Minniti}}},
\oauthor{\binits{M.} \bsnm{{Rejkuba}}},
\oauthor{\binits{R.K.} \bsnm{{Saito}}},
\oauthor{\binits{E.} \bsnm{{Valenti}}},
\oauthor{\binits{M.} \bsnm{{Zoccali}}},
\oauthor{\binits{R.} \bsnm{{Poleski}}},
\oauthor{\binits{G.} \bsnm{{Pietrzynski}}},
\oauthor{\binits{J.} \bsnm{{Skowron}}},
\oauthor{\binits{I.} \bsnm{{Soszynski}}},
\oauthor{\binits{M.K.} \bsnm{{Szymanski}}},
\oauthor{\binits{A.} \bsnm{{Udalski}}},
\oauthor{\binits{K.} \bsnm{{Ulaczyk}}},
\oauthor{\binits{L.} \bsnm{{Wyrzykowski}}},
{Interstellar Extinction Curve Variations Toward the Inner Milky Way: A
  Challenge to Observational Cosmology}.
ArXiv e-prints
(2015)
\end{botherref}
\endbibitem

\bibitem[\protect\citeauthoryear{{Niino}}{2011}]{Niino+2011b}
\begin{barticle}
\bauthor{\binits{Y.} \bsnm{{Niino}}},
\batitle{{Revisiting the metallicity of long-duration gamma-ray burst host
  galaxies: the role of chemical inhomogeneity within galaxies}}.
\bjtitle{\mnras}
\bvolume{417},
\bfpage{567}--\blpage{572}
(\byear{2011}).
doi:\doiurl{10.1111/j.1365-2966.2011.19299.x}
\end{barticle}
\endbibitem

\bibitem[\protect\citeauthoryear{{Niino} et~al.}{2015}]{Niino+2015}
\begin{barticle}
\bauthor{\binits{Y.} \bsnm{{Niino}}},
\bauthor{\binits{K.} \bsnm{{Nagamine}}},
\bauthor{\binits{B.} \bsnm{{Zhang}}},
\batitle{{Metallicity measurements of gamma-ray burst and supernova explosion
  sites: lessons from H II regions in M31}}.
\bjtitle{\mnras}
\bvolume{449},
\bfpage{2706}--\blpage{2717}
(\byear{2015}).
doi:\doiurl{10.1093/mnras/stv436}
\end{barticle}
\endbibitem

\bibitem[\protect\citeauthoryear{{Niino} et~al.}{2011}]{Niino+2011a}
\begin{barticle}
\bauthor{\binits{Y.} \bsnm{{Niino}}},
\bauthor{\binits{J.} \bsnm{{Choi}}},
\bauthor{\binits{M.A.R.} \bsnm{{Kobayashi}}},
\bauthor{\binits{K.} \bsnm{{Nagamine}}},
\bauthor{\binits{T.} \bsnm{{Totani}}},
\bauthor{\binits{B.} \bsnm{{Zhang}}},
\batitle{{Luminosity Distribution of Gamma-ray Burst Host Galaxies at Redshift
  z = 1 in Cosmological Smoothed Particle Hydrodynamic Simulations:
  Implications for the Metallicity Dependence of GRBs}}.
\bjtitle{\apj}
\bvolume{726},
\bfpage{88}
(\byear{2011}).
doi:\doiurl{10.1088/0004-637X/726/2/88}
\end{barticle}
\endbibitem

\bibitem[\protect\citeauthoryear{{Nishiyama} et~al.}{2008}]{Nishiyama2008}
\begin{barticle}
\bauthor{\binits{S.} \bsnm{{Nishiyama}}},
\bauthor{\binits{T.} \bsnm{{Nagata}}},
\bauthor{\binits{M.} \bsnm{{Tamura}}},
\bauthor{\binits{R.} \bsnm{{Kandori}}},
\bauthor{\binits{H.} \bsnm{{Hatano}}},
\bauthor{\binits{S.} \bsnm{{Sato}}},
\bauthor{\binits{K.} \bsnm{{Sugitani}}},
\batitle{{The Interstellar Extinction Law toward the Galactic Center. II. V, J,
  H, and K$_{s}$ Bands}}.
\bjtitle{\apj}
\bvolume{680},
\bfpage{1174}--\blpage{1179}
(\byear{2008}).
doi:\doiurl{10.1086/587791}
\end{barticle}
\endbibitem

\bibitem[\protect\citeauthoryear{{Nishiyama} et~al.}{2009}]{Nishiyama2009}
\begin{barticle}
\bauthor{\binits{S.} \bsnm{{Nishiyama}}},
\bauthor{\binits{M.} \bsnm{{Tamura}}},
\bauthor{\binits{H.} \bsnm{{Hatano}}},
\bauthor{\binits{D.} \bsnm{{Kato}}},
\bauthor{\binits{T.} \bsnm{{Tanab{\'e}}}},
\bauthor{\binits{K.} \bsnm{{Sugitani}}},
\bauthor{\binits{T.} \bsnm{{Nagata}}},
\batitle{{Interstellar Extinction Law Toward the Galactic Center III: J, H,
  K$_{S}$ Bands in the 2MASS and the MKO Systems}}.
\bjtitle{\apj}
\bvolume{696},
\bfpage{1407}--\blpage{1417}
(\byear{2009}).
doi:\doiurl{10.1088/0004-637X/696/2/1407}
\end{barticle}
\endbibitem

\bibitem[\protect\citeauthoryear{{Nuza} et~al.}{2007}]{Nuza+2007a}
\begin{barticle}
\bauthor{\binits{S.E.} \bsnm{{Nuza}}},
\bauthor{\binits{P.B.} \bsnm{{Tissera}}},
\bauthor{\binits{L.J.} \bsnm{{Pellizza}}},
\bauthor{\binits{D.G.} \bsnm{{Lambas}}},
\bauthor{\binits{C.} \bsnm{{Scannapieco}}},
\bauthor{\binits{M.E.} \bsnm{{de Rossi}}},
\batitle{{The host galaxies of long-duration gamma-ray bursts in a cosmological
  hierarchical scenario}}.
\bjtitle{\mnras}
\bvolume{375},
\bfpage{665}--\blpage{672}
(\byear{2007}).
doi:\doiurl{10.1111/j.1365-2966.2006.11324.x}
\end{barticle}
\endbibitem

\bibitem[\protect\citeauthoryear{{Perley} and
  {Kemper}}{2008}]{Perley+2008grbox}
\begin{bchapter}
\bauthor{\binits{D.} \bsnm{{Perley}}},
\bauthor{\binits{Y.} \bsnm{{Kemper}}},
\bctitle{{GRBOX: A New Online Catalog of GRBs}},
in \bbtitle{American Institute of Physics Conference Series},
ed. by \beditor{\binits{M.} \bsnm{{Galassi}}},
\beditor{\binits{D.} \bsnm{{Palmer}}},
\beditor{\binits{E.} \bsnm{{Fenimore}}}
\bsertitle{American Institute of Physics Conference Series},
vol. \bseriesno{1000},
\byear{2008},
pp. \bfpage{631}--\blpage{634}.
doi:\doiurl{10.1063/1.2943549}
\end{bchapter}
\endbibitem

\bibitem[\protect\citeauthoryear{{Perley} et~al.}{2008}]{Perley+2008}
\begin{barticle}
\bauthor{\binits{D.A.} \bsnm{{Perley}}},
\bauthor{\binits{J.S.} \bsnm{{Bloom}}},
\bauthor{\binits{N.R.} \bsnm{{Butler}}},
\bauthor{\binits{L.K.} \bsnm{{Pollack}}},
\bauthor{\binits{J.} \bsnm{{Holtzman}}},
\bauthor{\binits{C.H.} \bsnm{{Blake}}},
\bauthor{\binits{D.} \bsnm{{Kocevski}}},
\bauthor{\binits{W.T.} \bsnm{{Vestrand}}},
\bauthor{\binits{W.} \bsnm{{Li}}},
\bauthor{\binits{R.J.} \bsnm{{Foley}}},
\bauthor{\binits{E.} \bsnm{{Bellm}}},
\bauthor{\binits{H.-W.} \bsnm{{Chen}}},
\bauthor{\binits{J.X.} \bsnm{{Prochaska}}},
\bauthor{\binits{D.} \bsnm{{Starr}}},
\bauthor{\binits{A.V.} \bsnm{{Filippenko}}},
\bauthor{\binits{E.E.} \bsnm{{Falco}}},
\bauthor{\binits{A.H.} \bsnm{{Szentgyorgyi}}},
\bauthor{\binits{J.} \bsnm{{Wren}}},
\bauthor{\binits{P.R.} \bsnm{{Wozniak}}},
\bauthor{\binits{R.} \bsnm{{White}}},
\bauthor{\binits{J.} \bsnm{{Pergande}}},
\batitle{{The Troublesome Broadband Evolution of GRB 061126: Does a Gray Burst
  Imply Gray Dust?}}
\bjtitle{\apj}
\bvolume{672},
\bfpage{449}--\blpage{464}
(\byear{2008}).
doi:\doiurl{10.1086/523929}
\end{barticle}
\endbibitem

\bibitem[\protect\citeauthoryear{{Perley} et~al.}{2009}]{Perley+2009a}
\begin{barticle}
\bauthor{\binits{D.A.} \bsnm{{Perley}}},
\bauthor{\binits{S.B.} \bsnm{{Cenko}}},
\bauthor{\binits{J.S.} \bsnm{{Bloom}}},
\bauthor{\binits{H.-W.} \bsnm{{Chen}}},
\bauthor{\binits{N.R.} \bsnm{{Butler}}},
\bauthor{\binits{D.} \bsnm{{Kocevski}}},
\bauthor{\binits{J.X.} \bsnm{{Prochaska}}},
\bauthor{\binits{M.} \bsnm{{Brodwin}}},
\bauthor{\binits{K.} \bsnm{{Glazebrook}}},
\bauthor{\binits{M.M.} \bsnm{{Kasliwal}}},
\bauthor{\binits{S.R.} \bsnm{{Kulkarni}}},
\bauthor{\binits{S.} \bsnm{{Lopez}}},
\bauthor{\binits{E.O.} \bsnm{{Ofek}}},
\bauthor{\binits{M.} \bsnm{{Pettini}}},
\bauthor{\binits{A.M.} \bsnm{{Soderberg}}},
\bauthor{\binits{D.} \bsnm{{Starr}}},
\batitle{{The Host Galaxies of Swift Dark Gamma-ray Bursts: Observational
  Constraints on Highly Obscured and Very High Redshift GRBs}}.
\bjtitle{\aj}
\bvolume{138},
\bfpage{1690}--\blpage{1708}
(\byear{2009}).
doi:\doiurl{10.1088/0004-6256/138/6/1690}
\end{barticle}
\endbibitem

\bibitem[\protect\citeauthoryear{{Perley} et~al.}{2010}]{Perley+2010}
\begin{barticle}
\bauthor{\binits{D.A.} \bsnm{{Perley}}},
\bauthor{\binits{J.S.} \bsnm{{Bloom}}},
\bauthor{\binits{C.R.} \bsnm{{Klein}}},
\bauthor{\binits{S.} \bsnm{{Covino}}},
\bauthor{\binits{T.} \bsnm{{Minezaki}}},
\bauthor{\binits{P.} \bsnm{{Wo{\'z}niak}}},
\bauthor{\binits{W.T.} \bsnm{{Vestrand}}},
\bauthor{\binits{G.G.} \bsnm{{Williams}}},
\bauthor{\binits{P.} \bsnm{{Milne}}},
\bauthor{\binits{N.R.} \bsnm{{Butler}}},
\bauthor{\binits{A.C.} \bsnm{{Updike}}},
\bauthor{\binits{T.} \bsnm{{Kr{\"u}hler}}},
\bauthor{\binits{P.} \bsnm{{Afonso}}},
\bauthor{\binits{A.} \bsnm{{Antonelli}}},
\bauthor{\binits{L.} \bsnm{{Cowie}}},
\bauthor{\binits{P.} \bsnm{{Ferrero}}},
\bauthor{\binits{J.} \bsnm{{Greiner}}},
\bauthor{\binits{D.H.} \bsnm{{Hartmann}}},
\bauthor{\binits{Y.} \bsnm{{Kakazu}}},
\bauthor{\binits{A.} \bsnm{{K{\"u}pc{\"u} Yolda{\c s}}}},
\bauthor{\binits{A.N.} \bsnm{{Morgan}}},
\bauthor{\binits{P.A.} \bsnm{{Price}}},
\bauthor{\binits{J.X.} \bsnm{{Prochaska}}},
\bauthor{\binits{Y.} \bsnm{{Yoshii}}},
\batitle{{Evidence for supernova-synthesized dust from the rising afterglow of
  GRB071025 at z \~{} 5}}.
\bjtitle{\mnras}
\bvolume{406},
\bfpage{2473}--\blpage{2487}
(\byear{2010}).
doi:\doiurl{10.1111/j.1365-2966.2010.16772.x}
\end{barticle}
\endbibitem

\bibitem[\protect\citeauthoryear{{Perley} et~al.}{2013}]{Perley+2013a}
\begin{barticle}
\bauthor{\binits{D.A.} \bsnm{{Perley}}},
\bauthor{\binits{A.J.} \bsnm{{Levan}}},
\bauthor{\binits{N.R.} \bsnm{{Tanvir}}},
\bauthor{\binits{S.B.} \bsnm{{Cenko}}},
\bauthor{\binits{J.S.} \bsnm{{Bloom}}},
\bauthor{\binits{J.} \bsnm{{Hjorth}}},
\bauthor{\binits{T.} \bsnm{{Kr{\"u}hler}}},
\bauthor{\binits{A.V.} \bsnm{{Filippenko}}},
\bauthor{\binits{A.} \bsnm{{Fruchter}}},
\bauthor{\binits{J.P.U.} \bsnm{{Fynbo}}},
\bauthor{\binits{P.} \bsnm{{Jakobsson}}},
\bauthor{\binits{J.} \bsnm{{Kalirai}}},
\bauthor{\binits{B.} \bsnm{{Milvang-Jensen}}},
\bauthor{\binits{A.N.} \bsnm{{Morgan}}},
\bauthor{\binits{J.X.} \bsnm{{Prochaska}}},
\bauthor{\binits{J.M.} \bsnm{{Silverman}}},
\batitle{{A Population of Massive, Luminous Galaxies Hosting Heavily
  Dust-obscured Gamma-Ray Bursts: Implications for the Use of GRBs as Tracers
  of Cosmic Star Formation}}.
\bjtitle{\apj}
\bvolume{778},
\bfpage{128}
(\byear{2013}).
doi:\doiurl{10.1088/0004-637X/778/2/128}
\end{barticle}
\endbibitem

\bibitem[\protect\citeauthoryear{{Perley} et~al.}{2015a}]{Perley+2015a}
\begin{barticle}
\bauthor{\binits{D.A.} \bsnm{{Perley}}},
\bauthor{\binits{R.A.} \bsnm{{Perley}}},
\bauthor{\binits{J.} \bsnm{{Hjorth}}},
\bauthor{\binits{M.J.} \bsnm{{Micha{\l}owski}}},
\bauthor{\binits{S.B.} \bsnm{{Cenko}}},
\bauthor{\binits{P.} \bsnm{{Jakobsson}}},
\bauthor{\binits{T.} \bsnm{{Kr{\"u}hler}}},
\bauthor{\binits{A.J.} \bsnm{{Levan}}},
\bauthor{\binits{D.} \bsnm{{Malesani}}},
\bauthor{\binits{N.R.} \bsnm{{Tanvir}}},
\batitle{{Connecting GRBs and ULIRGs: A Sensitive, Unbiased Survey for Radio
  Emission from Gamma-Ray Burst Host Galaxies at 0{$<$}z{$<$}2.5}}.
\bjtitle{\apj}
\bvolume{801},
\bfpage{102}
(\byear{2015}a)
\end{barticle}
\endbibitem

\bibitem[\protect\citeauthoryear{{Perley} et~al.}{2016a}]{Perley+2015b}
\begin{barticle}
\bauthor{\binits{D.A.} \bsnm{{Perley}}},
\bauthor{\binits{T.} \bsnm{{Kr{\"u}hler}}},
\bauthor{\binits{S.} \bsnm{{Schulze}}},
\bauthor{\binits{A.} \bsnm{{de Ugarte Postigo}}},
\bauthor{\binits{J.} \bsnm{{Hjorth}}},
\bauthor{\binits{E.} \bsnm{{Berger}}},
\bauthor{\binits{S.B.} \bsnm{{Cenko}}},
\bauthor{\binits{R.} \bsnm{{Chary}}},
\bauthor{\binits{A.} \bsnm{{Cucchiara}}},
\bauthor{\binits{R.} \bsnm{{Ellis}}},
\bauthor{\binits{W.} \bsnm{{Fong}}},
\bauthor{\binits{J.P.U.} \bsnm{{Fynbo}}},
\bauthor{\binits{J.} \bsnm{{Gorosabel}}},
\bauthor{\binits{J.} \bsnm{{Greiner}}},
\bauthor{\binits{P.} \bsnm{{Jakobsson}}},
\bauthor{\binits{T.} \bsnm{{Laskar}}},
\bauthor{\binits{A.J.} \bsnm{{Levan}}},
\bauthor{\binits{M.J.} \bsnm{{Micha{\l}owski}}},
\bauthor{\binits{B.} \bsnm{{Milvang-Jensen}}},
\bauthor{\binits{N.R.} \bsnm{{Tanvir}}},
\bauthor{\binits{C.C.} \bsnm{{Th{\"o}ne}}},
\bauthor{\binits{K.} \bsnm{{Wiersema}}},
\batitle{The Swift Gamma-Ray Burst Host Galaxy Legacy Survey - I. Sample Selection and
  Redshift Distribution}.
\bjtitle{\apj}
\bvolume{817},
\bfpage{7}
(\byear{2016})
doi:\doiurl{10.3847/0004-637X/817/1/7}
\end{barticle}
\endbibitem

\bibitem[\protect\citeauthoryear{{Perley} et~al.}{2016b}]{Perley+2015c}
\begin{barticle}
\bauthor{\binits{D.A.} \bsnm{{Perley}}},
\bauthor{\binits{N.R.} \bsnm{{Tanvir}}},
\bauthor{\binits{J.} \bsnm{{Hjorth}}},
\bauthor{\binits{T.} \bsnm{{Laskar}}},
\bauthor{\binits{E.} \bsnm{{Berger}}},
\bauthor{\binits{R.} \bsnm{{Chary}}},
\bauthor{\binits{A.} \bsnm{{de Ugarte Postigo}}},
\bauthor{\binits{J.P.U.} \bsnm{{Fynbo}}},
\bauthor{\binits{T.} \bsnm{{Kr{\"u}hler}}},
\bauthor{\binits{A.J.} \bsnm{{Levan}}},
\bauthor{\binits{M.J.} \bsnm{{Micha{\l}owski}}},
\bauthor{\binits{S.} \bsnm{{Schulze}}},
\batitle{The Swift GRB Host Galaxy Legacy Survey - II. Rest-Frame NIR Luminosity
  Distribution and Evidence for a Near-Solar Metallicity Threshold}.
\bjtitle{\apj}
\bvolume{817},
\bfpage{8}
(\byear{2016})
doi:\doiurl{10.3847/0004-637X/817/1/8}
\end{barticle}
\endbibitem

\bibitem[\protect\citeauthoryear{{Perna} and {Lazzati}}{2002}]{Perna+2002}
\begin{barticle}
\bauthor{\binits{R.} \bsnm{{Perna}}},
\bauthor{\binits{D.} \bsnm{{Lazzati}}},
\batitle{{Time-Dependent Photozionization in a Dusty Medium. I. Code
  Description and General Results}}.
\bjtitle{\apj}
\bvolume{570},
\bfpage{271}--\blpage{277}
(\byear{2002})
\end{barticle}
\endbibitem

\bibitem[\protect\citeauthoryear{{Piranomonte} et~al.}{2015}]{Piranomonte+2015}
\begin{barticle}
\bauthor{\binits{S.} \bsnm{{Piranomonte}}},
\bauthor{\binits{J.} \bsnm{{Japelj}}},
\bauthor{\binits{S.D.} \bsnm{{Vergani}}},
\bauthor{\binits{S.} \bsnm{{Savaglio}}},
\bauthor{\binits{E.} \bsnm{{Palazzi}}},
\bauthor{\binits{S.} \bsnm{{Covino}}},
\bauthor{\binits{H.} \bsnm{{Flores}}},
\bauthor{\binits{P.} \bsnm{{Goldoni}}},
\bauthor{\binits{G.} \bsnm{{Cupani}}},
\bauthor{\binits{T.} \bsnm{{Kr{\"u}hler}}},
\bauthor{\binits{F.} \bsnm{{Mannucci}}},
\bauthor{\binits{F.} \bsnm{{Onori}}},
\bauthor{\binits{A.} \bsnm{{Rossi}}},
\bauthor{\binits{V.} \bsnm{{D'Elia}}},
\bauthor{\binits{E.} \bsnm{{Pian}}},
\bauthor{\binits{P.} \bsnm{{D'Avanzo}}},
\bauthor{\binits{A.} \bsnm{{Gomboc}}},
\bauthor{\binits{F.} \bsnm{{Hammer}}},
\bauthor{\binits{S.} \bsnm{{Randich}}},
\bauthor{\binits{F.} \bsnm{{Fiore}}},
\bauthor{\binits{L.} \bsnm{{Stella}}},
\bauthor{\binits{G.} \bsnm{{Tagliaferri}}},
\batitle{{GRB host galaxies with VLT/X-Shooter: properties at z=0.8--1.3}}.
\bjtitle{\mnras}
\bvolume{452},
\bfpage{3293}--\blpage{3303}
(\byear{2015}).
doi:\doiurl{10.1093/mnras/stv1569}
\end{barticle}
\endbibitem

\bibitem[\protect\citeauthoryear{{Podsiadlowski}
  et~al.}{2010}]{Podsiadlowski+2010}
\begin{barticle}
\bauthor{\binits{P.} \bsnm{{Podsiadlowski}}},
\bauthor{\binits{N.} \bsnm{{Ivanova}}},
\bauthor{\binits{S.} \bsnm{{Justham}}},
\bauthor{\binits{S.} \bsnm{{Rappaport}}},
\batitle{{Explosive common-envelope ejection: implications for gamma-ray bursts
  and low-mass black-hole binaries}}.
\bjtitle{\mnras}
\bvolume{406},
\bfpage{840}--\blpage{847}
(\byear{2010}).
doi:\doiurl{10.1111/j.1365-2966.2010.16751.x}
\end{barticle}
\endbibitem

\bibitem[\protect\citeauthoryear{{Pontzen} et~al.}{2010}]{Pontzen+2010}
\begin{barticle}
\bauthor{\binits{A.} \bsnm{{Pontzen}}},
\bauthor{\binits{A.} \bsnm{{Deason}}},
\bauthor{\binits{F.} \bsnm{{Governato}}},
\bauthor{\binits{M.} \bsnm{{Pettini}}},
\bauthor{\binits{J.} \bsnm{{Wadsley}}},
\bauthor{\binits{T.} \bsnm{{Quinn}}},
\bauthor{\binits{A.} \bsnm{{Brooks}}},
\bauthor{\binits{J.} \bsnm{{Bellovary}}},
\bauthor{\binits{J.P.U.} \bsnm{{Fynbo}}},
\batitle{{The nature of HI absorbers in gamma-ray burst afterglows: clues from
  hydrodynamic simulations}}.
\bjtitle{\mnras}
\bvolume{402},
\bfpage{1523}--\blpage{1535}
(\byear{2010}).
doi:\doiurl{10.1111/j.1365-2966.2009.16017.x}
\end{barticle}
\endbibitem

\bibitem[\protect\citeauthoryear{{Prochaska} and {Wolfe}}{2002}]{Prochaska2002}
\begin{barticle}
\bauthor{\binits{J.X.} \bsnm{{Prochaska}}},
\bauthor{\binits{A.M.} \bsnm{{Wolfe}}},
\batitle{{The UCSD HIRES/Keck I Damped Ly{$\alpha$} Abundance Database. II. The
  Implications}}.
\bjtitle{\apj}
\bvolume{566},
\bfpage{68}--\blpage{92}
(\byear{2002}).
doi:\doiurl{10.1086/338080}
\end{barticle}
\endbibitem

\bibitem[\protect\citeauthoryear{{Prochaska} et~al.}{2006}]{Prochaska2006}
\begin{barticle}
\bauthor{\binits{J.X.} \bsnm{{Prochaska}}},
\bauthor{\binits{H.-W.} \bsnm{{Chen}}},
\bauthor{\binits{J.S.} \bsnm{{Bloom}}},
\batitle{{Dissecting the Circumstellar Environment of {$\gamma$}-Ray Burst
  Progenitors}}.
\bjtitle{\apj}
\bvolume{648},
\bfpage{95}--\blpage{110}
(\byear{2006}).
doi:\doiurl{10.1086/505737}
\end{barticle}
\endbibitem

\bibitem[\protect\citeauthoryear{{Prochaska} et~al.}{2008}]{Prochaska2008}
\begin{barticle}
\bauthor{\binits{J.X.} \bsnm{{Prochaska}}},
\bauthor{\binits{M.} \bsnm{{Dessauges-Zavadsky}}},
\bauthor{\binits{E.} \bsnm{{Ramirez-Ruiz}}},
\bauthor{\binits{H.-W.} \bsnm{{Chen}}},
\batitle{{A Survey for N V Absorption at z$\sim${}z$_{GRB}$ in GRB Afterglow
  Spectra: Clues to Gas Near the Progenitor Star}}.
\bjtitle{\apj}
\bvolume{685},
\bfpage{344}--\blpage{353}
(\byear{2008}).
doi:\doiurl{10.1086/590529}
\end{barticle}
\endbibitem

\bibitem[\protect\citeauthoryear{{Prochaska} et~al.}{2009}]{Prochaska+2009}
\begin{barticle}
\bauthor{\binits{J.X.} \bsnm{{Prochaska}}},
\bauthor{\binits{Y.} \bsnm{{Sheffer}}},
\bauthor{\binits{D.A.} \bsnm{{Perley}}},
\bauthor{\binits{J.S.} \bsnm{{Bloom}}},
\bauthor{\binits{L.A.} \bsnm{{Lopez}}},
\bauthor{\binits{M.} \bsnm{{Dessauges-Zavadsky}}},
\bauthor{\binits{H.-W.} \bsnm{{Chen}}},
\bauthor{\binits{A.V.} \bsnm{{Filippenko}}},
\bauthor{\binits{M.} \bsnm{{Ganeshalingam}}},
\bauthor{\binits{W.} \bsnm{{Li}}},
\bauthor{\binits{A.A.} \bsnm{{Miller}}},
\bauthor{\binits{D.} \bsnm{{Starr}}},
\batitle{{The First Positive Detection of Molecular Gas in a GRB Host Galaxy}}.
\bjtitle{\apjl}
\bvolume{691},
\bfpage{27}--\blpage{32}
(\byear{2009}).
doi:\doiurl{10.1088/0004-637X/691/1/L27}
\end{barticle}
\endbibitem

\bibitem[\protect\citeauthoryear{{Raskin} et~al.}{2008}]{Raskin+2008}
\begin{barticle}
\bauthor{\binits{C.} \bsnm{{Raskin}}},
\bauthor{\binits{E.} \bsnm{{Scannapieco}}},
\bauthor{\binits{J.} \bsnm{{Rhoads}}},
\bauthor{\binits{M.} \bsnm{{Della Valle}}},
\batitle{{Using Spatial Distributions to Constrain Progenitors of Supernovae
  and Gamma-Ray Bursts}}.
\bjtitle{\apj}
\bvolume{689},
\bfpage{358}--\blpage{370}
(\byear{2008}).
doi:\doiurl{10.1086/592495}
\end{barticle}
\endbibitem

\bibitem[\protect\citeauthoryear{{Ritchey} et~al.}{2014}]{Ritchey14}
\begin{botherref}
\oauthor{\binits{A.M.} \bsnm{{Ritchey}}},
\oauthor{\binits{D.E.} \bsnm{{Welty}}},
\oauthor{\binits{J.A.} \bsnm{{Dahlstrom}}},
\oauthor{\binits{D.G.} \bsnm{{York}}},
{Diffuse Atomic and Molecular Gas in the Interstellar Medium of M82 toward SN
  2014J}.
ArXiv e-prints
(2014)
\end{botherref}
\endbibitem

\bibitem[\protect\citeauthoryear{{Rol} et~al.}{2005}]{Rol+2005}
\begin{barticle}
\bauthor{\binits{E.} \bsnm{{Rol}}},
\bauthor{\binits{R.A.M.J.} \bsnm{{Wijers}}},
\bauthor{\binits{C.} \bsnm{{Kouveliotou}}},
\bauthor{\binits{L.} \bsnm{{Kaper}}},
\bauthor{\binits{Y.} \bsnm{{Kaneko}}},
\batitle{{How Special Are Dark Gamma-Ray Bursts: A Diagnostic Tool}}.
\bjtitle{\apj}
\bvolume{624},
\bfpage{868}--\blpage{879}
(\byear{2005}).
doi:\doiurl{10.1086/429082}
\end{barticle}
\endbibitem

\bibitem[\protect\citeauthoryear{{Rol} et~al.}{2007}]{Rol+2007}
\begin{barticle}
\bauthor{\binits{E.} \bsnm{{Rol}}},
\bauthor{\binits{A.} \bsnm{{van der Horst}}},
\bauthor{\binits{K.} \bsnm{{Wiersema}}},
\bauthor{\binits{S.K.} \bsnm{{Patel}}},
\bauthor{\binits{A.} \bsnm{{Levan}}},
\bauthor{\binits{M.} \bsnm{{Nysewander}}},
\bauthor{\binits{C.} \bsnm{{Kouveliotou}}},
\bauthor{\binits{R.A.M.J.} \bsnm{{Wijers}}},
\bauthor{\binits{N.} \bsnm{{Tanvir}}},
\bauthor{\binits{D.} \bsnm{{Reichart}}},
\bauthor{\binits{A.S.} \bsnm{{Fruchter}}},
\bauthor{\binits{J.} \bsnm{{Graham}}},
\bauthor{\binits{J.-E.} \bsnm{{Ovaldsen}}},
\bauthor{\binits{A.O.} \bsnm{{Jaunsen}}},
\bauthor{\binits{P.} \bsnm{{Jonker}}},
\bauthor{\binits{W.} \bsnm{{van Ham}}},
\bauthor{\binits{J.} \bsnm{{Hjorth}}},
\bauthor{\binits{R.L.C.} \bsnm{{Starling}}},
\bauthor{\binits{P.T.} \bsnm{{O'Brien}}},
\bauthor{\binits{J.} \bsnm{{Fynbo}}},
\bauthor{\binits{D.N.} \bsnm{{Burrows}}},
\bauthor{\binits{R.} \bsnm{{Strom}}},
\batitle{{GRB 051022: Physical Parameters and Extinction of a Prototype Dark
  Burst}}.
\bjtitle{\apj}
\bvolume{669},
\bfpage{1098}--\blpage{1106}
(\byear{2007}).
doi:\doiurl{10.1086/521336}
\end{barticle}
\endbibitem

\bibitem[\protect\citeauthoryear{{Rossi} et~al.}{2012}]{Rossi+2012}
\begin{barticle}
\bauthor{\binits{A.} \bsnm{{Rossi}}},
\bauthor{\binits{S.} \bsnm{{Klose}}},
\bauthor{\binits{P.} \bsnm{{Ferrero}}},
\bauthor{\binits{J.} \bsnm{{Greiner}}},
\bauthor{\binits{L.A.} \bsnm{{Arnold}}},
\bauthor{\binits{E.} \bsnm{{Gonsalves}}},
\bauthor{\binits{D.H.} \bsnm{{Hartmann}}},
\bauthor{\binits{A.C.} \bsnm{{Updike}}},
\bauthor{\binits{D.A.} \bsnm{{Kann}}},
\bauthor{\binits{T.} \bsnm{{Kr{\"u}hler}}},
\bauthor{\binits{E.} \bsnm{{Palazzi}}},
\bauthor{\binits{S.} \bsnm{{Savaglio}}},
\bauthor{\binits{S.} \bsnm{{Schulze}}},
\bauthor{\binits{P.M.J.} \bsnm{{Afonso}}},
\bauthor{\binits{L.} \bsnm{{Amati}}},
\bauthor{\binits{A.J.} \bsnm{{Castro-Tirado}}},
\bauthor{\binits{C.} \bsnm{{Clemens}}},
\bauthor{\binits{R.} \bsnm{{Filgas}}},
\bauthor{\binits{J.} \bsnm{{Gorosabel}}},
\bauthor{\binits{L.K.} \bsnm{{Hunt}}},
\bauthor{\binits{A.} \bsnm{{K{\"u}pc{\"u} Yolda{\c s}}}},
\bauthor{\binits{N.} \bsnm{{Masetti}}},
\bauthor{\binits{M.} \bsnm{{Nardini}}},
\bauthor{\binits{A.} \bsnm{{Nicuesa Guelbenzu}}},
\bauthor{\binits{F.E.} \bsnm{{Olivares}}},
\bauthor{\binits{E.} \bsnm{{Pian}}},
\bauthor{\binits{A.} \bsnm{{Rau}}},
\bauthor{\binits{P.} \bsnm{{Schady}}},
\bauthor{\binits{S.} \bsnm{{Schmidl}}},
\bauthor{\binits{A.} \bsnm{{Yolda{\c s}}}},
\bauthor{\binits{A.} \bsnm{{de Ugarte Postigo}}},
\batitle{{A deep search for the host galaxies of gamma-ray bursts with no
  detected optical afterglow}}.
\bjtitle{\aap}
\bvolume{545},
\bfpage{77}
(\byear{2012}).
doi:\doiurl{10.1051/0004-6361/201117201}
\end{barticle}
\endbibitem

\bibitem[\protect\citeauthoryear{{Salvaterra} et~al.}{2012}]{Salvaterra+2012}
\begin{barticle}
\bauthor{\binits{R.} \bsnm{{Salvaterra}}},
\bauthor{\binits{S.} \bsnm{{Campana}}},
\bauthor{\binits{S.D.} \bsnm{{Vergani}}},
\bauthor{\binits{S.} \bsnm{{Covino}}},
\bauthor{\binits{P.} \bsnm{{D'Avanzo}}},
\bauthor{\binits{D.} \bsnm{{Fugazza}}},
\bauthor{\binits{G.} \bsnm{{Ghirlanda}}},
\bauthor{\binits{G.} \bsnm{{Ghisellini}}},
\bauthor{\binits{A.} \bsnm{{Melandri}}},
\bauthor{\binits{L.} \bsnm{{Nava}}},
\bauthor{\binits{B.} \bsnm{{Sbarufatti}}},
\bauthor{\binits{H.} \bsnm{{Flores}}},
\bauthor{\binits{S.} \bsnm{{Piranomonte}}},
\bauthor{\binits{G.} \bsnm{{Tagliaferri}}},
\batitle{{A Complete Sample of Bright Swift Long Gamma-Ray Bursts. I. Sample
  Presentation, Luminosity Function and Evolution}}.
\bjtitle{\apj}
\bvolume{749},
\bfpage{68}
(\byear{2012}).
doi:\doiurl{10.1088/0004-637X/749/1/68}
\end{barticle}
\endbibitem

\bibitem[\protect\citeauthoryear{{Salvaterra} et~al.}{2013}]{Salvaterra+2013a}
\begin{barticle}
\bauthor{\binits{R.} \bsnm{{Salvaterra}}},
\bauthor{\binits{U.} \bsnm{{Maio}}},
\bauthor{\binits{B.} \bsnm{{Ciardi}}},
\bauthor{\binits{M.A.} \bsnm{{Campisi}}},
\batitle{{Simulating high-z gamma-ray burst host galaxies}}.
\bjtitle{\mnras}
\bvolume{429},
\bfpage{2718}--\blpage{2726}
(\byear{2013}).
doi:\doiurl{10.1093/mnras/sts541}
\end{barticle}
\endbibitem

\bibitem[\protect\citeauthoryear{{Sanders} et~al.}{2012}]{Sanders+2012c}
\begin{barticle}
\bauthor{\binits{N.E.} \bsnm{{Sanders}}},
\bauthor{\binits{N.} \bsnm{{Caldwell}}},
\bauthor{\binits{J.} \bsnm{{McDowell}}},
\bauthor{\binits{P.} \bsnm{{Harding}}},
\batitle{{The Metallicity Profile of M31 from Spectroscopy of Hundreds of H II
  Regions and PNe}}.
\bjtitle{\apj}
\bvolume{758},
\bfpage{133}
(\byear{2012}).
doi:\doiurl{10.1088/0004-637X/758/2/133}
\end{barticle}
\endbibitem

\bibitem[\protect\citeauthoryear{{Savage} and {Sembach}}{1996}]{Savage1996}
\begin{barticle}
\bauthor{\binits{B.D.} \bsnm{{Savage}}},
\bauthor{\binits{K.R.} \bsnm{{Sembach}}},
\batitle{{Interstellar Gas-Phase Abundances and Physical Conditions toward Two
  Distant High-Latitude Halo Stars}}.
\bjtitle{\apj}
\bvolume{470},
\bfpage{893}
(\byear{1996}).
doi:\doiurl{10.1086/177919}
\end{barticle}
\endbibitem

\bibitem[\protect\citeauthoryear{{Savaglio} and {Fall}}{2004}]{Savaglio04}
\begin{barticle}
\bauthor{\binits{S.} \bsnm{{Savaglio}}},
\bauthor{\binits{S.M.} \bsnm{{Fall}}},
\batitle{{Dust Depletion and Extinction in a Gamma-Ray Burst Afterglow}}.
\bjtitle{\apj}
\bvolume{614},
\bfpage{293}--\blpage{300}
(\byear{2004}).
doi:\doiurl{10.1086/423447}
\end{barticle}
\endbibitem

\bibitem[\protect\citeauthoryear{{Savaglio} et~al.}{2003}]{Savaglio2003}
\begin{barticle}
\bauthor{\binits{S.} \bsnm{{Savaglio}}},
\bauthor{\binits{S.M.} \bsnm{{Fall}}},
\bauthor{\binits{F.} \bsnm{{Fiore}}},
\batitle{{Heavy-Element Abundances and Dust Depletions in the Host Galaxies of
  Three Gamma-Ray Bursts}}.
\bjtitle{\apj}
\bvolume{585},
\bfpage{638}--\blpage{646}
(\byear{2003}).
doi:\doiurl{10.1086/346225}
\end{barticle}
\endbibitem

\bibitem[\protect\citeauthoryear{{Savaglio} et~al.}{2009}]{Savaglio+2009}
\begin{barticle}
\bauthor{\binits{S.} \bsnm{{Savaglio}}},
\bauthor{\binits{K.} \bsnm{{Glazebrook}}},
\bauthor{\binits{D.} \bsnm{{Le Borgne}}},
\batitle{{The Galaxy Population Hosting Gamma-Ray Bursts}}.
\bjtitle{\apj}
\bvolume{691},
\bfpage{182}--\blpage{211}
(\byear{2009}).
doi:\doiurl{10.1088/0004-637X/691/1/182}
\end{barticle}
\endbibitem

\bibitem[\protect\citeauthoryear{{Schady} et~al.}{2010}]{Schady2010}
\begin{barticle}
\bauthor{\binits{P.} \bsnm{{Schady}}},
\bauthor{\binits{M.J.} \bsnm{{Page}}},
\bauthor{\binits{S.R.} \bsnm{{Oates}}},
\bauthor{\binits{M.} \bsnm{{Still}}},
\bauthor{\binits{M.} \bsnm{{de Pasquale}}},
\bauthor{\binits{T.} \bsnm{{Dwelly}}},
\bauthor{\binits{N.P.M.} \bsnm{{Kuin}}},
\bauthor{\binits{S.T.} \bsnm{{Holland}}},
\bauthor{\binits{F.E.} \bsnm{{Marshall}}},
\bauthor{\binits{P.W.A.} \bsnm{{Roming}}},
\batitle{{Dust and metal column densities in gamma-ray burst host galaxies}}.
\bjtitle{\mnras}
\bvolume{401},
\bfpage{2773}--\blpage{2792}
(\byear{2010}).
doi:\doiurl{10.1111/j.1365-2966.2009.15861.x}
\end{barticle}
\endbibitem

\bibitem[\protect\citeauthoryear{{Schady} et~al.}{2012}]{Schady2012}
\begin{barticle}
\bauthor{\binits{P.} \bsnm{{Schady}}},
\bauthor{\binits{T.} \bsnm{{Dwelly}}},
\bauthor{\binits{M.J.} \bsnm{{Page}}},
\bauthor{\binits{T.} \bsnm{{Kr{\"u}hler}}},
\bauthor{\binits{J.} \bsnm{{Greiner}}},
\bauthor{\binits{S.R.} \bsnm{{Oates}}},
\bauthor{\binits{M.} \bsnm{{de Pasquale}}},
\bauthor{\binits{M.} \bsnm{{Nardini}}},
\bauthor{\binits{P.W.A.} \bsnm{{Roming}}},
\bauthor{\binits{A.} \bsnm{{Rossi}}},
\bauthor{\binits{M.} \bsnm{{Still}}},
\batitle{{The dust extinction curves of gamma-ray burst host galaxies}}.
\bjtitle{\aap}
\bvolume{537},
\bfpage{15}
(\byear{2012}).
doi:\doiurl{10.1051/0004-6361/201117414}
\end{barticle}
\endbibitem

\bibitem[\protect\citeauthoryear{{Schady} et~al.}{2014}]{Schady+2014}
\begin{barticle}
\bauthor{\binits{P.} \bsnm{{Schady}}},
\bauthor{\binits{S.} \bsnm{{Savaglio}}},
\bauthor{\binits{T.} \bsnm{{M{\"u}ller}}},
\bauthor{\binits{T.} \bsnm{{Kr{\"u}hler}}},
\bauthor{\binits{T.} \bsnm{{Dwelly}}},
\bauthor{\binits{E.} \bsnm{{Palazzi}}},
\bauthor{\binits{L.K.} \bsnm{{Hunt}}},
\bauthor{\binits{J.} \bsnm{{Greiner}}},
\bauthor{\binits{H.} \bsnm{{Linz}}},
\bauthor{\binits{M.J.} \bsnm{{Micha{\l}owski}}},
\bauthor{\binits{D.} \bsnm{{Pierini}}},
\bauthor{\binits{S.} \bsnm{{Piranomonte}}},
\bauthor{\binits{S.D.} \bsnm{{Vergani}}},
\bauthor{\binits{W.K.} \bsnm{{Gear}}},
\batitle{{Herschel observations of gamma-ray burst host galaxies: implications
  for the topology of the dusty interstellar medium}}.
\bjtitle{\aap}
\bvolume{570},
\bfpage{52}
(\byear{2014}).
doi:\doiurl{10.1051/0004-6361/201424092}
\end{barticle}
\endbibitem

\bibitem[\protect\citeauthoryear{{Schulze} et~al.}{2015}]{Schulze+2015}
\begin{barticle}
\bauthor{\binits{S.} \bsnm{{Schulze}}},
\bauthor{\binits{R.} \bsnm{{Chapman}}},
\bauthor{\binits{J.} \bsnm{{Hjorth}}},
\bauthor{\binits{A.J.} \bsnm{{Levan}}},
\bauthor{\binits{P.} \bsnm{{Jakobsson}}},
\bauthor{\binits{G.} \bsnm{{Bj{\"o}rnsson}}},
\bauthor{\binits{D.A.} \bsnm{{Perley}}},
\bauthor{\binits{T.} \bsnm{{Kr{\"u}hler}}},
\bauthor{\binits{J.} \bsnm{{Gorosabel}}},
\bauthor{\binits{N.R.} \bsnm{{Tanvir}}},
\bauthor{\binits{A.} \bsnm{{de Ugarte Postigo}}},
\bauthor{\binits{J.P.U.} \bsnm{{Fynbo}}},
\bauthor{\binits{B.} \bsnm{{Milvang-Jensen}}},
\bauthor{\binits{P.} \bsnm{{M{\o}ller}}},
\bauthor{\binits{D.J.} \bsnm{{Watson}}},
\batitle{{The Optically Unbiased GRB Host (TOUGH) Survey. VII. The Host Galaxy
  Luminosity Function: Probing the Relationship between GRBs and Star Formation
  to Redshift 6}}.
\bjtitle{\apj}
\bvolume{808},
\bfpage{73}
(\byear{2015}).
doi:\doiurl{10.1088/0004-637X/808/1/73}
\end{barticle}
\endbibitem

\bibitem[\protect\citeauthoryear{{Smartt} et~al.}{2001}]{Smartt+2001a}
\begin{barticle}
\bauthor{\binits{S.J.} \bsnm{{Smartt}}},
\bauthor{\binits{K.A.} \bsnm{{Venn}}},
\bauthor{\binits{P.L.} \bsnm{{Dufton}}},
\bauthor{\binits{D.J.} \bsnm{{Lennon}}},
\bauthor{\binits{W.R.J.} \bsnm{{Rolleston}}},
\bauthor{\binits{F.P.} \bsnm{{Keenan}}},
\batitle{{Chemical abundances in the inner 5 kpc of the Galactic disk}}.
\bjtitle{\aap}
\bvolume{367},
\bfpage{86}--\blpage{105}
(\byear{2001}).
doi:\doiurl{10.1051/0004-6361+20000479}
\end{barticle}
\endbibitem

\bibitem[\protect\citeauthoryear{{Soderberg} et~al.}{2006}]{Soderberg+2006}
\begin{barticle}
\bauthor{\binits{A.M.} \bsnm{{Soderberg}}},
\bauthor{\binits{E.} \bsnm{{Nakar}}},
\bauthor{\binits{E.} \bsnm{{Berger}}},
\bauthor{\binits{S.R.} \bsnm{{Kulkarni}}},
\batitle{{Late-Time Radio Observations of 68 Type Ibc Supernovae: Strong
  Constraints on Off-Axis Gamma-Ray Bursts}}.
\bjtitle{\apj}
\bvolume{638},
\bfpage{930}--\blpage{937}
(\byear{2006}).
doi:\doiurl{10.1086/499121}
\end{barticle}
\endbibitem

\bibitem[\protect\citeauthoryear{{Sparre} et~al.}{2014}]{Sparre+2014}
\begin{barticle}
\bauthor{\binits{M.} \bsnm{{Sparre}}},
\bauthor{\binits{O.E.} \bsnm{{Hartoog}}},
\bauthor{\binits{T.} \bsnm{{Kr{\"u}hler}}},
\bauthor{\binits{J.P.U.} \bsnm{{Fynbo}}},
\bauthor{\binits{D.J.} \bsnm{{Watson}}},
\bauthor{\binits{K.} \bsnm{{Wiersema}}},
\bauthor{\binits{V.} \bsnm{{D'Elia}}},
\bauthor{\binits{T.} \bsnm{{Zafar}}},
\bauthor{\binits{P.M.J.} \bsnm{{Afonso}}},
\bauthor{\binits{S.} \bsnm{{Covino}}},
\bauthor{\binits{A.} \bsnm{{de Ugarte Postigo}}},
\bauthor{\binits{H.} \bsnm{{Flores}}},
\bauthor{\binits{P.} \bsnm{{Goldoni}}},
\bauthor{\binits{J.} \bsnm{{Greiner}}},
\bauthor{\binits{J.} \bsnm{{Hjorth}}},
\bauthor{\binits{P.} \bsnm{{Jakobsson}}},
\bauthor{\binits{L.} \bsnm{{Kaper}}},
\bauthor{\binits{S.} \bsnm{{Klose}}},
\bauthor{\binits{A.J.} \bsnm{{Levan}}},
\bauthor{\binits{D.} \bsnm{{Malesani}}},
\bauthor{\binits{B.} \bsnm{{Milvang-Jensen}}},
\bauthor{\binits{M.} \bsnm{{Nardini}}},
\bauthor{\binits{S.} \bsnm{{Piranomonte}}},
\bauthor{\binits{J.} \bsnm{{Sollerman}}},
\bauthor{\binits{R.} \bsnm{{S{\'a}nchez-Ram{\'{\i}}rez}}},
\bauthor{\binits{S.} \bsnm{{Schulze}}},
\bauthor{\binits{N.R.} \bsnm{{Tanvir}}},
\bauthor{\binits{S.D.} \bsnm{{Vergani}}},
\bauthor{\binits{R.A.M.J.} \bsnm{{Wijers}}},
\batitle{{The Metallicity and Dust Content of a Redshift 5 Gamma-Ray Burst Host
  Galaxy}}.
\bjtitle{\apj}
\bvolume{785},
\bfpage{150}
(\byear{2014}).
doi:\doiurl{10.1088/0004-637X/785/2/150}
\end{barticle}
\endbibitem

\bibitem[\protect\citeauthoryear{{Stanek} et~al.}{2006}]{Stanek+2006}
\begin{barticle}
\bauthor{\binits{K.Z.} \bsnm{{Stanek}}},
\bauthor{\binits{O.Y.} \bsnm{{Gnedin}}},
\bauthor{\binits{J.F.} \bsnm{{Beacom}}},
\bauthor{\binits{A.P.} \bsnm{{Gould}}},
\bauthor{\binits{J.A.} \bsnm{{Johnson}}},
\bauthor{\binits{J.A.} \bsnm{{Kollmeier}}},
\bauthor{\binits{M.} \bsnm{{Modjaz}}},
\bauthor{\binits{M.H.} \bsnm{{Pinsonneault}}},
\bauthor{\binits{R.} \bsnm{{Pogge}}},
\bauthor{\binits{D.H.} \bsnm{{Weinberg}}},
\batitle{{Protecting Life in the Milky Way: Metals Keep the GRBs Away}}.
\bjtitle{ActaA}
\bvolume{56},
\bfpage{333}--\blpage{345}
(\byear{2006})
\end{barticle}
\endbibitem

\bibitem[\protect\citeauthoryear{{Stanway} et~al.}{2011}]{Stanway+2011}
\begin{barticle}
\bauthor{\binits{E.R.} \bsnm{{Stanway}}},
\bauthor{\binits{M.N.} \bsnm{{Bremer}}},
\bauthor{\binits{N.R.} \bsnm{{Tanvir}}},
\bauthor{\binits{A.J.} \bsnm{{Levan}}},
\bauthor{\binits{L.J.M.} \bsnm{{Davies}}},
\batitle{{Constraining the molecular gas in the environs of a z=8 gamma-ray
  burst host galaxy}}.
\bjtitle{\mnras}
\bvolume{410},
\bfpage{1496}--\blpage{1502}
(\byear{2011}).
doi:\doiurl{10.1111/j.1365-2966.2010.17534.x}
\end{barticle}
\endbibitem

\bibitem[\protect\citeauthoryear{{Starling} et~al.}{2007}]{Starling07}
\begin{barticle}
\bauthor{\binits{R.L.C.} \bsnm{{Starling}}},
\bauthor{\binits{R.A.M.J.} \bsnm{{Wijers}}},
\bauthor{\binits{K.} \bsnm{{Wiersema}}},
\bauthor{\binits{E.} \bsnm{{Rol}}},
\bauthor{\binits{P.A.} \bsnm{{Curran}}},
\bauthor{\binits{C.} \bsnm{{Kouveliotou}}},
\bauthor{\binits{A.J.} \bsnm{{van der Horst}}},
\bauthor{\binits{M.H.M.} \bsnm{{Heemskerk}}},
\batitle{{Gamma-Ray Burst Afterglows as Probes of Environment and Blast Wave
  Physics. I. Absorption by Host-Galaxy Gas and Dust}}.
\bjtitle{\apj}
\bvolume{661},
\bfpage{787}--\blpage{800}
(\byear{2007}).
doi:\doiurl{10.1086/511953}
\end{barticle}
\endbibitem

\bibitem[\protect\citeauthoryear{{Starling} et~al.}{2011}]{Starling+2011}
\begin{barticle}
\bauthor{\binits{R.L.C.} \bsnm{{Starling}}},
\bauthor{\binits{K.} \bsnm{{Wiersema}}},
\bauthor{\binits{A.J.} \bsnm{{Levan}}},
\bauthor{\binits{T.} \bsnm{{Sakamoto}}},
\bauthor{\binits{D.} \bsnm{{Bersier}}},
\bauthor{\binits{P.} \bsnm{{Goldoni}}},
\bauthor{\binits{S.R.} \bsnm{{Oates}}},
\bauthor{\binits{A.} \bsnm{{Rowlinson}}},
\bauthor{\binits{S.} \bsnm{{Campana}}},
\bauthor{\binits{J.} \bsnm{{Sollerman}}},
\bauthor{\binits{N.R.} \bsnm{{Tanvir}}},
\bauthor{\binits{D.} \bsnm{{Malesani}}},
\bauthor{\binits{J.P.U.} \bsnm{{Fynbo}}},
\bauthor{\binits{S.} \bsnm{{Covino}}},
\bauthor{\binits{P.} \bsnm{{D'Avanzo}}},
\bauthor{\binits{P.T.} \bsnm{{O'Brien}}},
\bauthor{\binits{K.L.} \bsnm{{Page}}},
\bauthor{\binits{J.P.} \bsnm{{Osborne}}},
\bauthor{\binits{S.D.} \bsnm{{Vergani}}},
\bauthor{\binits{S.} \bsnm{{Barthelmy}}},
\bauthor{\binits{D.N.} \bsnm{{Burrows}}},
\bauthor{\binits{Z.} \bsnm{{Cano}}},
\bauthor{\binits{P.A.} \bsnm{{Curran}}},
\bauthor{\binits{M.} \bsnm{{de Pasquale}}},
\bauthor{\binits{V.} \bsnm{{D'Elia}}},
\bauthor{\binits{P.A.} \bsnm{{Evans}}},
\bauthor{\binits{H.} \bsnm{{Flores}}},
\bauthor{\binits{A.S.} \bsnm{{Fruchter}}},
\bauthor{\binits{P.} \bsnm{{Garnavich}}},
\bauthor{\binits{N.} \bsnm{{Gehrels}}},
\bauthor{\binits{J.} \bsnm{{Gorosabel}}},
\bauthor{\binits{J.} \bsnm{{Hjorth}}},
\bauthor{\binits{S.T.} \bsnm{{Holland}}},
\bauthor{\binits{A.J.} \bsnm{{van der Horst}}},
\bauthor{\binits{C.P.} \bsnm{{Hurkett}}},
\bauthor{\binits{P.} \bsnm{{Jakobsson}}},
\bauthor{\binits{A.P.} \bsnm{{Kamble}}},
\bauthor{\binits{C.} \bsnm{{Kouveliotou}}},
\bauthor{\binits{N.P.M.} \bsnm{{Kuin}}},
\bauthor{\binits{L.} \bsnm{{Kaper}}},
\bauthor{\binits{P.A.} \bsnm{{Mazzali}}},
\bauthor{\binits{P.E.} \bsnm{{Nugent}}},
\bauthor{\binits{E.} \bsnm{{Pian}}},
\bauthor{\binits{M.} \bsnm{{Stamatikos}}},
\bauthor{\binits{C.C.} \bsnm{{Th{\"o}ne}}},
\bauthor{\binits{S.E.} \bsnm{{Woosley}}},
\batitle{{Discovery of the nearby long, soft GRB 100316D with an associated
  supernova}}.
\bjtitle{\mnras}
\bvolume{411},
\bfpage{2792}--\blpage{2803}
(\byear{2011}).
doi:\doiurl{10.1111/j.1365-2966.2010.17879.x}
\end{barticle}
\endbibitem

\bibitem[\protect\citeauthoryear{{Starling} et~al.}{2013}]{Starling2013}
\begin{barticle}
\bauthor{\binits{R.L.C.} \bsnm{{Starling}}},
\bauthor{\binits{R.} \bsnm{{Willingale}}},
\bauthor{\binits{N.R.} \bsnm{{Tanvir}}},
\bauthor{\binits{A.E.} \bsnm{{Scott}}},
\bauthor{\binits{K.} \bsnm{{Wiersema}}},
\bauthor{\binits{P.T.} \bsnm{{O'Brien}}},
\bauthor{\binits{A.J.} \bsnm{{Levan}}},
\bauthor{\binits{G.C.} \bsnm{{Stewart}}},
\batitle{{X-ray absorption evolution in gamma-ray bursts: intergalactic medium
  or evolutionary signature of their host galaxies}}.
\bjtitle{\mnras}
\bvolume{431},
\bfpage{3159}--\blpage{3176}
(\byear{2013}).
doi:\doiurl{10.1093/mnras/stt400}
\end{barticle}
\endbibitem

\bibitem[\protect\citeauthoryear{{Sumi}}{2004}]{Sumi2004}
\begin{barticle}
\bauthor{\binits{T.} \bsnm{{Sumi}}},
\batitle{{Extinction map of the Galactic centre: OGLE-II Galactic bulge
  fields}}.
\bjtitle{\mnras}
\bvolume{349},
\bfpage{193}--\blpage{204}
(\byear{2004}).
doi:\doiurl{10.1111/j.1365-2966.2004.07482.x}
\end{barticle}
\endbibitem

\bibitem[\protect\citeauthoryear{{Svensson} et~al.}{2010}]{Svensson+2010}
\begin{barticle}
\bauthor{\binits{K.M.} \bsnm{{Svensson}}},
\bauthor{\binits{A.J.} \bsnm{{Levan}}},
\bauthor{\binits{N.R.} \bsnm{{Tanvir}}},
\bauthor{\binits{A.S.} \bsnm{{Fruchter}}},
\bauthor{\binits{L.-G.} \bsnm{{Strolger}}},
\batitle{{The host galaxies of core-collapse supernovae and gamma-ray bursts}}.
\bjtitle{\mnras}
\bvolume{405},
\bfpage{57}--\blpage{76}
(\byear{2010}).
doi:\doiurl{10.1111/j.1365-2966.2010.16442.x}
\end{barticle}
\endbibitem

\bibitem[\protect\citeauthoryear{{Svensson} et~al.}{2012}]{Svensson+2012}
\begin{barticle}
\bauthor{\binits{K.M.} \bsnm{{Svensson}}},
\bauthor{\binits{A.J.} \bsnm{{Levan}}},
\bauthor{\binits{N.R.} \bsnm{{Tanvir}}},
\bauthor{\binits{D.A.} \bsnm{{Perley}}},
\bauthor{\binits{M.J.} \bsnm{{Michalowski}}},
\bauthor{\binits{K.L.} \bsnm{{Page}}},
\bauthor{\binits{J.S.} \bsnm{{Bloom}}},
\bauthor{\binits{S.B.} \bsnm{{Cenko}}},
\bauthor{\binits{J.} \bsnm{{Hjorth}}},
\bauthor{\binits{P.} \bsnm{{Jakobsson}}},
\bauthor{\binits{D.} \bsnm{{Watson}}},
\bauthor{\binits{P.J.} \bsnm{{Wheatley}}},
\batitle{{The dark GRB 080207 in an extremely red host and the implications for
  gamma-ray bursts in highly obscured environments}}.
\bjtitle{\mnras}
\bvolume{421},
\bfpage{25}--\blpage{35}
(\byear{2012}).
doi:\doiurl{10.1111/j.1365-2966.2011.19811.x}
\end{barticle}
\endbibitem

\bibitem[\protect\citeauthoryear{{Tanvir} et~al.}{2004}]{Tanvir+2004}
\begin{barticle}
\bauthor{\binits{N.R.} \bsnm{{Tanvir}}},
\bauthor{\binits{V.E.} \bsnm{{Barnard}}},
\bauthor{\binits{A.W.} \bsnm{{Blain}}},
\bauthor{\binits{A.S.} \bsnm{{Fruchter}}},
\bauthor{\binits{C.} \bsnm{{Kouveliotou}}},
\bauthor{\binits{P.} \bsnm{{Natarajan}}},
\bauthor{\binits{E.} \bsnm{{Ramirez-Ruiz}}},
\bauthor{\binits{E.} \bsnm{{Rol}}},
\bauthor{\binits{I.A.} \bsnm{{Smith}}},
\bauthor{\binits{R.P.J.} \bsnm{{Tilanus}}},
\bauthor{\binits{R.A.M.J.} \bsnm{{Wijers}}},
\batitle{{The submillimetre properties of gamma-ray burst host galaxies}}.
\bjtitle{\mnras}
\bvolume{352},
\bfpage{1073}--\blpage{1080}
(\byear{2004}).
doi:\doiurl{10.1111/j.1365-2966.2004.08001.x}
\end{barticle}
\endbibitem

\bibitem[\protect\citeauthoryear{{Tanvir} et~al.}{2012}]{Tanvir+2012}
\begin{barticle}
\bauthor{\binits{N.R.} \bsnm{{Tanvir}}},
\bauthor{\binits{A.J.} \bsnm{{Levan}}},
\bauthor{\binits{A.S.} \bsnm{{Fruchter}}},
\bauthor{\binits{J.P.U.} \bsnm{{Fynbo}}},
\bauthor{\binits{J.} \bsnm{{Hjorth}}},
\bauthor{\binits{K.} \bsnm{{Wiersema}}},
\bauthor{\binits{M.N.} \bsnm{{Bremer}}},
\bauthor{\binits{J.} \bsnm{{Rhoads}}},
\bauthor{\binits{P.} \bsnm{{Jakobsson}}},
\bauthor{\binits{P.T.} \bsnm{{O'Brien}}},
\bauthor{\binits{E.R.} \bsnm{{Stanway}}},
\bauthor{\binits{D.} \bsnm{{Bersier}}},
\bauthor{\binits{P.} \bsnm{{Natarajan}}},
\bauthor{\binits{J.} \bsnm{{Greiner}}},
\bauthor{\binits{D.} \bsnm{{Watson}}},
\bauthor{\binits{A.J.} \bsnm{{Castro-Tirado}}},
\bauthor{\binits{R.A.M.J.} \bsnm{{Wijers}}},
\bauthor{\binits{R.L.C.} \bsnm{{Starling}}},
\bauthor{\binits{K.} \bsnm{{Misra}}},
\bauthor{\binits{J.F.} \bsnm{{Graham}}},
\bauthor{\binits{C.} \bsnm{{Kouveliotou}}},
\batitle{{Star Formation in the Early Universe: Beyond the Tip of the
  Iceberg}}.
\bjtitle{\apj}
\bvolume{754},
\bfpage{46}
(\byear{2012}).
doi:\doiurl{10.1088/0004-637X/754/1/46}
\end{barticle}
\endbibitem

\bibitem[\protect\citeauthoryear{{Th{\"o}ne} et~al.}{2008}]{Thoene+2008}
\begin{barticle}
\bauthor{\binits{C.C.} \bsnm{{Th{\"o}ne}}},
\bauthor{\binits{J.P.U.} \bsnm{{Fynbo}}},
\bauthor{\binits{G.} \bsnm{{{\"O}stlin}}},
\bauthor{\binits{B.} \bsnm{{Milvang-Jensen}}},
\bauthor{\binits{K.} \bsnm{{Wiersema}}},
\bauthor{\binits{D.} \bsnm{{Malesani}}},
\bauthor{\binits{D.} \bsnm{{Della Monica Ferreira}}},
\bauthor{\binits{J.} \bsnm{{Gorosabel}}},
\bauthor{\binits{D.A.} \bsnm{{Kann}}},
\bauthor{\binits{D.} \bsnm{{Watson}}},
\bauthor{\binits{M.J.} \bsnm{{Micha{\l}owski}}},
\bauthor{\binits{A.S.} \bsnm{{Fruchter}}},
\bauthor{\binits{A.J.} \bsnm{{Levan}}},
\bauthor{\binits{J.} \bsnm{{Hjorth}}},
\bauthor{\binits{J.} \bsnm{{Sollerman}}},
\batitle{{Spatially Resolved Properties of the GRB 060505 Host: Implications
  for the Nature of the Progenitor}}.
\bjtitle{\apj}
\bvolume{676},
\bfpage{1151}--\blpage{1161}
(\byear{2008}).
doi:\doiurl{10.1086/528943}
\end{barticle}
\endbibitem

\bibitem[\protect\citeauthoryear{{Th{\"o}ne} et~al.}{2013}]{Thoene+2013}
\begin{barticle}
\bauthor{\binits{C.C.} \bsnm{{Th{\"o}ne}}},
\bauthor{\binits{J.P.U.} \bsnm{{Fynbo}}},
\bauthor{\binits{P.} \bsnm{{Goldoni}}},
\bauthor{\binits{A.P.} \bsnm{{de Ugarte}}},
\bauthor{\binits{S.} \bsnm{{Campana}}},
\bauthor{\binits{S.D.} \bsnm{{Vergani}}},
\bauthor{\binits{S.} \bsnm{{Covino}}},
\bauthor{\binits{T.} \bsnm{{Kr{\"u}hler}}},
\bauthor{\binits{L.} \bsnm{{Kaper}}},
\bauthor{\binits{N.} \bsnm{{Tanvir}}},
\bauthor{\binits{T.} \bsnm{{Zafar}}},
\bauthor{\binits{V.} \bsnm{{D'Elia}}},
\bauthor{\binits{J.} \bsnm{{Gorosabel}}},
\bauthor{\binits{J.} \bsnm{{Greiner}}},
\bauthor{\binits{P.} \bsnm{{Groot}}},
\bauthor{\binits{F.} \bsnm{{Hammer}}},
\bauthor{\binits{P.} \bsnm{{Jakobsson}}},
\bauthor{\binits{S.} \bsnm{{Klose}}},
\bauthor{\binits{A.J.} \bsnm{{Levan}}},
\bauthor{\binits{B.} \bsnm{{Milvang-Jensen}}},
\bauthor{\binits{A.G.} \bsnm{{Nicuesa}}},
\bauthor{\binits{E.} \bsnm{{Palazzi}}},
\bauthor{\binits{S.} \bsnm{{Piranomonte}}},
\bauthor{\binits{G.} \bsnm{{Tagliaferri}}},
\bauthor{\binits{D.} \bsnm{{Watson}}},
\bauthor{\binits{K.} \bsnm{{Wiersema}}},
\bauthor{\binits{R.A.M.J.} \bsnm{{Wijers}}},
\batitle{{GRB 100219A with X-shooter - abundances in a galaxy at z =4.7}}.
\bjtitle{\mnras}
\bvolume{428},
\bfpage{3590}--\blpage{3606}
(\byear{2013}).
doi:\doiurl{10.1093/mnras/sts303}
\end{barticle}
\endbibitem

\bibitem[\protect\citeauthoryear{{Totani} et~al.}{2014}]{Totani+2014a}
\begin{barticle}
\bauthor{\binits{T.} \bsnm{{Totani}}},
\bauthor{\binits{K.} \bsnm{{Aoki}}},
\bauthor{\binits{T.} \bsnm{{Hattori}}},
\bauthor{\binits{G.} \bsnm{{Kosugi}}},
\bauthor{\binits{Y.} \bsnm{{Niino}}},
\bauthor{\binits{T.} \bsnm{{Hashimoto}}},
\bauthor{\binits{N.} \bsnm{{Kawai}}},
\bauthor{\binits{K.} \bsnm{{Ohta}}},
\bauthor{\binits{T.} \bsnm{{Sakamoto}}},
\bauthor{\binits{T.} \bsnm{{Yamada}}},
\batitle{{Probing intergalactic neutral hydrogen by the Lyman alpha red damping
  wing of gamma-ray burst 130606A afterglow spectrum at z = 5.913}}.
\bjtitle{\pasj}
\bvolume{66},
\bfpage{63}
(\byear{2014}).
doi:\doiurl{10.1093/pasj/psu032}
\end{barticle}
\endbibitem

\bibitem[\protect\citeauthoryear{{Tremonti} et~al.}{2004}]{Tremonti+2004}
\begin{barticle}
\bauthor{\binits{C.A.} \bsnm{{Tremonti}}},
\bauthor{\binits{T.M.} \bsnm{{Heckman}}},
\bauthor{\binits{G.} \bsnm{{Kauffmann}}},
\bauthor{\binits{J.} \bsnm{{Brinchmann}}},
\bauthor{\binits{S.} \bsnm{{Charlot}}},
\bauthor{\binits{S.D.M.} \bsnm{{White}}},
\bauthor{\binits{M.} \bsnm{{Seibert}}},
\bauthor{\binits{E.W.} \bsnm{{Peng}}},
\bauthor{\binits{D.J.} \bsnm{{Schlegel}}},
\bauthor{\binits{A.} \bsnm{{Uomoto}}},
\bauthor{\binits{M.} \bsnm{{Fukugita}}},
\bauthor{\binits{J.} \bsnm{{Brinkmann}}},
\batitle{{The Origin of the Mass-Metallicity Relation: Insights from 53,000
  Star-forming Galaxies in the Sloan Digital Sky Survey}}.
\bjtitle{\apj}
\bvolume{613},
\bfpage{898}--\blpage{913}
(\byear{2004}).
doi:\doiurl{10.1086/423264}
\end{barticle}
\endbibitem

\bibitem[\protect\citeauthoryear{{Trenti} et~al.}{2015}]{Trenti+2015}
\begin{barticle}
\bauthor{\binits{M.} \bsnm{{Trenti}}},
\bauthor{\binits{R.} \bsnm{{Perna}}},
\bauthor{\binits{R.} \bsnm{{Jimenez}}},
\batitle{{The Luminosity and Stellar Mass Functions of GRB Host Galaxies:
  Insight Into the Metallicity Bias}}.
\bjtitle{\apj}
\bvolume{802},
\bfpage{103}
(\byear{2015}).
doi:\doiurl{10.1088/0004-637X/802/2/103}
\end{barticle}
\endbibitem

\bibitem[\protect\citeauthoryear{{Trenti} et~al.}{2012}]{Trenti+2012}
\begin{barticle}
\bauthor{\binits{M.} \bsnm{{Trenti}}},
\bauthor{\binits{R.} \bsnm{{Perna}}},
\bauthor{\binits{E.M.} \bsnm{{Levesque}}},
\bauthor{\binits{J.M.} \bsnm{{Shull}}},
\bauthor{\binits{J.T.} \bsnm{{Stocke}}},
\batitle{{Gamma-Ray Burst Host Galaxy Surveys at Redshift z {$>$}\~{} 4: Probes
  of Star Formation Rate and Cosmic Reionization}}.
\bjtitle{\apjl}
\bvolume{749},
\bfpage{38}
(\byear{2012}).
doi:\doiurl{10.1088/2041-8205/749/2/L38}
\end{barticle}
\endbibitem

\bibitem[\protect\citeauthoryear{{van den Heuvel} and {Portegies
  Zwart}}{2013}]{vandenHeuvel+2013}
\begin{barticle}
\bauthor{\binits{E.P.J.} \bsnm{{van den Heuvel}}},
\bauthor{\binits{S.F.} \bsnm{{Portegies Zwart}}},
\batitle{{Are Superluminous Supernovae and Long GRBs the Products of Dynamical
  Processes in Young Dense Star Clusters?}}
\bjtitle{\apj}
\bvolume{779},
\bfpage{114}
(\byear{2013}).
doi:\doiurl{10.1088/0004-637X/779/2/114}
\end{barticle}
\endbibitem

\bibitem[\protect\citeauthoryear{{van der Horst}
  et~al.}{2009}]{vanderHorst+2009}
\begin{barticle}
\bauthor{\binits{A.J.} \bsnm{{van der Horst}}},
\bauthor{\binits{C.} \bsnm{{Kouveliotou}}},
\bauthor{\binits{N.} \bsnm{{Gehrels}}},
\bauthor{\binits{E.} \bsnm{{Rol}}},
\bauthor{\binits{R.A.M.J.} \bsnm{{Wijers}}},
\bauthor{\binits{J.K.} \bsnm{{Cannizzo}}},
\bauthor{\binits{J.} \bsnm{{Racusin}}},
\bauthor{\binits{D.N.} \bsnm{{Burrows}}},
\batitle{{Optical Classification of Gamma-Ray Bursts in the Swift Era}}.
\bjtitle{\apj}
\bvolume{699},
\bfpage{1087}--\blpage{1091}
(\byear{2009}).
doi:\doiurl{10.1088/0004-637X/699/2/1087}
\end{barticle}
\endbibitem

\bibitem[\protect\citeauthoryear{{Vergani} et~al.}{2011}]{Vergani2011a}
\begin{barticle}
\bauthor{\binits{S.D.} \bsnm{{Vergani}}},
\bauthor{\binits{S.} \bsnm{{Piranomonte}}},
\bauthor{\binits{P.} \bsnm{{Petitjean}}},
\bauthor{\binits{H.} \bsnm{{Flores}}},
\bauthor{\binits{P.} \bsnm{{Goldoni}}},
\bauthor{\binits{M.} \bsnm{{Rodrigues}}},
\bauthor{\binits{F.} \bsnm{{Hammer}}},
\bauthor{\binits{S.} \bsnm{{Covino}}},
\batitle{{GRB 021004 host galaxy and environment with X-shooter}}.
\bjtitle{Astronomische Nachrichten}
\bvolume{332},
\bfpage{292}
(\byear{2011}).
doi:\doiurl{10.1002/asna.201111538}
\end{barticle}
\endbibitem

\bibitem[\protect\citeauthoryear{{Vergani} et~al.}{2015}]{Vergani+2015}
\begin{barticle}
\bauthor{\binits{S.D.} \bsnm{{Vergani}}},
\bauthor{\binits{R.} \bsnm{{Salvaterra}}},
\bauthor{\binits{J.} \bsnm{{Japelj}}},
\bauthor{\binits{E.} \bsnm{{Le Floc'h}}},
\bauthor{\binits{P.} \bsnm{{D'Avanzo}}},
\bauthor{\binits{A.} \bsnm{{Fernandez-Soto}}},
\bauthor{\binits{T.} \bsnm{{Kr{\"u}hler}}},
\bauthor{\binits{A.} \bsnm{{Melandri}}},
\bauthor{\binits{S.} \bsnm{{Boissier}}},
\bauthor{\binits{S.} \bsnm{{Covino}}},
\bauthor{\binits{M.} \bsnm{{Puech}}},
\bauthor{\binits{J.} \bsnm{{Greiner}}},
\bauthor{\binits{L.K.} \bsnm{{Hunt}}},
\bauthor{\binits{D.} \bsnm{{Perley}}},
\bauthor{\binits{P.} \bsnm{{Petitjean}}},
\bauthor{\binits{T.} \bsnm{{Vinci}}},
\bauthor{\binits{F.} \bsnm{{Hammer}}},
\bauthor{\binits{A.} \bsnm{{Levan}}},
\bauthor{\binits{F.} \bsnm{{Mannucci}}},
\bauthor{\binits{S.} \bsnm{{Campana}}},
\bauthor{\binits{H.} \bsnm{{Flores}}},
\bauthor{\binits{A.} \bsnm{{Gomboc}}},
\bauthor{\binits{G.} \bsnm{{Tagliaferri}}},
\batitle{{Are long gamma-ray bursts biased tracers of star formation? Clues
  from the host galaxies of the Swift/BAT6 complete sample of LGRBs . I.
  Stellar mass at z {$<$} 1}}.
\bjtitle{\aap}
\bvolume{581},
\bfpage{102}
(\byear{2015}).
doi:\doiurl{10.1051/0004-6361/201425013}
\end{barticle}
\endbibitem

\bibitem[\protect\citeauthoryear{{Vernet} et~al.}{2011}]{Vernet+2011}
\begin{barticle}
\bauthor{\binits{J.} \bsnm{{Vernet}}},
\bauthor{\binits{H.} \bsnm{{Dekker}}},
\bauthor{\binits{S.} \bsnm{{D'Odorico}}},
\bauthor{\binits{L.} \bsnm{{Kaper}}},
\bauthor{\binits{P.} \bsnm{{Kjaergaard}}},
\bauthor{\binits{F.} \bsnm{{Hammer}}},
\bauthor{\binits{S.} \bsnm{{Randich}}},
\bauthor{\binits{F.} \bsnm{{Zerbi}}},
\bauthor{\binits{P.J.} \bsnm{{Groot}}},
\bauthor{\binits{J.} \bsnm{{Hjorth}}},
\bauthor{\binits{I.} \bsnm{{Guinouard}}},
\bauthor{\binits{R.} \bsnm{{Navarro}}},
\bauthor{\binits{T.} \bsnm{{Adolfse}}},
\bauthor{\binits{P.W.} \bsnm{{Albers}}},
\bauthor{\binits{J.-P.} \bsnm{{Amans}}},
\bauthor{\binits{J.J.} \bsnm{{Andersen}}},
\bauthor{\binits{M.I.} \bsnm{{Andersen}}},
\bauthor{\binits{P.} \bsnm{{Binetruy}}},
\bauthor{\binits{P.} \bsnm{{Bristow}}},
\bauthor{\binits{R.} \bsnm{{Castillo}}},
\bauthor{\binits{F.} \bsnm{{Chemla}}},
\bauthor{\binits{L.} \bsnm{{Christensen}}},
\bauthor{\binits{P.} \bsnm{{Conconi}}},
\bauthor{\binits{R.} \bsnm{{Conzelmann}}},
\bauthor{\binits{J.} \bsnm{{Dam}}},
\bauthor{\binits{V.} \bsnm{{de Caprio}}},
\bauthor{\binits{A.} \bsnm{{de Ugarte Postigo}}},
\bauthor{\binits{B.} \bsnm{{Delabre}}},
\bauthor{\binits{P.} \bsnm{{di Marcantonio}}},
\bauthor{\binits{M.} \bsnm{{Downing}}},
\bauthor{\binits{E.} \bsnm{{Elswijk}}},
\bauthor{\binits{G.} \bsnm{{Finger}}},
\bauthor{\binits{G.} \bsnm{{Fischer}}},
\bauthor{\binits{H.} \bsnm{{Flores}}},
\bauthor{\binits{P.} \bsnm{{Fran{\c c}ois}}},
\bauthor{\binits{P.} \bsnm{{Goldoni}}},
\bauthor{\binits{L.} \bsnm{{Guglielmi}}},
\bauthor{\binits{R.} \bsnm{{Haigron}}},
\bauthor{\binits{H.} \bsnm{{Hanenburg}}},
\bauthor{\binits{I.} \bsnm{{Hendriks}}},
\bauthor{\binits{M.} \bsnm{{Horrobin}}},
\bauthor{\binits{D.} \bsnm{{Horville}}},
\bauthor{\binits{N.C.} \bsnm{{Jessen}}},
\bauthor{\binits{F.} \bsnm{{Kerber}}},
\bauthor{\binits{L.} \bsnm{{Kern}}},
\bauthor{\binits{M.} \bsnm{{Kiekebusch}}},
\bauthor{\binits{P.} \bsnm{{Kleszcz}}},
\bauthor{\binits{J.} \bsnm{{Klougart}}},
\bauthor{\binits{J.} \bsnm{{Kragt}}},
\bauthor{\binits{H.H.} \bsnm{{Larsen}}},
\bauthor{\binits{J.-L.} \bsnm{{Lizon}}},
\bauthor{\binits{C.} \bsnm{{Lucuix}}},
\bauthor{\binits{V.} \bsnm{{Mainieri}}},
\bauthor{\binits{R.} \bsnm{{Manuputy}}},
\bauthor{\binits{C.} \bsnm{{Martayan}}},
\bauthor{\binits{E.} \bsnm{{Mason}}},
\bauthor{\binits{R.} \bsnm{{Mazzoleni}}},
\bauthor{\binits{N.} \bsnm{{Michaelsen}}},
\bauthor{\binits{A.} \bsnm{{Modigliani}}},
\bauthor{\binits{S.} \bsnm{{Moehler}}},
\bauthor{\binits{P.} \bsnm{{M{\o}ller}}},
\bauthor{\binits{A.} \bsnm{{Norup S{\o}rensen}}},
\bauthor{\binits{P.} \bsnm{{N{\o}rregaard}}},
\bauthor{\binits{C.} \bsnm{{P{\'e}roux}}},
\bauthor{\binits{F.} \bsnm{{Patat}}},
\bauthor{\binits{E.} \bsnm{{Pena}}},
\bauthor{\binits{J.} \bsnm{{Pragt}}},
\bauthor{\binits{C.} \bsnm{{Reinero}}},
\bauthor{\binits{F.} \bsnm{{Rigal}}},
\bauthor{\binits{M.} \bsnm{{Riva}}},
\bauthor{\binits{R.} \bsnm{{Roelfsema}}},
\bauthor{\binits{F.} \bsnm{{Royer}}},
\bauthor{\binits{G.} \bsnm{{Sacco}}},
\bauthor{\binits{P.} \bsnm{{Santin}}},
\bauthor{\binits{T.} \bsnm{{Schoenmaker}}},
\bauthor{\binits{P.} \bsnm{{Spano}}},
\bauthor{\binits{E.} \bsnm{{Sweers}}},
\bauthor{\binits{R.} \bsnm{{Ter Horst}}},
\bauthor{\binits{M.} \bsnm{{Tintori}}},
\bauthor{\binits{N.} \bsnm{{Tromp}}},
\bauthor{\binits{P.} \bsnm{{van Dael}}},
\bauthor{\binits{H.} \bsnm{{van der Vliet}}},
\bauthor{\binits{L.} \bsnm{{Venema}}},
\bauthor{\binits{M.} \bsnm{{Vidali}}},
\bauthor{\binits{J.} \bsnm{{Vinther}}},
\bauthor{\binits{P.} \bsnm{{Vola}}},
\bauthor{\binits{R.} \bsnm{{Winters}}},
\bauthor{\binits{D.} \bsnm{{Wistisen}}},
\bauthor{\binits{G.} \bsnm{{Wulterkens}}},
\bauthor{\binits{A.} \bsnm{{Zacchei}}},
\batitle{{X-shooter, the new wide band intermediate resolution spectrograph at
  the ESO Very Large Telescope}}.
\bjtitle{\aap}
\bvolume{536},
\bfpage{105}
(\byear{2011}).
doi:\doiurl{10.1051/0004-6361/201117752}
\end{barticle}
\endbibitem

\bibitem[\protect\citeauthoryear{{Vink} and {de Koter}}{2005}]{Vink+2005}
\begin{barticle}
\bauthor{\binits{J.S.} \bsnm{{Vink}}},
\bauthor{\binits{A.} \bsnm{{de Koter}}},
\batitle{{On the metallicity dependence of Wolf-Rayet winds}}.
\bjtitle{\aap}
\bvolume{442},
\bfpage{587}--\blpage{596}
(\byear{2005}).
doi:\doiurl{10.1051/0004-6361+20052862}
\end{barticle}
\endbibitem

\bibitem[\protect\citeauthoryear{{Vladilo} et~al.}{2006}]{Vladilo2006}
\begin{barticle}
\bauthor{\binits{G.} \bsnm{{Vladilo}}},
\bauthor{\binits{M.} \bsnm{{Centuri{\'o}n}}},
\bauthor{\binits{S.A.} \bsnm{{Levshakov}}},
\bauthor{\binits{C.} \bsnm{{P{\'e}roux}}},
\bauthor{\binits{P.} \bsnm{{Khare}}},
\bauthor{\binits{V.P.} \bsnm{{Kulkarni}}},
\bauthor{\binits{D.G.} \bsnm{{York}}},
\batitle{{Extinction and metal column density of HI regions up to redshift z
  $\simeq 2$}}.
\bjtitle{\aap}
\bvolume{454},
\bfpage{151}--\blpage{164}
(\byear{2006}).
doi:\doiurl{10.1051/0004-6361+20054742}
\end{barticle}
\endbibitem

\bibitem[\protect\citeauthoryear{{Vreeswijk} et~al.}{2008}]{Vreeswijk2008}
\begin{botherref}
\oauthor{\binits{P.M.} \bsnm{{Vreeswijk}}},
\oauthor{\binits{A.} \bsnm{{Smette}}},
\oauthor{\binits{D.} \bsnm{{Malesani}}},
\oauthor{\binits{J.P.U.} \bsnm{{Fynbo}}},
\oauthor{\binits{B.} \bsnm{{Milvang-Jensen}}},
\oauthor{\binits{P.} \bsnm{{Jakobsson}}},
\oauthor{\binits{A.O.} \bsnm{{Jaunsen}}},
\oauthor{\binits{U.} \bsnm{{Oslo}}},
\oauthor{\binits{C.} \bsnm{{Ledoux}}},
{VLT/UVES redshift of GRB 080319B.}
GRB Coordinates Network
\textbf{7444}
(2008)
\end{botherref}
\endbibitem

\bibitem[\protect\citeauthoryear{{Vreeswijk} et~al.}{2012}]{Vreeswijk2012}
\begin{barticle}
\bauthor{\binits{P.M.} \bsnm{{Vreeswijk}}},
\bauthor{\binits{C.} \bsnm{{Ledoux}}},
\bauthor{\binits{A.} \bsnm{{De Cia}}},
\bauthor{\binits{A.} \bsnm{{Smette}}},
\batitle{{Probing the environment of GRBs with absorption-line spectroscopy.}}
\bjtitle{Memorie della Societa Astronomica Italiana Supplementi}
\bvolume{21},
\bfpage{14}
(\byear{2012})
\end{barticle}
\endbibitem

\bibitem[\protect\citeauthoryear{{Wainwright} et~al.}{2007}]{Wainwright+2007}
\begin{barticle}
\bauthor{\binits{C.} \bsnm{{Wainwright}}},
\bauthor{\binits{E.} \bsnm{{Berger}}},
\bauthor{\binits{B.E.} \bsnm{{Penprase}}},
\batitle{{A Morphological Study of Gamma-Ray Burst Host Galaxies}}.
\bjtitle{\apj}
\bvolume{657},
\bfpage{367}--\blpage{377}
(\byear{2007}).
doi:\doiurl{10.1086/510794}
\end{barticle}
\endbibitem

\bibitem[\protect\citeauthoryear{{Wang} and {Dai}}{2011}]{Wang+2011}
\begin{barticle}
\bauthor{\binits{F.Y.} \bsnm{{Wang}}},
\bauthor{\binits{Z.G.} \bsnm{{Dai}}},
\batitle{{An Evolving Stellar Initial Mass Function and the Gamma-ray Burst
  Redshift Distribution}}.
\bjtitle{\apjl}
\bvolume{727},
\bfpage{34}
(\byear{2011}).
doi:\doiurl{10.1088/2041-8205/727/2/L34}
\end{barticle}
\endbibitem

\bibitem[\protect\citeauthoryear{{Wang} et~al.}{2012}]{Wang+2012}
\begin{barticle}
\bauthor{\binits{W.-H.} \bsnm{{Wang}}},
\bauthor{\binits{H.-W.} \bsnm{{Chen}}},
\bauthor{\binits{K.-Y.} \bsnm{{Huang}}},
\batitle{{ALMA Submillimeter Continuum Imaging of the Host Galaxies of GRB
  021004 and GRB 080607}}.
\bjtitle{\apjl}
\bvolume{761},
\bfpage{32}
(\byear{2012}).
doi:\doiurl{10.1088/2041-8205/761/2/L32}
\end{barticle}
\endbibitem

\bibitem[\protect\citeauthoryear{{Watson}}{2009}]{Watson2009}
\begin{bchapter}
\bauthor{\binits{D.} \bsnm{{Watson}}},
\bctitle{{Probing Dust with Gamma-Ray Bursts}},
in \bbtitle{Cosmic Dust - Near and Far},
ed. by \beditor{\binits{T.} \bsnm{{Henning}}},
\beditor{\binits{E.} \bsnm{{Gr{\"u}n}}},
\beditor{\binits{J.} \bsnm{{Steinacker}}}
\bsertitle{Astronomical Society of the Pacific Conference Series},
vol. \bseriesno{414},
\byear{2009},
p. \bfpage{277}
\end{bchapter}
\endbibitem

\bibitem[\protect\citeauthoryear{{Watson} and {Jakobsson}}{2012}]{Watson12}
\begin{barticle}
\bauthor{\binits{D.} \bsnm{{Watson}}},
\bauthor{\binits{P.} \bsnm{{Jakobsson}}},
\batitle{{Dust Extinction Bias in the Column Density Distribution of Gamma-Ray
  Bursts: High Column Density, Low-redshift GRBs are More Heavily Obscured}}.
\bjtitle{\apj}
\bvolume{754},
\bfpage{89}
(\byear{2012}).
doi:\doiurl{10.1088/0004-637X/754/2/89}
\end{barticle}
\endbibitem

\bibitem[\protect\citeauthoryear{{Watson} et~al.}{2006}]{Watson2006}
\begin{barticle}
\bauthor{\binits{D.} \bsnm{{Watson}}},
\bauthor{\binits{J.P.U.} \bsnm{{Fynbo}}},
\bauthor{\binits{C.} \bsnm{{Ledoux}}},
\bauthor{\binits{P.} \bsnm{{Vreeswijk}}},
\bauthor{\binits{J.} \bsnm{{Hjorth}}},
\bauthor{\binits{A.} \bsnm{{Smette}}},
\bauthor{\binits{A.C.} \bsnm{{Andersen}}},
\bauthor{\binits{K.} \bsnm{{Aoki}}},
\bauthor{\binits{T.} \bsnm{{Augusteijn}}},
\bauthor{\binits{A.P.} \bsnm{{Beardmore}}},
\bauthor{\binits{D.} \bsnm{{Bersier}}},
\bauthor{\binits{J.M.} \bsnm{{Castro Cer{\'o}n}}},
\bauthor{\binits{P.} \bsnm{{D'Avanzo}}},
\bauthor{\binits{D.} \bsnm{{Diaz-Fraile}}},
\bauthor{\binits{J.} \bsnm{{Gorosabel}}},
\bauthor{\binits{P.} \bsnm{{Hirst}}},
\bauthor{\binits{P.} \bsnm{{Jakobsson}}},
\bauthor{\binits{B.L.} \bsnm{{Jensen}}},
\bauthor{\binits{N.} \bsnm{{Kawai}}},
\bauthor{\binits{G.} \bsnm{{Kosugi}}},
\bauthor{\binits{P.} \bsnm{{Laursen}}},
\bauthor{\binits{A.} \bsnm{{Levan}}},
\bauthor{\binits{J.} \bsnm{{Masegosa}}},
\bauthor{\binits{J.} \bsnm{{N{\"a}r{\"a}nen}}},
\bauthor{\binits{K.L.} \bsnm{{Page}}},
\bauthor{\binits{K.} \bsnm{{Pedersen}}},
\bauthor{\binits{A.} \bsnm{{Pozanenko}}},
\bauthor{\binits{J.N.} \bsnm{{Reeves}}},
\bauthor{\binits{V.} \bsnm{{Rumyantsev}}},
\bauthor{\binits{T.} \bsnm{{Shahbaz}}},
\bauthor{\binits{D.} \bsnm{{Sharapov}}},
\bauthor{\binits{J.} \bsnm{{Sollerman}}},
\bauthor{\binits{R.L.C.} \bsnm{{Starling}}},
\bauthor{\binits{N.} \bsnm{{Tanvir}}},
\bauthor{\binits{K.} \bsnm{{Torstensson}}},
\bauthor{\binits{K.} \bsnm{{Wiersema}}},
\batitle{{A logN$_{HI}$ = 22.6 Damped Ly{$\alpha$} Absorber in a Dark Gamma-Ray
  Burst: The Environment of GRB 050401}}.
\bjtitle{\apj}
\bvolume{652},
\bfpage{1011}--\blpage{1019}
(\byear{2006}).
doi:\doiurl{10.1086/508049}
\end{barticle}
\endbibitem

\bibitem[\protect\citeauthoryear{{Watson} et~al.}{2013}]{Watson13}
\begin{barticle}
\bauthor{\binits{D.} \bsnm{{Watson}}},
\bauthor{\binits{T.} \bsnm{{Zafar}}},
\bauthor{\binits{A.C.} \bsnm{{Andersen}}},
\bauthor{\binits{J.P.U.} \bsnm{{Fynbo}}},
\bauthor{\binits{J.} \bsnm{{Gorosabel}}},
\bauthor{\binits{J.} \bsnm{{Hjorth}}},
\bauthor{\binits{P.} \bsnm{{Jakobsson}}},
\bauthor{\binits{T.} \bsnm{{Kr{\"u}hler}}},
\bauthor{\binits{P.} \bsnm{{Laursen}}},
\bauthor{\binits{G.} \bsnm{{Leloudas}}},
\bauthor{\binits{D.} \bsnm{{Malesani}}},
\batitle{{Helium in Natal H II Regions: The Origin of the X-Ray Absorption in
  Gamma-Ray Burst Afterglows}}.
\bjtitle{\apj}
\bvolume{768},
\bfpage{23}
(\byear{2013}).
doi:\doiurl{10.1088/0004-637X/768/1/23}
\end{barticle}
\endbibitem

\bibitem[\protect\citeauthoryear{{Waxman} and {Draine}}{2000}]{Waxman2000}
\begin{barticle}
\bauthor{\binits{E.} \bsnm{{Waxman}}},
\bauthor{\binits{B.T.} \bsnm{{Draine}}},
\batitle{{Dust Sublimation by Gamma-ray Bursts and Its Implications}}.
\bjtitle{\apj}
\bvolume{537},
\bfpage{796}--\blpage{802}
(\byear{2000}).
doi:\doiurl{10.1086/309053}
\end{barticle}
\endbibitem

\bibitem[\protect\citeauthoryear{{Wolf} and {Podsiadlowski}}{2007}]{Wolf+2007}
\begin{barticle}
\bauthor{\binits{C.} \bsnm{{Wolf}}},
\bauthor{\binits{P.} \bsnm{{Podsiadlowski}}},
\batitle{{The metallicity dependence of the long-duration gamma-ray burst rate
  from host galaxy luminosities}}.
\bjtitle{\mnras}
\bvolume{375},
\bfpage{1049}--\blpage{1058}
(\byear{2007}).
doi:\doiurl{10.1111/j.1365-2966.2006.11373.x}
\end{barticle}
\endbibitem

\bibitem[\protect\citeauthoryear{{Woosley} and {Heger}}{2006}]{Woosley+2006}
\begin{barticle}
\bauthor{\binits{S.E.} \bsnm{{Woosley}}},
\bauthor{\binits{A.} \bsnm{{Heger}}},
\batitle{{The Progenitor Stars of Gamma-Ray Bursts}}.
\bjtitle{\apj}
\bvolume{637},
\bfpage{914}--\blpage{921}
(\byear{2006}).
doi:\doiurl{10.1086/498500}
\end{barticle}
\endbibitem

\bibitem[\protect\citeauthoryear{{Yoon} and {Langer}}{2005}]{Yoon+2005}
\begin{barticle}
\bauthor{\binits{S.-C.} \bsnm{{Yoon}}},
\bauthor{\binits{N.} \bsnm{{Langer}}},
\batitle{{Evolution of rapidly rotating metal-poor massive stars towards
  gamma-ray bursts}}.
\bjtitle{\aap}
\bvolume{443},
\bfpage{643}--\blpage{648}
(\byear{2005}).
doi:\doiurl{10.1051/0004-6361+20054030}
\end{barticle}
\endbibitem

\bibitem[\protect\citeauthoryear{{Yoon} et~al.}{2006}]{Yoon+2006}
\begin{barticle}
\bauthor{\binits{S.-C.} \bsnm{{Yoon}}},
\bauthor{\binits{N.} \bsnm{{Langer}}},
\bauthor{\binits{C.} \bsnm{{Norman}}},
\batitle{{Single star progenitors of long gamma-ray bursts. I. Model grids and
  redshift dependent GRB rate}}.
\bjtitle{\aap}
\bvolume{460},
\bfpage{199}--\blpage{208}
(\byear{2006}).
doi:\doiurl{10.1051/0004-6361+20065912}
\end{barticle}
\endbibitem

\bibitem[\protect\citeauthoryear{{Zafar} and {Watson}}{2013}]{Zafar2013}
\begin{barticle}
\bauthor{\binits{T.} \bsnm{{Zafar}}},
\bauthor{\binits{D.} \bsnm{{Watson}}},
\batitle{{The metals-to-dust ratio to very low metallicities using GRB and QSO
  absorbers; extremely rapid dust formation}}.
\bjtitle{\aap}
\bvolume{560},
\bfpage{26}
(\byear{2013}).
doi:\doiurl{10.1051/0004-6361/201321413}
\end{barticle}
\endbibitem

\bibitem[\protect\citeauthoryear{{Zafar} et~al.}{2011}]{Zafar2011}
\begin{barticle}
\bauthor{\binits{T.} \bsnm{{Zafar}}},
\bauthor{\binits{D.} \bsnm{{Watson}}},
\bauthor{\binits{J.P.U.} \bsnm{{Fynbo}}},
\bauthor{\binits{D.} \bsnm{{Malesani}}},
\bauthor{\binits{P.} \bsnm{{Jakobsson}}},
\bauthor{\binits{A.} \bsnm{{de Ugarte Postigo}}},
\batitle{{The extinction curves of star-forming regions from z = 0.1 to 6.7
  using GRB afterglow spectroscopy}}.
\bjtitle{\aap}
\bvolume{532},
\bfpage{143}
(\byear{2011}).
doi:\doiurl{10.1051/0004-6361/201116663}
\end{barticle}
\endbibitem

\bibitem[\protect\citeauthoryear{{Zahid} et~al.}{2014}]{Zahid+2014}
\begin{barticle}
\bauthor{\binits{H.J.} \bsnm{{Zahid}}},
\bauthor{\binits{G.I.} \bsnm{{Dima}}},
\bauthor{\binits{R.-P.} \bsnm{{Kudritzki}}},
\bauthor{\binits{L.J.} \bsnm{{Kewley}}},
\bauthor{\binits{M.J.} \bsnm{{Geller}}},
\bauthor{\binits{H.S.} \bsnm{{Hwang}}},
\bauthor{\binits{J.D.} \bsnm{{Silverman}}},
\bauthor{\binits{D.} \bsnm{{Kashino}}},
\batitle{{The Universal Relation of Galactic Chemical Evolution: The Origin of
  the Mass-Metallicity Relation}}.
\bjtitle{\apj}
\bvolume{791},
\bfpage{130}
(\byear{2014}).
doi:\doiurl{10.1088/0004-637X/791/2/130}
\end{barticle}
\endbibitem

\end{thebibliography}


\end{document}